\documentstyle[12pt]{article}   

\large

\makeatletter
\@addtoreset{equation}{section}
\makeatother

\newtheorem{Theorem}{Theorem}[section]
\newtheorem{Definition}{Definition}[section]
\newtheorem{Lemma}{Lemma}[section]
\newtheorem{Corollary}{Corollary}[section]

\def\be{\begin{equation}}
\def\ee{\end{equation}}
\def\ba{\begin{eqnarray}}
\def\ea{\end{eqnarray}}

\def\a{{\cal A}}

\def\ab{\overline{\a}}

\def\Co{{\mathchoice
{\setbox0=\hbox{$\displaystyle\rm C$}\hbox{\hbox to0pt
{\kern0.4\wd0\vrule height0.9\ht0\hss}\box0}}
{\setbox0=\hbox{$\textstyle\rm C$}\hbox{\hbox to0pt
{\kern0.4\wd0\vrule height0.9\ht0\hss}\box0}}
{\setbox0=\hbox{$\scriptstyle\rm C$}\hbox{\hbox to0pt
{\kern0.4\wd0\vrule height0.9\ht0\hss}\box0}}
{\setbox0=\hbox{$\scriptscriptstyle\rm C$}\hbox{\hbox to0pt
{\kern0.4\wd0\vrule height0.9\ht0\hss}\box0}}}}

\def\Rl{{\mathchoice
{\setbox0=\hbox{$\displaystyle\rm R$}\hbox{\hbox to0pt
{\kern0.4\wd0\vrule height0.9\ht0\hss}\box0}}
{\setbox0=\hbox{$\textstyle\rm R$}\hbox{\hbox to0pt
{\kern0.4\wd0\vrule height0.9\ht0\hss}\box0}}
{\setbox0=\hbox{$\scriptstyle\rm R$}\hbox{\hbox to0pt
{\kern0.4\wd0\vrule height0.9\ht0\hss}\box0}}
{\setbox0=\hbox{$\scriptscriptstyle\rm R$}\hbox{\hbox to0pt
{\kern0.4\wd0\vrule height0.9\ht0\hss}\box0}}}}

\def\Nl{{\mathchoice
{\setbox0=\hbox{$\displaystyle\rm N$}\hbox{\hbox to0pt
{\kern0.4\wd0\vrule height0.9\ht0\hss}\box0}}
{\setbox0=\hbox{$\textstyle\rm N$}\hbox{\hbox to0pt
{\kern0.4\wd0\vrule height0.9\ht0\hss}\box0}}
{\setbox0=\hbox{$\scriptstyle\rm N$}\hbox{\hbox to0pt
{\kern0.4\wd0\vrule height0.9\ht0\hss}\box0}}
{\setbox0=\hbox{$\scriptscriptstyle\rm N$}\hbox{\hbox to0pt
{\kern0.4\wd0\vrule height0.9\ht0\hss}\box0}}}}

\def\Zl{{\mathchoice
{\setbox0=\hbox{$\displaystyle\rm Z$}\hbox{\hbox to0pt
{\kern0.4\wd0\vrule height0.9\ht0\hss}\box0}}
{\setbox0=\hbox{$\textstyle\rm Z$}\hbox{\hbox to0pt
{\kern0.4\wd0\vrule height0.9\ht0\hss}\box0}}
{\setbox0=\hbox{$\scriptstyle\rm Z$}\hbox{\hbox to0pt
{\kern0.4\wd0\vrule height0.9\ht0\hss}\box0}}
{\setbox0=\hbox{$\scriptscriptstyle\rm Z$}\hbox{\hbox to0pt
{\kern0.4\wd0\vrule height0.9\ht0\hss}\box0}}}}

\def\dprime{{\prime\prime}}

\title{Gauge Field Theory Coherent States (GCS) : IV.\\ 
Infinite Tensor Product\\ and Thermodynamical Limit}
\author{T. Thiemann\thanks{thiemann@aei-potsdam.mpg.de}, 
O. Winkler\thanks{winkler@aei-potsdam.mpg.de}\\
       MPI f. Gravitationsphysik, Albert-Einstein-Institut, \\
           Am M\"uhlenberg 1, 14476 Golm near Potsdam, Germany} 
\date{{\small Preprint AEI-2000-030}}   

\begin{document}

\maketitle

\begin{abstract}
In the canonical approach to Lorentzian Quantum General Relativity in four 
spacetime dimensions an important step forward has been made by Ashtekar,
Isham and Lewandowski some eight years ago through the introduction of
a Hilbert space structure which was later proved to be a 
faithful representation of the canonical commutation and adjointness 
relations of the quantum field algebra of diffeomorphism invariant
gauge field theories by Ashtekar, Lewandowski, Marolf, Mour\~ao and Thiemann.

This Hilbert space, together with its generalization due to Baez and Sawin, 
is appropriate for semi-classical quantum general relativity
if the spacetime is spatially compact. In the spatially non-compact case, 
however, an extension of the Hilbert space is needed in order to 
approximate metrics that are macroscopically nowhere degenerate. 

For this purpose, in this paper we apply the theory of the Infinite Tensor 
Product (ITP) of 
Hilbert Spaces, developed by von Neumann more than sixty years ago, to 
Quantum General Relativity. The cardinality of the number of tensor product 
factors can take the value of any possible Cantor
aleph, making this mathematical theory well suited to our problem in which
a Hilbert space is attached to each edge of an arbitrarily complicated, 
generally infinite graph.
 
The new framework opens a pandora's box full of techniques, appropriate to 
pose fascinating physical questions such as quantum topology change, 
semi-classical quantum gravity, effective low energy physics
etc. from the universal point of view of the ITP. In particular, the study
of photons and gravitons propagating on fluctuating quantum spacetimes is now
in reach, the topic of the next paper in this series.
\end{abstract}

\section{Introduction}
\label{s1}

Quantum General Relativity (QGR) has matured over the past decade to a 
mathematically well-defined theory of quantum gravity. 
In contrast to string theory, by definition QGR is a
manifestly background independent, diffeomorphism 
invariant and non-perturbative theory.
The obvious advantage is that one will never have to postulate the
existence of a non-perturbative extension of the theory,
which in string theory has been called the still unknown 
M(ystery)-Theory.

The disadvantage of a non-perturbative and background independent
formulation is, of course, that one is faced with new and interesting 
mathematical problems so that one cannot just go ahead and 
``start calculating scattering amplitudes'': 
As there is no background around which one could perturb, rather the full 
metric is fluctuating, one is not
doing quantum field theory on a spacetime but only on a differential
manifold. Once there is no (Minkowski) metric at our disposal, one loses
familiar notions such as causality structure, locality, Poincar\'e group 
and so forth, in other words, the theory is not a theory to which
the Wightman axioms apply. Therefore, one must build an entirely
new mathematical apparatus to treat the resulting quantum field theory 
which is {\it drastically different from the Fock space picture 
to which particle physicists are used to}.

As a consequence, the mathematical formulation of the theory was the main 
focus of research in the field over the past decade. The main 
achievements to date are the following (more or less in chronological 
order) : 
\begin{itemize}
\item[i)] {\it Kinematical Framework}\\
The starting 
point was the introduction of new field variables \cite{1} for the 
gravitational field which are better suited to a background  
independent formulation of the quantum theory than the ones employed
until that time. In its original version these variables were
complex valued, however, currently their real valued version,  
considered first in \cite{1a} for {\it classical} Euclidean gravity and 
later in \cite{1b} for {\it classical} Lorentzian gravity, is preferred 
because 
to date it seems that it is only with these variables that one can rigorously
define the kinematics andf dynamics of Euclidean or Lorentzian  {\it 
quantum} gravity \cite{1c}. \\
These variables are coordinates for the infinite dimensional phase
space of an $SU(2)$ gauge theory subject to further constraints 
besides the Gauss law, that is, a connection and a canonically
conjugate electric field. As such, it is very natural to introduce
smeared functions of these variables, specifically Wilson loop and 
electric flux functions. (Notice that one does not need a metric 
to define these functions, that is, they are background independent).
This had been done for ordinary gauge fields already before in \cite{2} 
and was then reconsidered for gravity (see e.g. \cite{3}).\\
The next step was the choice of a representation of the canonical
commutation relations between the electric and magnetic degrees
of freedom. This involves the choice of a suitable space of 
distributional connections \cite{4} and a faithful measure thereon \cite{5}
which, as one can show \cite{6}, is $\sigma$-additive. The corresponding
$L_2$ Hilbert space and its generalization \cite{6a}
will be henceforth called the Ashtekar-Isham-Lewandowski-Baez-Sawin 
(AILBS) Hilbert space.
The proof that the AILBS Hilbert space indeed solves the adjointness 
relations induced by the reality structure of the classical theory
as well as the canonical commutation relations induced by the symplectic 
structure of the classical theory can be found in \cite{7}.
Independently, a second representation
of the canonical commutation relations, called the loop representation, 
had been advocated (see e.g. \cite{8} and especially \cite{8a} and references
therein)
but both representations were shown to be unitarily equivalent in
\cite{9} (see also \cite{10} for a different method of proof).\\
This is then the first major achievement : The theory is based on
a rigorously defined kinematical framework.
\item[ii)] {\it Geometrical Operators}\\
The second major achievement concerns the spectra of positive 
semi-definite, self-adjoint geometrical
operators measuring lengths \cite{11}, areas \cite{12,13}
and volumes \cite{12,14,15,16,8} of curves, surfaces and regions
in spacetime. These spectra are pure point (discete) and imply a discrete
Planck scale structure. It should be pointed out that the discreteness
is, in contrast to other approaches to quantum gravity, not put in
by hand but it is a {\it prediction} !
\item[iii)] {\it Regularization- and Renormalization Techniques}\\
The third major achievement is that there is a new 
regularization and renormalization technique \cite{17,18}
for diffeomorphism covariant, density-one-valued operators at our disposal
which was successfully tested in model theories \cite{19}. This
technique can be applied, in particular, to the standard model
coupled to gravity \cite{20,21} and to the Poincar\'e generators at 
spatial infinity \cite{22}. In particular, it works for {\it Lorentzian}
gravity while all earlier proposals could at best work in the Euclidean 
context only (see, e.g. \cite{8a} and references therein).
The algebra of important operators of the
resulting quantum field theories was shown to be consistent \cite{23}. 
Most surprisingly, these operators are {\it UV and IR finite} !
Notice that, at least as far as these operators are concerned, this result 
is stronger than the believed but 
unproved finiteness of scattering amplitudes
order by order in perturbation theory of the five critical
string theories, in a sense we claim that the perturbation series converges.
The absence of the divergences that usually plague interacting quantum fields
propagating on a Minkowski background can be understood intuitively
from the diffeomorphism invariance of the theory : ``short and long distances
are gauge equivalent''. We will elaborate more on this point in future 
publications. 
\item[iv)] {\it Spin Foam Models}\\
After the construction of the densely defined Hamiltonian constraint
operator of \cite{17,18}, a formal, Euclidean functional integral was
constructed in \cite{23a} and gave rise to the so-called spin foam 
models   
(a spin foam is a history of a graph with faces as the history of edges)
\cite{23b}. Spin foam models are in close connection with causal
spin-network evolutions \cite{23c}, state sum models \cite{23d} and
topological quantum field theory, in particular BF theory \cite{23e}. To
date most results are at a formal level and for the Euclidean version of the
theory only but the programme is exciting since it may restore manifest
four-dimensional diffeomorphism invariance which in the Hamiltonian
formulation is somewhat hidden.
\item[v)]
Finally, the fifth major achievement is the existence of a rigorous and 
satisfactory framework \cite{24,25,26,27,28,29,30} for the quantum 
statistical description of black holes
which reproduces the Bekenstein-Hawking Entropy-Area relation and applies,
in particular, to physical Schwarzschild black holes while stringy black 
holes so far are under control only for extremal charged black holes.
\end{itemize}
Summarizing, the work of the past decade has now 
culminated in a promising starting point for a quantum theory of the 
gravitational field plus matter and the stage is set to pose and answer 
physical questions. 

The most basic and most important question that one should ask is :
{\it Does the theory have classical general relativity as its classical
limit ?} Notice that even if the answer is negative, the existence
of a consistent, interacting, diffeomorphism invariant quantum field theory 
in four dimensions is already a quite non-trivial result. However, we can 
claim to have a satisfactory quantum theory of Einstein's theory
only if the answer is positive. 

In order to address this question with a mathematically well-defined 
procedure we have developed in \cite{31,32} a theory of coherent states 
for the matter content of the standard model (with possible
supersymmetric extensions) coupled to gravity. These states are labelled 
by classical solutions to the field equations and have the property that
a) the expectation values of densely defined field operators with 
respect to these states take the value prescribed by the classical 
solution and b) they saturate the Heisenberg uncertainty bound without 
quenching.

The way this has been achieved so far is the following :
The degrees of freedom of, say, the gravitational field, are labelled by
piecewise analytic (smooth) graphs (webs) composed of a {\it finite} 
number of edges (paths) only.
For each such graph one finds a subspace of the AILBS
Hilbert space which is the {\it finite} tensor product of mutually 
isomorphic Hilbert spaces, one for each edge (path) of the graph (web). 
The closure
of finite linear combinations of vectors from these subspaces labelled 
by graphs (webs), which turn out to be mutually orthogonal, is 
forms the AILBS Hilbert space. What has been done in 
\cite{31,32} is to develop a theory of coherent states for each of these
Hilbert spaces labelled by a finite graph $\gamma$. More precisely, one 
constructs coherent states $\psi^s_e$ for each of the Hilbert spaces 
labelled by a single edge $e$ of $\gamma$ and the classicality parameter 
$s$ ($s\to 0$ is the classical limit) and then the coherent state
for the whole graph $\gamma$ is simply the tensor product of those for 
each of its edges. 

This framework is sufficient if the initial data hypersurface $\Sigma$ is 
compact since one can describe the quantum metric as precisely as one 
wishes in terms of finite graphs by taking the graph to be finer
and finer, filling $\Sigma$ more and more densely. However, if $\Sigma$ is 
non-compact, 
say of the topology of $\Rl^3$ as required for Minkwoski space or the Kruskal
extension of the Schwarzschild spacetime which in turn are the most 
important spacetimes if we want to make contact with the low energy 
physics of the standard model, scattering theory, Hawking radiation and thus 
the semiclassical approximation
of quantum gravity by the theory of ordinary Quantum Field Theory on 
(curved) backgrounds, then the above framework is insufficient. What one 
needs in this case is an infinite graph no matter how coarse the graph is,
that is, no matter whether the lattice spacing is 1mm or of the order of the
Planck length, in order to 
fill $\Sigma$ everywhere we need an infinite graph, no region of $\Sigma$
of infinite volume must be empty if we wish to approximate a 
non-degenerate metric as all the classical metrics are. 

One may think that one can get away by
taking an infinite superposition of states labelled by
mutually different finite graphs. However, such states have infinite norm
with respect to the AILBS scalar product as the following simple example 
shows : Namely, let $\gamma_\infty$ be a cubic lattice,
an infinte graph filling all of $\Sigma:=\Rl^3$ as densely as we wish and 
construct the state $\psi^s:=\sum_e z_e T_e$ where the sum runs over 
all edges of $\gamma_\infty$, $z_e$ are complex coefficients and $T_e$
is some linear combination of spin-network states over $e$. Obviously, this
state is an infinite linear combination of states over finite graphs.
Then, because of homogenity, this state produces the correct classical 
limit, corresponding to, say, Minkowski space, for each of the holonomy 
operators $\hat{h}_e$ at most if $z_e=z$ is independent of $e$  and 
$T_e(h_e)=T(h_e)$ is the same linear combination of spin-network states 
for each $e$. But then the norm of the state is formally 
$|z|^2 ||T||^2 \sum_e 1=\infty$ and badly diverges.

On the other hand, we will show that one can give meaning to states of the 
form $\psi^s_{\gamma_\infty}:=\otimes_e \psi^s_e$ where 
$\gamma_\infty$ is an infinite graph and if one defines the 
inner product to be the product of the inner products of the tensor product 
factors, then $||\psi^s_{\gamma_\infty}||=1$ while the semiclassical 
behaviour with respect to every possible operator over $\gamma_\infty$ is 
preserved and identical to the one for finite tensor products.  
Notice that in our case the dimension of the Hilbert space over each edge is 
countably infinite. If $\gamma_\infty$ is countably infinite,
the case to which we restrict in this paper, then the direct 
sum Hilbert space is still separable. But even if the Hilbert space over 
each edge would be only two-dimensional then the countably infinite tensor
product Hilbert space is non-separable !\\
\\
This article is organized as follows :\\

In section two we recall the basic kinematical structure of canonical  
Quantum General Relativity.

In section three we list the essential properties of our family of coherent 
states for finite tensor products as needed for the purpose of the 
present paper.

In section four we give an account of von Neumann's theory of the 
Infinite Tensor Product \cite{33} for the general case, in particular
the occurance of von Neumann algebras of different factor types
induced by the operator algebras on each tensor product factor. 

Section five contains the new results of this paper. We apply the general 
ITP theory to our situation focussing on general and abstract 
properties only. We extend the quantum kinematical framework of Ashtekar,
Isham and Lewandowski to piecewise analytical, infinite graphs,
connect it with the (semi)classical analysis for canonical quantum field 
theories over non-compact initial data hypersurfaces and finally 
discuss the transfer of dynamical results as obtained earlier for finite
graphs.

In particular, 
for any possible solution of the Einstein field equations we are able to 
identify an element of the ITP Hilbert space, a so-called $C_0$-vector
$\Omega$ in von Neumann's terminology, which in the theory of quantum fields
propagating on curved background spacetimes, plays the role
of the vacuum or ground state and which can be constructed purely
in terms of our coherent states. Perturbations of this vacuum, which in 
von Neumann's terminology lie in the subspace of the ITP Hilbert space 
generated by the strong equivalence class of the $C_0$-vector $\Omega$, 
can naturally be identified with the usual Fock states of QFT on 
the curved background that $\Omega$ approximates. This opens the 
possibility to make contact with the usual perturbation theory defined
in terms of Fock states. 

In fact, in \cite{60} we show that it is possible to map a precisely defined 
subspace of the ITP Hilbert space 
for Einstein-Maxwell theory, to the Fock space defined in terms of, say, 
$n-$Photon states propagating
on Minkowski spacetime up to corrections due to pure quantum gravity 
effects caused by the fluctuating nature of the quantum metric and which one 
hopes to measure in experiment.
More precisely, this subspace is generated by the operator algebra of the 
Maxwell field acting on a $C_0$-vector $\Omega$ of the Einstein-Maxwell
ITP Hilbert space and which is cyclic for that subspace. This vector 
$\Omega$ is a minimal uncertainty vector for Einstein-Maxwell theory
approximating the Minkowski metric and vanishing electromagnetic field
respectively. It should be noted, however, that all the states so 
constructed are states of the fully interacting Einstein-Maxwell theory and 
not only of the free Maxwell theory propagating on Minkowski space 
(an example of a free quantum field theory on a fixed curved background).
The two
sets of states so constructed are in a one-to one and onto correspondence,
leading to expectation values for physical operators which coincide 
to lowest order in the Planck length. As far as quantum gravity corrections 
are concerned, however, these states are physically very 
different, the states of the interacting theory give rise to the so-called
$\gamma-$ray-burst effect \cite{34} which is just one way to measure the 
Poincar\'e non-invariance of the present state of our universe at the 
fundamental level.
In \cite{35} we will explicitly compute the size of this 
effect from first principles by a down-to-the-ground-computation, thereby
significantly improving the results of \cite{36}.

\section{Kinematical Structure of Diffeomorphism Invariant Quantum
Gauge Theories}
\label{s2}

In this section we will recall the main ingredients of the mathematical
formulation of (Lorentzian) diffeomorphism invariant classical and quantum 
field theories of 
connections with local degrees of freedom in any dimension and for
any compact gauge group. See \cite{31,7} and references therein
for more details.

\subsection{Classical Theory}
\label{s2.1}

Let $G$ be a compact gauge group, $\Sigma$ a $D-$dimensional manifold 
and consider a principal $G-$bundle with connection over $\Sigma$.
Let us denote the pull-back (by local sections)
to $\Sigma$ of the connection by $A_a^i$
where $a,b,c,..=1,..,D$ denote tensorial indices and $i,j,k,..=1,..,
\dim(G)$ denote indices for the Lie algebra of $G$. 
Likewise, consider a vector bundle of electric fields, whose
projection to $\Sigma$ is a Lie algebra valued 
vector density of weight one. We will denote the set of generators
of the rank $N-1$ Lie algebra of $G$ by $\tau_i$ which are normalized
according to $\mbox{tr}(\tau_i\tau_j)=-N\delta_{ij}$ and 
$[\tau_i,\tau_j]=2f_{ij}\;^k\tau_k$ defines the structure constants 
of $Lie(G)$. 

Let $F^a_i$ be a Lie algebra valued vector density test field of weight one 
and let $f_a^i$ be a Lie algebra valued covector test field. 
We consider the smeared quantities
\be \label{2.1}
F(A):=\int_\Sigma d^Dx F^a_i A_a^i\mbox{ and } 
E(f):=\int_\Sigma d^Dx E^a_i f_a^i 
\ee
While both objects are diffeomorphism covariant, only the latter is gauge 
covariant, one reason to introduce the singular smearing discussed below.
The choice of the space of pairs of test fields $(F,f)\in\tilde{{\cal S}}$ 
depends on the boundary conditions on
the space of connections and electric fields which in turn depends on the 
topology of $\Sigma$ and will not be specified in what follows. 

Consider the set $M$
of all pairs of smooth functions $(A,E)$ on $\Sigma$ such that (\ref{2.1}) is 
well defined for any $(F,f)\in {\cal S}$.
We define a topology on $M$ through the globally defined metric :
\ba \label{2.2}
&& d_{\rho,\sigma}[(A,E),(A',E')]\\
&:=&\sqrt{-\frac{1}{N}\int_\Sigma d^Dx 
[\sqrt{\det(\rho)} \rho^{ab} \mbox{tr}([A_a-A'_a][A_b-A'_b])
+
\frac{[\sigma_{ab} \mbox{tr}([E^a-E^{a\prime}][E^b-E^{b\prime}])}
{\sqrt{\det(\sigma)}}]} 
\nonumber
\ea
where $\rho_{ab},\sigma_{ab}$ are fiducial metrics on $\Sigma$ of 
everywhere Euclidean signature. Their fall-off behaviour has to be suited
to the boundary conditions of the fields $A,E$ at spatial infinity
(if $\Sigma$ is spatially non-compact).
Notice that the metric (\ref{2.2}) on $M$ is gauge invariant. It can be used   
in the usual way to equip $M$ with the structure of a smooth,
infinite dimensional differential
manifold modelled on a Banach (in fact Hilbert) space $\cal E$
where ${\cal S}\times {\cal S}\subset {\cal E}$. (It is the 
weighted Sobolev space $H^2_{0,\rho}\times H^2_{0,\sigma^{-1}}$ in the 
notation of \cite{37}). 

Finally, we equip $M$ with the structure of an infinite dimensional 
symplectic manifold through the following strong (in the sense of 
\cite{38}) symplectic structure 
\be \label{2.3}
\Omega((f,F),(f',F'))_m:=\int_\Sigma d^Dx [F^a_i f^{i\prime}_a
-F^{a\prime}_i f^i_a](x)
\ee
for any $(f,F),(f',F')\in {\cal E}$. We have abused the notation by 
identifying the tangent space to $M$ at $m$ with $\cal E$. To see 
that $\Omega$ is a strong symplectic structure one uses standard 
Banach space techniques. Computing the Hamiltonian vector fields
(with respect to $\Omega$) of the functions $E(f),F(A)$ we obtain the
following elementary Poisson brackets
\be \label{2.4}
\{E(f),E(f')\}=\{F(A),F'(A)\}=0,\;\{E(f),A(F)\}=F(f)
\ee
As a first step towards quantization of the symplectic manifold
$(M,\Omega)$ one must choose a polarization. As usual in gauge theories,
we will use connections as the configuration variables and electric fields 
as canonically conjugate momenta. As a second step one must decide
on a complete set of coordinates of $M$ which are to become the elementary
quantum operators. The analysis just outlined suggests to use the 
coordinates $E(f),F(A)$. However, the well-known immediate problem is that 
these coordinates are not gauge covariant. Thus, we proceed as follows :

The idea is to construct the theory from smaller building blocks, labelled
by graphs embedded into $\Sigma$. In the literature, two sets of graphs, 
labelling the so-called
cylindrical functions, have been proposed : the set of {\it finite} piecewise 
analytical graphs $\Gamma^\omega_0$ in \cite{5} and in \cite{6a,39}
the restriction $\Gamma^\infty_0$ to so-called ``webs'' of the 
set of {\it all} piecewise smooth graphs $\Gamma^\infty$.
(We do not discuss here a third 
alternative, the set of finite piecewise linear graphs \cite{39a}).
Here we call a graph $\gamma$ finite 
if its sets of oriented edges $e$ and vertices $v$ respectively,
denoted by $E(\gamma)$ and $V(\gamma)$ respectively, have finite 
cardinality. A web is a special kind of a piecewise smooth graph 
which may not be finite but which can be obtained as the union
of a {\it finite} number of smooth curves with {\it finite range}
(the diffeomorphic image in $\Sigma$ of a closed interval in $\Rl$)
and such that its vertex set has a finite number of accumulation points.
(In fact, this is the essential difference between $\Gamma^\omega_0$ and
$\Gamma^\infty_0$ since a graph generated by a finite number of 
analytical curves is a piecewise analytical, finite graph which cannot have 
any accumulation points).
There are some additional restrictions on the common intersections of the
curves in a web which we do not need to explain here, see \cite{6a}
for all details. It is not difficult to prove that 
both $\Gamma^\omega_0,\Gamma^\infty_0$ are closed under forming {\it finite} 
numbers of intersections and unions.

In this paper we are going to extend the framework to {\it truly infinite}
graphs. That is, a priori, we do not impose any finiteness restriction 
neither on the 
number of edges or vertices of a graph nor on the range of its edges.
Various extensions are possible. A simple possibility is the set
$\Gamma^\omega$ of piecewise analytic graphs with possibly a 
countably infinite number of edges. Such graphs can still have accumulation
points of edges and vertices (e.g. the graph which looks like a ladder in a 
two-plane whose spokes are mutually parallel and come arbitrarily close to 
each other). An even simpler choice is the set $\Gamma^\omega_\sigma$ of 
piecewise analytic, $\sigma$-finite graphs which can be considered 
in locally compact manifolds $\Sigma$ (every point has a compact 
neighbourhood) which, of course, is satisfied for any finite-dimensional 
manifold that we have in mind here. They are characterized by the fact that 
$\gamma\cup U\in\Gamma^\omega_0$, i.e. the restriction of $\gamma$ to any 
compact set  
is a piecewise analytic finite graph whose number of edges is uniformly 
bounded. More precisely :
\begin{Definition} \label{def2.0}
Let $\Sigma$ be a locally compact manifold. A graph 
$\gamma\in\Gamma^\omega_\sigma$ is said to be a piecewise analytic, 
$\sigma$-finite graph, if for each compact subset $U\subset\Sigma$
the restriction of the graph is a finite graph,
$\gamma\cup U\in \Gamma^\omega_0$. Moreover,
for any compact cover $\cal U$ of $\Sigma$ the set  
$\{|E(\gamma\cap U)|;\;U\in{\cal U}\}$ is bounded.
\end{Definition}
Clearly, truly infinite piecewise analytic graphs exist only if 
$\Sigma$ is not compact and in this case $\Gamma^\omega_0$ is a
proper subset of $\Gamma^\omega$. In order to obtain maximally nice graphs 
we will make the further restriction that $\Sigma$ is paracompact,
see section \ref{s5.1}.

The next simple choice is the set 
$\Gamma^\infty$ of all piecewise smooth graphs with possibly a countable 
number of edges and possibly a countable number of accumulation points.
More properly, we should call them the set of {\it infinite webs}, that is,
the web $\gamma$ is allowed to be generated by a countably infinite number 
of smooth curves such that for each accumulation point 
$p_i,\;i=1,..,N\le\infty$ there exists a neighbourhood $U_i$ such that the 
$U_i$ are mutually disjoint and such that $\gamma$ restricted to $U_i$
is an element of $\Gamma^\infty_0$. It is a non-trivial task
to decide whether any of the three sets $\Gamma^\omega,\Gamma^\omega_\sigma,
\Gamma^\infty$ are closed under taking finite unions and we will do
this in this paper only for $\Gamma^\omega_\sigma$, leaving the remaining
cases for future publications.\\ 

Finally, we could consider  
$\Gamma$, the set of {\it all} piecewise smooth, oriented graphs $\gamma$ 
embedded into 
$\Sigma$. That is, we do not impose any restriction on the cardinality of 
the sets $E(\gamma), V(\gamma)$, or on the nature of the accumulation points.
This set is trivially closed under arbitrary unions but it is beyond present 
analytical control, furthermore, it is not clear whether $\Gamma$ and 
$\Gamma^\infty$ are 
really different and to analyze these questions is beyond the scope of 
the present paper, too.

Suffice it to say that for the purposes that we have in mind, to take the
classical limit, it is sufficient to work with the set 
$\Gamma^\omega_\sigma$ that 
is technically much easier to handle. Thus, from now on we
will assume that $\gamma\in\Gamma^\omega_\sigma$, the typical graph that we 
will need in our applications and that is good to have in mind as an example
is a regular cubic lattice in $\Rl^3$. 

Let $\gamma$ be a graph and $e$ an edge of $\gamma$. We denote by $h_e(A)$ 
the holonomy
of $A$ along $e$ and say that a function $f$ on $\a$ is cylindrical with 
respect to $\gamma$ if there exists a function $f_\gamma$ on 
$G^{|E(\gamma)|}$ such that $f=p_\gamma^\ast f_\gamma=f_\gamma\circ 
p_\gamma$ where $p_\gamma(A)=\{h_e(A)\}_{e\in E(\gamma)}$. The set
of functions cylindrical over $\gamma$ is denoted by Cyl$_\gamma$.
Holonomies are invariant under
reparameterizations of the edge and in this article we assume that
the edges are always analyticity preserving diffeomorphic images from 
$[0,1]$ to a one-dimensional submanifold of $\Sigma$ if it has compact range
and from $[0,1),\;(0,1],\;(0,1)$ if it has semi-finite
or infinite range. Gauge transformations are functions
$g:\;\Sigma\mapsto G;\;x\mapsto g(x)$ and they act on
holonomies as $h_e\mapsto g(e(0))h_e g(e(1))^{-1}$ where in the (semi)finite 
case $e(0)$ or $e(1)$ or both are not points in $\Sigma$ and we simply
set $g(e(0))=1$ or $g(e(1))=1$, which is justified by the the boundary 
conditions, restricting gauge transformations to be trivial at spatial
infinity.

Next, given a graph $\gamma$ we choose a polyhedronal decomposition
$P_\gamma$ of $\Sigma$ dual to $\gamma$. The precise definition
of a dual polyhedronal decomposition can be found in \cite{36a} but
for the purposes of the present paper it is sufficient to know that
$P_\gamma$ assigns to each edge $e$ of $\gamma$ an open ``face''
$S_e$ (a polyhedron of codimension one embedded into $\Sigma$) with 
the following properties :\\ 
(1) the surfaces $S_e$ are mutually non-intersecting,\\ 
(2) only the edge $e$ intersects $S_e$, the intersection is transversal
and consists only of one point which is an interiour point of both
$e$ and $S_e$,\\
(3) $S_e$ carries the normal orientation which agrees with the orientation 
of $e$.\\
Furthermore, we choose a system $\Pi_\gamma$ of paths $\rho_e(x) \subset
S_e,\; x\in S_e,\; e\in E(\gamma)$ connecting the intersection point 
$p_e=e\cap S_e$ with $x$. The paths vary smoothly with
$x$ and the triples $(\gamma,P_\gamma,\Pi_\gamma)$
have the property that if $\gamma,\gamma'$ are diffeomorphic, so
are $P_\gamma,P_{\gamma'}$ and $\Pi_\gamma,\Pi_{\gamma'}$.

With these structures we define the following function on $(M,\Omega)$
\be \label{2.5}
P^e_i(A,E):=-\frac{1}{N}
\mbox{tr}(\tau_i h_e(0,1/2)[\int_{S_e} h_{\rho_e(x)} \ast E(x) 
h_{\rho_e(x)}^{-1}] h_e(0,1/2)^{-1})
\ee
where $h_e(s,t)$ denotes the holonomy of $A$ along $e$ between the 
parameter values $s<t$, $\ast$ denotes the Hodge dual, that is,
$\ast E$ is a $(D-1)-$form on $\Sigma$, $E^a:=E^a_i\tau_i$ and
we have chosen a parameterization of $e$ such that $p_e=e(1/2)$.

Notice that in contrast to similar variables used earlier in the literature
the function $P^e_i$ is {\it gauge covariant}. Namely, under gauge 
transformations it transforms as $P^e\mapsto g(e(0)) P^e g(e(0))^{-1}$,
the price to pay being that $P^e$ depends on both $A$ and $E$ and not 
only on $E$. The idea is therefore to use the variables $h_e,P^e_i$
for all possible graphs $\gamma$ as the coordinates of $M$.

The problem with the functions $h_e(A)$ and $P^e_i(A,E)$ on $M$ is that 
they are not differentiable on $M$, that is, $Dh_e, DP^e_i$ are nowhere  
bounded operators on $\cal E$ as one can easily see. The reason for this is,
of course, that these are functions on $M$ which are not properly smeared 
with functions from $\tilde{{\cal S}}$, rather they are smeared with 
distributional
test functions with support on $e$ or $S_e$ respectively. Nevertheless,
one would like to base the quantization of the theory on these functions 
as basic variables because of their gauge and diffeomorphism covariance.
Indeed, under diffeomorphisms $h_e\mapsto h_{\varphi^{-1}(e)},
P^e_i\mapsto P^{\varphi^{-1}(e)}$ where we abuse notation since 
$P^e$ depends also on $S_e,\rho_e$, see \cite{36a} for more details.
We proceed as follows. %
\begin{Definition} \label{def2.1}
By $\bar{M}_\gamma$ we denote the direct product 
$[G\times Lie(G)]^{|E(\gamma)|}$. 
The subset of $\bar{M}_\gamma$ of pairs 
$(h_e(A),P^e_i(A,E))_{e\in E(\gamma)}$ as 
$(A,E)$ varies over $M$ will be denoted by $(\bar{M}_\gamma)_{|M}$. We have a 
corresponding map $p_\gamma:\;M\mapsto \bar{M}_\gamma$ which
maps $M$ onto $(\bar{M}_{\gamma})_{|M}$.
\end{Definition}
Notice that the set $(\bar{M}_\gamma)_{|M}$ is, in general, a proper subset 
of $\bar{M}_\gamma$,
depending on the boundary conditions on $(A,E)$, the topology of $\Sigma$ 
and the ``size'' of $e,S_e$. For instance, in the limit of $e,S_e\to  
e\cap S_e$ but holding the number of edges fixed, $(\bar{M}_\gamma)_{|M}$  
will consist of only one point in $\bar{M}_\gamma$. This follows from the 
smoothness of the $(A,E)$. 

We equip a subset $M_\gamma$ of $\bar{M}_\gamma$ with the structure of a 
differentiable manifold modelled
on the Banach space ${\cal E}_\gamma=\Rl^{2\dim(G)|E(\gamma)|}$
by using the natural direct product manifold structure of
$[G\times Lie(G)]^{|E(\gamma)|}$. While $\bar{M}_\gamma$ is a kind of 
distributional phase space, $M_\gamma$ has suitable regularity properties
similar to (\ref{2.2}).

In order to proceed and to give $M_\gamma$ a symplectic structure 
{\it derived from $(M,\Omega)$} one must 
regularize the elementary functions $h_e, P^e_i$ by writing them as limits  
(in which the regulator vanishes) of functions which can be expressed  
in terms of the $F(A),E(f)$. Then one can compute their Poisson
brackets with respect to the symplectic structure $\Omega$ at finite
regulator and then take the limit pointwise on $M$. The result is the 
following  
well-defined strong symplectic structure $\Omega_\gamma$ on $M_\gamma$. 
\ba \label{2.6}
\{h_e,h_{e'}\}_\gamma &=& 0\nonumber\\
\{P^e_i,h_{e'}\}_\gamma &=&
\delta^e_{e'} \frac{\tau_i}{2}h_e\nonumber\\
\{P^e_i,P^{e'}_j\}_\gamma &=&
-\delta^{ee'}f_{ij}\;^k P^e_k
\ea
Since $\Omega_\gamma$ is obviously block diagonal, each block standing
for one copy of $G\times Lie(G)$, to check that $\Omega_\gamma$ is 
non-degenerate and closed reduces to doing it for each factor together
with an appeal to well-known Hilbert space techniques to establish that
$\Omega_\gamma$ is a surjection of ${\cal E}_\gamma$.
This is done in \cite{36a} where it is shown that each copy is isomorphic
with the cotangent bundle $T^\ast G$ equipped with the symplectic structure
(\ref{2.6}) (choose $e=e'$ and delete the label $e$). \\
\\
Now that we have managed to assign to each graph $\gamma$ a symplectic
manifold $(M_\gamma,\Omega_\gamma)$ we can quantize it by using geometric
quantization. This can be done in a well-defined way because the relations
(\ref{2.6}) show that the corresponding operators are non-distributional.
This is therefore a clean starting point for the regularization of any 
operator
of quantum gauge field theory which can always be written in terms 
of the $\hat{h}_e,\hat{P}^e,\;e\in E(\gamma)$ if we apply this operator to
a function which depends only on the $h_e,\; e\in E(\gamma)$. 

The question is what $(M_\gamma,\Omega_\gamma)$ has to do with $(M,\Omega)$.
In \cite{36a} it is shown that there exists a partial order $\prec$ on the 
set of triples $(\gamma,P_\gamma,\Pi_\gamma)$ and one can form
a generalized projective limit $M_\infty$ of the $M_\gamma$
(in particular, $\gamma\prec\gamma'$ means $\gamma\subset\gamma'$).
Moreover, the family 
of symplectic structures $\Omega_\gamma$ is self-consistent
in the sense that if 
$(\gamma,P_\gamma,\Pi_\gamma)\prec (\gamma',P_{\gamma'},\Pi_{\gamma'})$ 
then $p_{\gamma'\gamma}^\ast\{f,g\}_\gamma
=\{p_{\gamma'\gamma}^\ast f,p_{\gamma'\gamma}^\ast g\}_{\gamma'}$
for any $f,g\in C^\infty(M_\gamma)$ and 
$p_{\gamma'\gamma}:\;M_{\gamma'}\mapsto M_\gamma$ is a natural projection. 

Now, via the maps $p_\gamma$ of definition \ref{def2.1} we can identify
$M$ with a subset of $M_\infty$. Moreover, in \cite{36a} it is shown that
there is a generalized projective sequence $(\gamma_n,P_{\gamma_n},
\Pi_{\gamma_n})$
such that $\lim_{n\to\infty}p_{\gamma_n}^\ast\Omega_{\gamma_n}=\Omega$
pointwise in $M$. This displays $(M,\Omega)$ as embedded into a 
generlized projective
limit of the $(M_\gamma,\Omega_\gamma)$, intuitively speaking, as $\gamma$ 
fills all of $\Sigma$, we recover $(M,\Omega)$ from the 
$(M_\gamma,\Omega_\gamma)$. On non-compact manifolds $\Sigma$ this is 
possible only if the label set $\Gamma^\omega_\sigma$ contains infinite 
graphs.

It follows that quantization of $(M,\Omega)$, and conversely taking the 
classical limit, can be studied purely in terms of $M_\gamma,\Omega_\gamma$
for {\it all} $\gamma$. The quantum kinematical framework for this will be 
given in the next subsection.

\subsection{Quantum Theory}
\label{s2.2}

At this point there is a clash with the previous subsection because 
the quantum kinematical structure has so far been defined only 
for the {\it finite} category of graphs $\Gamma^\omega_0$. We thus 
have to extend this framework which we will do in section \ref{s5.1}.
However, as the structure from $\Gamma^\omega_0$ can be nicely embedded into
the more general context, we will repeat here the relevant notions
for finite, piecewise analytical graphs $\gamma$.\\

Let us denote the set of all smooth connections by $\a$. This is our
classical configuration space and we will choose for its coordinates the
holonomies $h_e(A),\;e\in\gamma,\;\gamma\in\Gamma^\omega_0$. $\a$ is 
naturally equipped with a metric topology induced by (\ref{2.2}). 

Recall the notion of a function cylindrical over a graph from the 
previous subsection.
A particularly useful set of cylindrical functions are the so-called 
spin-netwok functions \cite{40,41,9} which so far have been introduced
only for $\Gamma^\omega_0$, in fact, it is unclear whether one can
define spin-network functions for all elements of $\Gamma^\infty_0$,
see \cite{39} for a discussion. As we will see in section \ref{s5},
the spin-network bases proves to be of modest practical use in the 
context of $\Gamma^\omega_\sigma$ only, to be replaced by what we will 
call {\it von Neumann bases based on $C_0$-vectors}. To see what
the problem is, we anyway have to introduce them here.

A spin-network function is 
labelled by a graph $\gamma\in \Gamma^\omega_0$, a set of non-trivial 
irreducible representations 
$\vec{\pi}=\{\pi_e\}_{e\in E(\gamma)}$ (choose from each equivalence 
class of equivalent
representations once and for all a fixed representant), one for each 
edge of $\gamma$, and a set $\vec{c}=\{c_v\}_{v\in V(\gamma)}$ of
contraction matrices, one for each vertex of $\gamma$, which 
contract the indices of the tensor product 
$\otimes_{e\in E(\gamma)} \pi_e(h_e)$ in such a way that the resulting
function is gauge invariant. We denote spin-network functions as
$T_I$ where $I=\{\gamma,\vec{\pi},\vec{c}\}$ is a compound label.
One can show that these functions are linearly independent.
From now on we denote by $\tilde{\Phi}_\gamma$ finite linear combinations of
spin-network functions over $\gamma$, by $\Phi_\gamma$ the finite linear 
combinations of elements from any possible $\tilde{\Phi}_{\gamma'},\;
\gamma'\subset\gamma$ a subgraph of $\gamma$  
and by $\Phi$ the finite linear 
combinations of spin-network functions from an arbitrary finite collection 
of graphs. Clearly $\tilde{\Phi}_{\gamma}$ is a subspace of 
$\Phi_\gamma$ which by itself is a proper subspace of the set
Cyl$^\infty_\gamma$ of smooth cylindrical functions over $\gamma$.  
To express this distinction we will say that functions 
in $\tilde{\Phi}_\gamma$ are labelled by ``coloured graphs'' $\gamma$
while functions in $\Phi_\gamma$ are labelled simply by graphs $\gamma$,
abusing the notation by using the same symbol $\gamma$.

The set $\Phi$ of finite linear combinations of spin-network functions 
forms an Abelian $^\ast$ algebra 
of functions on $\a$. By completing it with respect to the sup-norm 
topology it 
becomes an Abelian C$^\ast$ algebra $\cal B$ (here the compactness of $G$ is 
crucial). The spectrum $\ab$ of this algebra, 
that is, the set of all algebraic homomorphisms ${\cal B}\mapsto\Co$
is called the quantum configuration space. This space is equipped with
the Gel'fand topology, that is, the space of continuous functions
$C^0(\ab)$
on $\ab$ is given by the Gel'fand transforms of elements of $\cal B$.
Recall that the Gel'fand transform is given by $\hat{f}(\bar{A}):=
\bar{A}(f)\;\forall \bar{A}\in \ab$. It is a general result that $\ab$ with 
this topology is a compact Hausdorff space. Obviously, the elements of
$\a$ are contained in $\ab$ and one can show that $\a$ is even dense
\cite{36}. Generic elements of $\ab$ are, however, distributional.

The idea is now to construct a Hilbert space consisting of square
integrable functions on $\ab$ with respect to some measure $\mu$. Recall 
that one can define a measure on a locally compact Hausdorff space 
by prescribing a positive linear functional $\chi_\mu$ on the space 
of continuous functions thereon. The particular measure
we choose is given by $\chi_{\mu_0}(\hat{T}_I)=1$ if $I=\{\{p\},
\vec{0},\vec{1}\}$ and $\chi_{\mu_0}(\hat{T}_I)=0$ otherwise. Here
$p$ is any point in $\Sigma$, $0$ denotes the 
trivial representation and $1$ the trivial contraction matrix. In other 
words, (Gel'fand transforms of) spin-network functions play the same role 
for $\mu_0$ as 
Wick-polynomials do for Gaussian measures and like those they form
an orthonormal basis in the Hilbert space ${\cal H}:=L_2(\ab,d\mu_0)$ 
obtained by completing their finite linear span $\Phi$.\\
An equivalent definition of $\ab,\mu_0$ is as follows :\\ 
$\ab$ is in one to one correspondence, via the surjective map $H$ defined 
below, with the set $\ab':=\mbox{Hom}({\cal X},G)$
of homomorphisms from the groupoid $\cal X$ of composable, holonomically
independent, analytical paths
into the gauge group. The correspondence is explicitly given by
$\ab\ni\bar{A}\mapsto H_{\bar{A}}\in\mbox{Hom}({\cal X},G)$
where ${\cal X}\ni e\mapsto H_{\bar{A}}(e):=\bar{A}(h_e)=
\tilde{h}_e(\bar{A})\in G$ and $\tilde{h}_e$ is the Gel'fand transform
of the function $\a\ni A\mapsto h_e(A)\in G$. Consider now the restriction
of $\cal X$ to ${\cal X}_\gamma$, the groupoid of composable edges of  
the graph $\gamma$. One can then show that the projective limit of the 
corresponding {\it cylindrical sets} 
$\ab'_\gamma:=\mbox{Hom}({\cal X}_\gamma,G)$ coincides with $\ab'$.
Moreover, we have $\{\{H(e)\}_{e\in E(\gamma)};\;H\in\ab'_\gamma\}=
\{\{H_{\bar{A}}(e)\}_{e\in E(\gamma)};\;\bar{A}\in\ab\}=
G^{|E(\gamma)|}$.
Let now $f\in{\cal B}$ be a function cylindrical over $\gamma$ then 
$$
\chi_{\mu_0}(\tilde{f})=\int_{\ab} d\mu_0(\bar{A}) \tilde{f}(\bar{A})
=\int_{G^{|E(\gamma)|}} \otimes_{e\in E(\gamma)} d\mu_H(h_e)
f_\gamma(\{h_e\}_{e\in E(\gamma)})
$$
where $\mu_H$ is the Haar measure on $G$.
As usual, $\a$ turns out to be contained in a measurable subset of 
$\ab$ which has measure zero with respect to $\mu_0$. It turns out 
that it is this definition of the measure which can be extended to
the category of infinite graphs.

Let, as before, $\Phi_\gamma$ be the finite linear span of spin-network 
functions over $\gamma$ or any of its subgraphs and ${\cal H}_\gamma$ its 
completion with respect to
$\mu_0$. Clearly, $\cal H$ itself is the completion of the finite linear
span $\Phi$ of vectors from the mutually orthogonal $\tilde{\Phi}_\gamma$. 
Our basic coordinates of $M_\gamma$ are promoted to operators on ${\cal H}$ 
with dense domain $\Phi$. As $h_e$ is group-valued and $P^e$ is real-valued
we must check that the adjointness relations coming from these reality 
conditions as well as the Poisson brackets (\ref{2.6}) are implemented on
our ${\cal H}$. This turns out to be precisely the case if we choose
$\hat{h}_e$ to be a multiplication operator and 
$\hat{P}^e_j=i\hbar\kappa X^e_j/2$
where $\kappa$ is the gravitational constant,
$X^e_j=X_j(h_e)$ and $X^j(h),\;h\in G$ is the vector field on $G$
generating left translations into the $j-th$ coordinate direction of 
$Lie(G)\equiv T_h(G)$ (the tangent space of $G$ at $h$ can be identified 
with the Lie algebra of $G$) and $\kappa$ is the coupling constant of the 
theory. For details see \cite{7,31}.

The question is now whether all of this structure can be extended to
the infinite analytic category. In particular, in what sense does 
a spin-network function converge, what is the sup-norm for a function
which is a finite linear combination of infinite products of
holonomy functions etc. Obviously, at this point one must invoke 
the theory of the Infinite Tensor Product. We therefore have to postpone 
the answer to these questions to section \ref{s5}.

\section{Gauge Field Theory Coherent States}
\label{s3}

For a rather general idea of how to obtain coherent states for 
arbitrary canonically quantized quantum (field) theories and quantum gauge
field theories in particular, see \cite{31} which is based on \cite{43}. 
Here we will stick with the 
heat kernel family initialized by the mathematician Brian Hall \cite{44}
who proved that the associated Segal-Bargmann space is unitarily
equivalent with the usual $L_2$ space. These results were extended to
the Hilbert spaces underlying cylindrical functions of section \ref{s2.2}
in \cite{45}. However, the semiclassical properties of these states 
were only later analyzed in \cite{32}.

\subsection{Compact Group Coherent States}
\label{s3.1}

We will begin with only one copy of $G$ and consider the space of
square integrable functions over $G$ with respect to the Haar measure
$d\mu_H$, that is, the Hilbert space ${\cal H}_G=L_2(G,d\mu_H)$. Let 
$s$ be a positive real number, $\pi$ a (once and for all fixed, arbitrary
representant from its equivalence class) irreducible representation, 
$\chi_\pi$ its character and $d_\pi$ its dimension. Let $\Delta$ be the 
Laplacian on $G$, then it is well known that the $\dim_\pi^2$ functions
$\pi_{AB}$ are eigenfunctions of $-\Delta$ with eigenvalue 
$\lambda_\pi\ge 0$ which vanishes if and only if $\pi$ is the trivial
representation.

Let $h\in G$ denote an element of $G$ and $g\in G_\Co$ an element of its 
complexification (for instance, if $G=SU(2)$ then $G_\Co=SL(2,\Co)$).
Then the (non-normalized) coherent state $\psi^s_g$ at classicality 
parameter $s$
and phase space point $g$ (the reason for this notation will be explained 
shortly) is defined by
\be \label{3.1}
\psi^s_g(h):=\sum_\pi d_\pi e^{-s\lambda_\pi/2} \chi_\pi(gh^{-1})
=(e^{s\Delta/2}\delta_{h'})(h)_{|h'\to g}
\ee
that is, it is given by heat kernel evolution with time parameter $s$ 
of the $\delta$-distribution on $G$ followed by analytic continuation. 

On ${\cal H}_G$ we introduce multiplication and derivative operators  
on the dense domain ${\cal D}:=C^\infty(G)$ by
\be \label{3.2}
(\hat{h}_{AB}f)(h):=h_{AB}f(h) \mbox{ and } (\hat{p}_j f)(h)
=\frac{is}{2}(X_j f)(h)
\ee
where $h_{AB}$ denote the matrix elements of the defining representation
of $G$ and $i,j,k,..=1,2,..,\dim(G)$ and 
$X_j(h)=\mbox{tr}([\tau_j h]^T\partial/\partial h)$ denotes the generator
of right translations on $G$ into the $j$'th coordinate direction of
$Lie(G)$, the Lie algebra of $G$. We choose a basis $\tau_j$ in $Lie(G)$
with respect to which tr$(\tau_j \tau_k)=-N\delta_{jk},\;
[\tau_j,\tau_k]=2f_{jk}\;^l \tau_l$ where $N-1$ is the rank of $G$.
The operators ({3.2}) enjoy the canonical commutation relations
\be \label{3.3}
[\hat{h}_{AB},\hat{h}_{CD}]=0,\;
[\hat{p}_j,\hat{h}_{AB}]=\frac{is}{2}(\tau_j \hat{h})_{AB},\;
[\hat{p}_j,\hat{p}_k]=-is f_{jk}\;^l \hat{p}_l
\ee
mirroring the classical Poisson brackets
\be \label{3.4}
\{h_{AB},h_{CD}\}=0,\;
\{p_j,h_{AB}\}=\frac{1}{2}(\tau_j h)_{AB},\;
\{p_j,p_k\}= -f_{jk}\;^l p_l
\ee
on the phase space $T^\ast G$, the cotangent bundle over $G$, where
$s$ plays the role of Planck's constant. It is easy 
to check that the CCR of ({3.3}) and the adjointness relations
coming from $\overline{p}_j=p_j,\;\overline{h_{AB}}=f_{AB}(h)$
are faithfully implemented on ${\cal H}_G$. Here, $f_{AB}$ depends on the 
group, e.g. $f_{AB}(h)=(h^{-1})_{BA}$ for $G=SU(N)$, and $\hat{p}_j$
is essentially self-adjoint with core $\cal D$.

We now consider $G$ as a subgroup of some unitary group so that the 
$\tau_j$ are antihermitean. We then
identify $G_\Co$ with $T^\ast G$ by the diffeomorphism
\be \label{3.5}
\phi:\; T^\ast G \mapsto G^\Co;\; (h,p)\mapsto g:=e^{-ip^j\tau_j/2}h=:Hh
\ee
where the inverse is simply given by the unique polar decomposition
of $g\in G_\Co$. One can show that the symplectic structure (\ref{3.4}) 
is compatible with the complex structure of $G_\Co$, displaying the complex
manifold $G_\Co$ as a K\"ahler manifold.

Next we define on $\cal D$ the annihilation and creation operators 
\be \label{3.6}
\hat{g}_{AB}:=e^{s\Delta/2}\hat{h}_{AB}e^{-s\Delta/2} \mbox{ and }
(\hat{g}_{AB})^\dagger:=e^{-s\Delta/2}f(\hat{h})_{AB}e^{s\Delta/2} 
\ee
Then, as one can show, $\hat{g}=e^{Ns/4}e^{-i\hat{p}_j\tau_j}\hat{h}$
so that the operator $\hat{g}$ qualifies as a quantization of the 
polar decomposition of $g$.

To call these operators annihilation and creation operators 
is justified by the following list of properties with respect to the 
coherent states (\ref{3.1}).
\begin{itemize}
\item[i)] {\it Eigenstate Property}\\
The states (\ref{3.1}) are simultaneous eigenstates of the operators
$\hat{g}_{AB}$ with eigenvalue $g_{AB}$.
\be \label{3.7}
\hat{g}_{AB}\psi^s_g=g_{AB} \psi^s_g
\ee
\item[ii)] {\it Expectation Value Property}\\
From this it follows immediately that the 
expectation values of the operators (\ref{3.6}) with respect
to the states (\ref{3.1}) exactly equal their classical ones as 
prescribed by the phase space point $g$.
\be \label{3.8}
\frac{<\psi^s_g,\hat{g}_{AB}\psi^s_g>}{||\psi^s_g||^2}=g_{AB},\;
\frac{<\psi^s_g,(\hat{g}_{AB})^\dagger\psi^s_g>}{||\psi^s_g||^2}=
\overline{g_{AB}}
\ee
\item[iii)] {\it Saturation of the Heisenberg Uncertainty Bound}\\ 
Consider the self-adjoint operators
$\hat{x}_{AB}=(\hat{g}_{AB}+[\hat{g}_{AB}]^\dagger)/2,\;
\hat{y}_{AB}=(\hat{g}_{AB}-[\hat{g}_{AB}]^\dagger)/(2i)$ an their 
classical counterparts analogously built from $g_{AB}$. Then
these operators saturate the Heisenberg uncertainty obstruction bound,
moreover, the coherent states are {\it unquenched} for $\hat{x},\hat{y}$.
\be \label{3.9}
<[\hat{x}_{AB}-x_{AB}]^2>^s_g=<[\hat{y}_{AB}-y_{AB}]^2>^s_g=
\frac{|<[\hat{x}_{AB},\hat{y}_{AB}]>^s_g|}{2}
\ee
where $<.>^s_g$ denotes the expectation value with respect to
$\psi^s_g$. Thus they are minimal uncertainty states.
\item[iv)] {\it Completeness and Segal-Bargmann Hilbert Space}\\
There exists a measure $\nu_s$ on $G_\Co$ and a unitary map
\be \label{3.10}
\hat{U}_s :\; {\cal H}_G\mapsto{\cal H}_{G^\Co}:=\mbox{Hol}(G_\Co)\cap
L_2(G_\Co,d\nu_s);f(h)\mapsto (\hat{U}_s f)(g):=(e^{s\Delta/2}f)(h)_{h\to g}
\ee
between ${\cal H}_G$ and the space of $\nu_s$-square integrable, holomorphic 
functions, the Segal-Bargmann space. Moreover, the coherent
states satisfy the overcompleteness relation
\be \label{3.11}
1_{{\cal H}_G}=\int_{G_\Co} d\nu_s(g) \hat{P}_{\psi^s_g}
\ee
where $\hat{P}_f$ denotes the projection onto the one-dimensional subspace 
of ${\cal H}_G$ spanned by the element $f$. 
\item[v)] {\it Peakedness Properties}\\
As usual, semiclassical 
behaviour of the system is most conveniently studied in terms of 
${\cal H}_{G_\Co}$ because wave functions depend on phase space rather
than on configuration space only. For instance, we have the peakedness 
property of the overlap function
\be \label{3.12}
\frac{|<\psi^s_g,\psi^s_{g'}>|^2}{||\psi^s_g||^2 ||\psi^s_{g'}||^2}
=e^{-\frac{F(p,p')+G(\theta,\theta')}{s}}(1-K_s(g,g'))
\ee
where $g=e^{-i p_j\tau_j/2} e^{\theta_j \tau_j}$ (and similar for $g'$)
is the polar decomposition of $g$. $K_s$ is a positive function, uniformly 
bounded by a constant $K'_s$ independent of $g,g'$ that approaches
zero exponentially fast as $s\to 0$. $F,G$ are positive definite functions 
which take the value zero if and only if $p_j=p_j'$ and 
$\theta_j=\theta_j'$, moreover for small $p_j'-p_j,\theta_j'-\theta_j$ we 
have $F(p,p')\approx (p'_j-p_j)^2,\;
G(\theta,\theta')\approx (\theta'_j-\theta_j)^2$ which shows that these
states generalize the familiar $T^\ast \Rl$ coherent states to the 
non-linear setting of $T^\ast G$. It can be shown \cite{46,32} that 
(\ref{3.12}) is the probability density, with respect to
the Liouville measure on $T^\ast G$, to find the system at the phase space 
point $g'$ if it is in the state $\psi^s_g$ and that density equals  
unity at $g=g'$ and is otherwise strongly, Gaussian suppressed
as $s\to 0$ with width $\sqrt{s}$. Similar peakedness properties can be 
established in the configuration or momentum representation.
\item[vi)] {\it Ehrenfest Theorems}\\
The expectation value property holds for the operators $\hat{g}_{AB}$
and $\hat{g}_{AB}^\dagger$ at any value of $s$. For the remaining 
operators one can show 
\be \label{3.13}
\lim_{s\to 0} <\hat{h}_{AB}>^s_g=h_{AB} \mbox{ and }
\lim_{s\to 0} <\hat{p}_j>^s_g=p_j
\ee
where $g=e^{-i p_j \tau_j/2}h$ and the convergence is exponentially fast.
The result (\ref{3.13}) extends to arbitrary polynomials of 
$\hat{h}_{AB},\hat{p}_j$ and even to non-polynomial, non-analytic
functions of $\hat{p}_j$ of the type that occur in quantum gravity,
most importantly the volume operator mentioned in the introduction.
\end{itemize}
These beautiful properties of the states introduced by Hall will be 
extended to the gauge field theory case in the next subsection.

\subsection{Graph Coherent States}
\label{s3.2}

Let $\gamma\in \Gamma^\omega_0$ be a piecewise analytic, finite graph, 
that is, with a finite
number of edges $e\in E(\gamma)$. For each edge $e$ we introduce
the functions $h_e,P^e_j$ on $(M,\omega)$ as in subsection (\ref{s2.1}).
Furthermore, we introduce the dimensionless quantities
\be \label{3.14}
p^e_j:=\frac{P^e_j}{a^{n_D}} \mbox{ and } s:=\frac{\hbar\kappa}{a^{n_D}}
\ee
Here $n_D=n'_D$ if $n'_D\not=0$ and $n_D=1$ otherwise where $n'_D=D-3$
for Yang-Mills theory and $n'_D=D-1$ for general relativity. Furthermore,
if $n'_D\not=0$ then 
$a$ is some fixed, arbitrary parameter of the dimension of a length
(e.g. $a=$1cm), if $n'_D=0$ then $a$ is dimensionfree and 
$\hbar\kappa$ is 
the Feinstruktur constant. Then the Poisson bracket relations of 
(\ref{2.6}) become \ba \label{3.15}
{[}\hat{h}_e,\hat{h}_{e'}]_\gamma &=& 0\nonumber\\
{[}\hat{p}^e_j,\hat{h}_{e'}]_\gamma &=&
is \delta^e_{e'} \frac{\tau_j}{2}\hat{h}_e\nonumber\\
{[}\hat{p}^e_i,\hat{p}^{e'}_j]_\gamma &=&
-is\delta^{ee'}f_{ij}\;^k \hat{p}^e_k
\ea
where the notation $[.,.]_\gamma$ indicates that all operators are 
restricted to the subspace ${\cal H}_\gamma$ of $\cal H$. It is trivial 
to see that these relations classically carry over from the category 
$\Gamma^\omega_0$ to the category $\Gamma^\omega_\sigma$.

We can now introduce the graph coherent states
\be \label{3.16}
\psi^s_{\gamma,\vec{g}}(\vec{h}):=\prod_{e\in E(\gamma)} \psi^s_{g_e}(h_e)
\ee
which are obviously neither gauge invariant nor diffeomorphism invariant.
In \cite{31} it was indicated how to obtain diffeomorphism invariant
coherent states from those in (\ref{3.16}) and in \cite{32} the same was 
done in order to obtain gauge invariant ones, employing the group averaging
technique \cite{7}. Since at the moment we are interested in issues
related to the classical limit of the theory, in particular,
whether we obtain in the classical limit the classical
Einstein equations in an appropriate sense, we will not use those invariant 
states for two reasons :\\
1) In order to check the correctness of the classical 
limit we must verify, in particular, whether the quantum constraint algebra
of the the quantum theory becomes the Dirac algebra in the classical limit.
However, one cannot check an algebra on its kernel, see \cite{23} for a 
discussion.\\
2) As far as the gauge -- and diffeomorphism constraint are concerned,
it is perfectly fine to work with non-invariant coherent states because
the corresponding gauge groups are represented as unitarily on the 
Hilbert space. This implies that expectation values of gauge -- and
diffeomorphism invariant operators are automatically also gauge -- and
diffeomorphism invariant and so qualify as expectation values of the reduced 
theory. Famously, the time reparameterizations associated with the 
Hamiltonian constraint of the theory cannot be unitarily represented
and so the argument just given does not carry over to operators
commuting with the Hamiltonian constraint. Presumably, the Hamiltonian 
constraint cannot be exponentiated at all and one will then have to 
work with its infinitesimal version. To pass then to the reduced theory
one would need to work with coherent states that are annihilated by the 
Hamiltonian constraint (trivial representation of the ``would be 
time reparameterization group'').\\

The coherent states (\ref{3.16}) then form a valid starting point for
adressing semiclassical questions in the case that $\Sigma$ is
compact, say, in some cosmological situations. To cover the case that
$\Sigma$ is asymptotically flat we must blow up the framework and pass
to the Infinite Tensor Product.

\newpage

\section{The Abstract Infinite Tensor Product}
\label{s4}

\begin{verse} 
{\it Bei Systemen mit N Teilchen ist der Hilbertraum das Tensorprodukt
von den N Hilbertr\"aumen der einzelnen Teilchen.\\
Das unendliche Tensorprodukt \"offnet die T\"ur zu den
mathematischen Finessen der Feldtheorie.\\
~~~~~~~~~~~~~~~~~~~~~~~~~~~~~~~~~~~~~~~(Walter Thirring)}
\end{verse}

The Infinite Tensor Product (ITP) of Hilbert spaces is a standard construction 
in statistical physics (through the thermodynamic (or infinite volume) limit) 
as well as in Operator Theory (von Neumann Algebras). In fact,
the first examples of von Neumann algebras which are not of factor type
$I_n, I_\infty$ (isomorphic to an algebra of bounded operators on 
a (separable) Hilbert space) have been constructed by using the ITP.

On the other hand, since the concept of separable (Fock) Hilbert spaces plays 
such a dominant role in high energy physics, presumably many theoretical 
physicists belonging to that community have 
never come across the concept of the Infinite Tensor Product (ITP) of Hilbert 
spaces which produces a non-separable Hilbert space in general.
In fact, let us quote from
Streater\&Wightman, \cite{47} p. 86, 87 in that respect :\\
{\it ``...It is sometimes argued that in quantum field theory one is 
dealing with a system of an infinite number of degrees of freedom and so must 
use a non-separable Hilbert space..... Our next task is to explain why this 
is wrong, or at best is grossly misleading.....All these arguments make it
clear that that there is no evidence that separable Hilbert spaces are
not the natural state spaces for quantum field theory....''}\\

Because of this, we have decided to include here a rough account of the
most important concepts associated with the abstract Infinite Tensor Product.
As it will become clear shortly, the ITP decomposes into an uncountable 
direct sum of Hilbert spaces which in most applications are separable.
Each of these tiny subspaces of the complete ITP are isomorphic 
with the usual Fock spaces of quantum field theory on Minkowski space 
(or some other background). Presumably, the 
fact that one
can do with separable Hilbert spaces in ordinary QFT is directly related to 
the fact that one {\it fixes the background} since this fixes the vacuum. 
The necessity to deal with the 
full ITP in quantum gravity could therefore be based on the fact that, in a 
sense, one {\it has to consider all possible backgrounds at once} ! 
More precisely, the metric cannot be fixed to equal a given background but 
becomes itself a fluctuating quantum operator. \\
\\
We follow the beautiful and comprehensive exposition by von Neumann 
\cite{33} who invented the Infinite Tensor Product (ITP) more than sixty 
years ago already. The reader is 
recommended to consult this work for more details.

\subsection{Definition of the Infinite Tensor Product of Hilbert Spaces}
\label{s4.1}

Let $\cal I$ be some set of indices $\alpha$. 
We will not restrict the cardinality $|{\cal I}|$, rather for the sake
of maximal generality we will allow $|{\cal I}|$ to take any possible value
in the set of Cantor's Alephs \cite{48}. The cardinality of the countably
infinite sets is given by the non-standard number $\aleph$. Then the 
cardinality of
any other infinite set can be written as a function of $\aleph$ (usually
exponentials (of exponentials of..) $\aleph$), e.g.
the set $\Rl$ has the cardinality $2^\aleph$. The mathematical 
justification for this amount of generality is because, following von 
Neumann \cite{33},\\
{\it ``...while the theory of enumerably infinite direct products 
$\otimes_{n=1}^\infty {\cal H}_n$ 
presents essentially new features, when compared with that of the finite
$\otimes_{n=1}^N {\cal H}_n$, the passage from 
$\otimes_{n=1}^\infty {\cal H}_n$ to the general
$\otimes_{\alpha\in{\cal I}} {\cal H}_\alpha$ presents no further 
difficulties...., the generalizations of the direct product lead
to higher set-theoretical powers (G. Cantor's ``Alephs''), and to 
{\it no measure problems at all}.''}
\begin{Definition} \label{def4.1}
Let $\{z_\alpha\}_{\alpha\in{\cal I}}$ be a collection of complex numbers. 
The infinite product 
\be \label{4.1}
\prod_{\alpha\in{\cal I}} z_\alpha
\ee
is said to converge to the number $z\in\Co$ \\
$\Leftrightarrow\;\forall\; \delta>0\;\exists\;{\cal I}_0(\delta)\subset
{\cal I},\;|{\cal I}_0(\delta)|<\infty\;\ni\;
|z-\prod_{\alpha\in{\cal J}} z_a|<\delta\;\forall\;
{\cal I}_0(\delta)\subset{\cal J}\subset{\cal I},\;|{\cal J}|<\infty$.
\end{Definition}
From the definition it is also straightforward to prove that if
$\prod_\alpha z_\alpha,\prod_\alpha z'_\alpha$ converge to $z,z'$ 
respectively then $\prod_\alpha z_\alpha z_\alpha'$ converges to $zz'$.

Recall that a series $\sum_\alpha z_\alpha$ converges if and only if
it converges absolutely which in turn is the case if and only if
$z_\alpha=0$ for all but countably infinitely many $\alpha\in{\cal I}$. 
The following theorem gives a useful convergence criterion for 
infinite products.
\begin{Theorem} \label{th4.1} ~\\
1)\\ 
Let $\rho_\alpha\ge 0$. \\
i) If $\exists\;\alpha_0\in {\cal I}\;\ni\;
\rho_{\alpha_0}=0$ then $\prod_\alpha \rho_\alpha=0$. \\
ii) If $\rho_\alpha>0\; \forall \alpha$ then $\prod_\alpha \rho_\alpha$
converges if and only if $\sum_\alpha \mbox{max}(\rho_\alpha-1,0)$ 
converges.\\
iii) If $\rho_\alpha>0\; \forall \alpha$ then $\prod_\alpha \rho_\alpha$
converges to $\rho>0$ if and only if $\sum_\alpha |\rho_\alpha-1|$
converges.\\
2)\\ 
Let $z_\alpha=\rho_\alpha e^{i\varphi_\alpha}\in\Co$ where 
$\rho_\alpha=|z_\alpha|,\;\varphi_\alpha\in[-\pi,\pi]$. Then
$\prod_\alpha z_\alpha$ converges if and only if\\ 
i) either $\prod_\alpha \rho_\alpha$ converges to zero in which case 
$\prod_\alpha z_\alpha=0$,\\
ii) or $\prod_\alpha \rho_\alpha$ converges to $\rho>0$ and
$\sum_\alpha |\varphi_\alpha|$ converges in which case
$\prod_\alpha z_\alpha=\rho e^{i\sum_\alpha \varphi_\alpha}$.
\end{Theorem}
In contrast to the case of an infinite series, absolute convergence of an 
infinite product does not imply convergence, the phases of the factors
could fluctuate too wildly. This motivates the following
definition.
\begin{Definition} \label{def4.2}
Let $z_\alpha\in \Co$. We say that $\prod_\alpha z_\alpha$ is 
quasi-convergent if $\prod_\alpha |z_\alpha|$ converges. In this case we 
define the value of $\prod_\alpha z_\alpha$ to equal 
$\prod_\alpha z_\alpha$ if $\prod_\alpha z_\alpha$ is even convergent and 
to equal zero otherwise. 
\end{Definition}
This definition assigns a value to the infinite product of numbers which
converge absolutely but not necessarily non-absolutely. As a corollary
of theorem \ref{th4.1} we have
\begin{Corollary} \label{col4.1}
Quasi-convergence of $\prod_\alpha z_\alpha$ to a non-vanishing value
is equivalent with convergence to the same value. A necessary and 
sufficient criterion is that $z_\alpha\not=0\;\forall \alpha$ and that 
$\sum_\alpha |z_\alpha-1|$ converges.
\end{Corollary}
After having defined convergence for infinite products of complex numbers
we are ready to turn to the ITP of Hilbert spaces.
\begin{Definition} \label{def4.2a}
Let ${\cal H}_\alpha,\;\alpha\in{\cal I}$ be an arbitrary collection of 
Hilbert spaces. For a sequence $f:=\{f_\alpha\}_{\alpha\in{\cal I}},\;
f_\alpha\in{\cal H}_\alpha$ the object 
\be \label{4.2}
\otimes_f:=\otimes_\alpha f_\alpha 
\ee
is called a $C$-vector provided that $\prod_\alpha ||f_\alpha||_\alpha$
converges, where $||.||_\alpha$ denotes the Hilbert norm of 
${\cal H}_\alpha$. The set of $C$-vectors will be called $V_C$.
\end{Definition}
The following property holds for $C$-vectors, enabling us to compute 
their inner products.
\begin{Lemma} \label{la4.1}
For two $C$-vectors $\otimes_f=\otimes_\alpha f_\alpha, 
\otimes_g=\otimes_\alpha g_\alpha$ the inner product 
\be \label{4.3}
<\otimes_f,\otimes_g>:=\prod_\alpha <f_\alpha,g_\alpha>_\alpha
\ee
is a quasi-convergent product of the individual inner products 
$<f_\alpha,g_\alpha>_\alpha$ on ${\cal H}_\alpha$.
\end{Lemma}
There are $C$-vectors $\otimes_f$ such that 
$\prod_\alpha ||f_\alpha||_\alpha=0$ although
$||f_\alpha||_\alpha>0 \forall \alpha$. Thus, it is 
conceivable that it happens that $<\Phi_f,\Phi_g>\not=0$ for some
$C$-vector $\Phi_g$. If that would be the case, the Schwarz inequality
would be violated for the inner product (\ref{4.3}) on $C$-vectors.
That this is not the case is the content of the following lemma.
\begin{Lemma} \label{la4.2}
Let $\otimes_f$ be a $C$-sequence with $\prod_\alpha ||f_\alpha||_\alpha=0$.
Then $<\otimes_f,\otimes_g>=0$ for any $C$-vector $\otimes_g$.
\end{Lemma}
To distinguish trivial $C$-vectors from non-trivial ones we define
\begin{Definition} \label{def4.3}
A sequence $(f_\alpha)$ defines a $C_0$-vector $\otimes_f=
\otimes_\alpha f_\alpha$ iff
\be \label{4.4}
\sum_\alpha |\;||f_\alpha||_\alpha-1|
\ee
converges. The set of $C_0$-vectors will be denoted by $V_0$.  
\end{Definition}
It is easy to prove by means of theorem \ref{th4.1} that every $C_0$-vector 
is a $C$-vector but only those $C$-vectors are $C_0$-vectors 
for which $<\otimes_f,.>$,
considered as a linear functional on $C$-vectors, does not equal zero which
by lemma \ref{la4.2} implies, in particular, that 
$\prod_\alpha ||f_\alpha||_\alpha\not=0$. It follows that the norm
of a $C_0$-vector does not vanish, as the following lemma reveals.
\begin{Lemma} \label{la4.3}
For any complex numbers, the convergence of one of 
$\sum_\alpha |\;|z_\alpha|-1|,\;
\sum_\alpha |\;|z_\alpha|^2-1|$ implies the convergence of the other.
\end{Lemma}
Thus, since by definition of a $C_0$-vector $\otimes_f$ and theorem
\ref{th4.1}1)iii) $z_\alpha=||f_\alpha||_\alpha$ satisfies the assumption
of lemma \ref{la4.3}, by that lemma and again theorem \ref{4.1}1iii) 
in the opposite direction we find that $||\otimes_f||>0$.

Obviously we will construct the ITP Hilbert space from the linear span
of $C_0$-vectors (we can ignore the $C-$vectors which are not $C_0$-vectors
by lemma \ref{la4.2}). For this it will be useful to know how the Hilbert 
space decomposes into orthogonal subspaces. The following definition
serves this purpose.
\begin{Definition} \label{def4.4}
If $\otimes_f$ is a $C_0$-vector, we will call the sequence $f=\{f_\alpha\}$
a $C_0$-sequence. We will call two $C_0$-sequences $f,g$ strongly 
equivalent, denoted $f\approx g$, provided that 
\be \label{4.5}
\sum_\alpha |<f_\alpha,g_\alpha>_\alpha-1|
\ee
converges.
\end{Definition}
\begin{Lemma} \label{la4.4}
Strong equivalence of $C_0$-sequences is an equivalence relation
(reflexive, symmetric, transitive).
\end{Lemma}
This lemma motivates the following definition.
\begin{Definition} \label{def4.5}
The strong eqivalence class of a $C_0$ sequence $f$ will be denoted
by $[f]$. The set of strong equivalence classes of $C_0$-sequences 
will be called ${\cal S}$.
\end{Definition}
The subsequent theorem justifies the notion of strong equivalence.
\begin{Theorem} \label{th4.2}
i) If $f^0\in [f]\not=[g]\ni g^0$ then $<\otimes_{f^0},\otimes_{g^0}>=0$.\\
ii) If $f^0,g^0\in [f]$ then $<\otimes_{f^0},\otimes_{g^0}>=0$ if and only if
there exists $\alpha\in {\cal I}$ such that $<f_\alpha,g_\alpha>_\alpha=0$.
\end{Theorem}
So, $C_0$-vectors from different strong equivalence classes are always 
orthogonal and those from the same strong equivalence class are orthogonal
if and only if they are orthogonal in at least one tensor product factor. 

The following theorem gives two useful criteria for strong equivalence.  
\begin{Theorem} \label{th4.3}
i) $[f]=[g]$ if and only if $\sum_\alpha ||f_\alpha^0-g_\alpha^0||_\alpha^2$
and $\sum_\alpha |\Im(<f_\alpha^0,g_\alpha^0>_\alpha)|$ converge
for some $f^0\in [f],g^0\in[g]$.\\
ii) If $f_\alpha=g_\alpha$ for all but finitely many $\alpha$ then
$f\approx g$.
\end{Theorem}
Obviously, it will be convenient to choose a representant $f^0\in [f]$
which is normalized in each tensor product factor. This is always 
possible.
\begin{Lemma} \label{la4.5}
For each $[f]\in {\cal S}$ there exists $f^0\approx f$ such that
$||f^0_\alpha||_\alpha=1$ for all $\alpha\in {\cal I}$.
\end{Lemma}
The next lemma reveals that caution is due when trying to extend 
multilinearity from the finite to the infinite tensor product.
\begin{Lemma} \label{la4.6}
Let $\prod_\alpha z_\alpha$ be quasi-convergent. Then\\
i) If $f$ is a $C$-sequence, so is $z\cdot f$ with $(z\cdot f)_\alpha:=
z_\alpha f_\alpha$.\\
ii) If moreover $\sum_\alpha ||z_\alpha|-1|$ converges and $f$ is a 
$C_0$sequence, so is $z\cdot f$.\\ 
iii) The product formula 
\be \label{4.6}
\otimes_{z\cdot f}=[\prod_\alpha z_\alpha] \otimes_f
\ee
fails to hold only if 1) $\prod_\alpha z_\alpha$ is not convergent and 
2) $<\otimes_f,.>\not=0$ considered as a linear functional on $C$-vectors.
In that case, $\{z_\alpha\},f$ satisfy the assumptions of ii), 
moreover $z_\alpha\not=0\;\forall \alpha$.\\
iv) If $\{z_\alpha\},f$ satisfy the assumptions of ii) then 
$[z\cdot f]=[f]$ iff $\sum_\alpha |z_\alpha -1|$ converges. If even 
$z_\alpha\not=0\;\forall \alpha$, the latter condition implies 
convergence of $\prod_\alpha z_\alpha$.
\end{Lemma}
An important conclusion that we can draw from this lemma is the following. 
If (\ref{4.6})
fails then, by iii), $f,z\cdot f$ are both $C_0$-sequences while
$\prod_\alpha z_\alpha$ is only quasi-convergent. Thus, both 
$\otimes_f,\otimes_{z\cdot f}\not=0$ while $\prod_\alpha z_\alpha=0$
by definition \ref{def4.2}. Thus, $[\prod_\alpha z_\alpha]\otimes_f
=0\not=\otimes_{z\cdot f}$.

Next, since, also by iii), 
$z_\alpha\not=0\;\forall \alpha$ we have from collorary \ref{col4.1}
that $\sum_\alpha |z_\alpha-1|$ cannot be convergent as otherwise
$\prod_\alpha z_\alpha$ would be convergent which cannot be the case as
$\prod_\alpha z_\alpha$ is only quasi-convergent. Thus, by iv)
$f,z\cdot f$ lie in different strong equivalence classes and  
therefore by theorem \ref{th4.2} $<\otimes_f,\otimes_{z\cdot f}>=0$.
\begin{Definition} \label{def4.6}
By ${\cal H}_C$ we denote the completion of the complex vector space of 
finite linear
combinations of elements from $V_C$ equipped with the sesquilinear form
$<.,.>$ obtained by extending (\ref{4.3}) from $V_C$ to ${\cal H}_C$
by sesquilinearity. 
\end{Definition}
Notice that for $C$-vectors which are not $C_0$-vectors we have 
$\otimes_f=0$ as an element of ${\cal H}_C$. 
\begin{Lemma} \label{la4.7}
$<\xi,\xi>\ge 0\;\forall \xi\in{\cal H}_C$ and we define
$||\xi||^2=<\xi,\xi>$. In particular, $<.,.>$ satisfies the Schwarz 
inequality and $||\xi||=0$ if and only if $\xi=0$.
\end{Lemma}
We can now give the definition of the ITP.
\begin{Definition} \label{def4.7}
We will denote by
\be \label{4.7}
{\cal H}^\otimes:=\otimes_\alpha {\cal H}_\alpha
\ee
the Cauchy-completion of the pre-Hilbert space ${\cal H}_C$. It 
is called the complete ITP of the ${\cal H}_\alpha$.
\end{Definition}
To analyze the structure ${\cal H}^\otimes$ in more detail, the strong
equivalence classes provide the basic tool.
\begin{Definition} \label{def4.8}
For a strong equivalence class $[f]\in {\cal S}$ we define 
the closed subspace of ${\cal H}^\otimes$
\be \label{4.8}
{\cal H}_{[f]}:=\overline{\{\sum_{k=1}^N z_k \otimes_{f^k};\;z_k\in \Co,\;
f^k\in [f],\; N<\infty\}}
\ee
by the closure of the finite linear combinations of $\otimes_{f'}$'s
with $f'\in [f]$. It is called the $[f]$-adic incomplete ITP of the 
${\cal H}_\alpha$'s.
\end{Definition}
Notice that we could absorb the $z_k$ in (\ref{4.8}) into one of the 
$f^k_\alpha$. Now we have the fundamental theorem which splits
${\cal H}^\otimes$ into simpler pieces.
\begin{Theorem} \label{th4.4}
The complete ITP decomposes as the direct sum over strong equivalence 
classes $[f]$ of $[f]$-adic ITP's, 
\be \label{4.9a}
{\cal H}^\otimes=\overline{\otimes_{[f]\in{\cal S}} {\cal H}_{[f]}}
\ee
\end{Theorem}
Also each $[f]$-adic ITP can be given a simple description.
\begin{Lemma} \label{la4.9}
For a given $[f]\in {\cal S}$, fix any $f^0\in [f]$. By lemma \ref{la4.5} 
we can choose an $f^0$ with $||f^0_\alpha||_\alpha=1$. Then 
${\cal H}_{[f]}$ is the closure of the vector space of finite linear
combinations of $\otimes_{f'}$'s where $f'\in [f]$ and 
$f'_\alpha=f^0_\alpha$ for all but finitely many $\alpha\in{\cal I}$.
\end{Lemma}
It is easy to provide a complete orthonormal basis for an $[f]$-adic
ITP if we know one in each ${\cal H}_\alpha$.
\begin{Lemma} \label{la4.10}
Let $f^0\in [f]\in{\cal S},\;||f^0_\alpha||_\alpha=1\;\forall \alpha$.
Let $d_\alpha=\dim({\cal H}_\alpha)$ (takes the value of some higher Cantor 
aleph if ${\cal H}_\alpha$ is not separable). Let 
${\cal J}_\alpha,\; 0\in {\cal J}_\alpha\;\forall\;\alpha\in {\cal I}$ 
be a set of indices of cardinality $d_\alpha$ and choose a complete 
orthonormal basis $e_\alpha^\beta,\;\beta\in {\cal J}_\alpha$ such that
$e_\alpha^0=f_\alpha^0$.

Consider the set $\cal F$ of functions 
\be \label{4.9}
\beta:\; {\cal I}\mapsto\times_\alpha {\cal J}_\alpha;\;
\alpha\mapsto \{\beta(\alpha)\}_{\alpha\in {\cal I}}
\ee
such that 1) $\beta(\alpha)\in {\cal J}_\alpha$ and 2) 
$\beta(\alpha)\not=0$ for finitely many $\alpha$ only.
Let 
\be \label{4.10}
\otimes_{e^\beta}:=\otimes_\alpha e_\alpha^{\beta(\alpha)}
\ee
Then $e^\beta\in[f]$ and the set of $C_0$-vectors 
$\{\otimes_{e^\beta};\;\beta\in {\cal F}\}$ forms a complete orthonormal
basis of ${\cal H}_{[f]}$, called a von Neumann basis.
\end{Lemma}
The following corollary establishes that the $[f]$-adic ITP's are mutually
isomorphic.
\begin{Corollary} \label{col4.2}
Each $[f]$-adic ITP is unitarily equivalent to the Hilbert space
${\cal H}_{{\cal F}}=L_2({\cal F},d\nu_0)$ of square summable functions on
$\cal F$, $\hat{\xi}:\;{\cal F}\mapsto \Co;\; \beta\mapsto 
\hat{\xi}(\beta)$, where $\nu_0$ is the counting measure. The unitary
map is given by
\be \label{4.11}
\hat{U}_{[f]}:\;{\cal H}_{{\cal F}}\mapsto {\cal H}_{[f]};\;
\hat{\xi}\mapsto \sum_{\beta\in {\cal F}} \hat{\xi}(\beta) \otimes_{e^\beta}
\ee
The inverse map is given by
\be \label{4.12}
\hat{U}^{-1}_{[f]}:\;{\cal H}_{[f]}\mapsto {\cal H}_{{\cal F}};\;
\xi\mapsto \hat{\xi}(\beta)=<\otimes_{e^\beta},\xi>
\ee
\end{Corollary}
In particular, since each $[f]$-adic ITP has a complete orthonormal basis 
labelled by $\cal F$ and since the ITP is the direct sum of (the 
mutually isomorphic) $[f]$-adic ITP's we have 
$\dim({\cal H}^\otimes)=|{\cal F}|\cdot |{\cal S}|$ where the appearing
cardinalities will be aleph-valued in general (already in the simplest
non-trivial case $\dim({\cal H}_\alpha)=2,\;{\cal I}=\Nl$).

Notice that the index set ${\cal I}$ is not required to have any
ordering structure, thus we have identities of the form
$\otimes_\alpha f_\alpha=f_{\alpha_0}\otimes [\otimes_{\alpha\not=\alpha_0}
f_\alpha]$, these are just different notations for the same object. 
However, is important to realize that the associative law generically 
does not hold for the ITP. By this we mean the following :\\
Let us decompose 
${\cal I}$ into mutually disjoint index sets ${\cal I}_l$ with $l\in 
{\cal L}$ then we can form the following Hilbert spaces :
${\cal H}^\otimes=\otimes_{\alpha\in {\cal I}} {\cal H}_\alpha$ and 
${\cal H}^{\otimes\prime}:=\otimes_{l\in {\cal L}} {\cal H}_l$ where  
${\cal H}_l:=\otimes_{\alpha\in {\cal I}_l} {\cal H}_\alpha$.
The $C_0$-vectors of ${\cal H}^\otimes$ are given by 
$\otimes_f=\otimes_{\alpha\in{\cal I}} f_\alpha$ while
the $C_0$-vectors of ${\cal H}^{\otimes\prime}$ are given by 
$\otimes_f'=\otimes_{l\in{\cal L}} f'_l$ where $f'_l\in {\cal H}_l$
is a (Cauchy limit of a) finite linear combination of vectors of the 
form $\otimes_f^l=\otimes_{\alpha\in {\cal I}_l} f_\alpha$. 
Inner products between $C_0$-vectors are computed as 
$<\otimes_f,\otimes_g>=\prod_\alpha <f_\alpha,g_\alpha>_\alpha$ and 
$<\otimes_f',\otimes_g'>=\prod_\alpha <f_l,g_l>_l$ respectively 
where $<\otimes_f^l,\otimes_g^l>_l=\prod_{\alpha\in{\cal I}_l}
<f_\alpha,g_\alpha>_\alpha$. 

It is easy to see that if 
$f=\{f_\alpha\}_{\alpha\in{\cal I}}$ is a $C_0$-sequence for  
${\cal H}^\otimes$ then $f'=\{f_l':=\otimes_f^l\}_{l\in{\cal L}}$
is a $C_0$-squence for ${\cal H}^{\otimes\prime}$.
However, the obvious map between $C_0$-sequences given by
\be \label{4.14}
C:\; f\mapsto f'
\ee
in general does not preserve 
the decomposition into strong equivalence classes of ${\cal H}^\otimes$
and ${\cal H}^{\otimes\prime}$ respectively. We will give a few examples to 
illustrate this point.
\begin{itemize}
\item[i)] ~~~~~\\
Let ${\cal I}={\cal L}=\Nl$ and ${\cal I}_l=\{2l-1,2l\}$
so that ${\cal I}=\cup_{l=1}^\infty {\cal I}_l$. Consider the following
two $C_0$-sequences : $f_\alpha,\;\alpha\in \Nl$ is just some normal 
vector in ${\cal H}_\alpha$, that is, $||f_\alpha||_\alpha=1$ and 
$g_\alpha=-f_\alpha$. Then certainly their strong equivalence classes
with respect to ${\cal H}^\otimes$ are different, $[f]\not=[g]$ since
$|<f_\alpha,g_\alpha>-1|=2$ so that (\ref{4.5}) blows up. On the other
hand we have $f'_l=\otimes_f^l=f_{2l-1}\otimes f_{2l}
=[-f_{2l-1}]\otimes [-f_{2l}]=\otimes_g^l=g'_l$. Thus, trivially
$[f']'=[g']'$ where the prime at the bracket indicates that the class
is with respect to ${\cal H}^{\otimes\prime}$.
\item[ii)] ~~~~~\\
Even multiplication by complex numbers is problematic :
Take the same index sets as in i) and consider the complex numbers
$z_\alpha=-1$. Then $\prod_\alpha z_\alpha$ is quasi-convergent but not 
convergent and therefore by definition $\prod_\alpha z_\alpha=0$.
Our map (\ref{4.14}) now sends $z\cdot f$ to $z'\cdot f'$ with $z'_l
=z_{2l-1} z_{2l}$.
Now $z'_l=1$ and thus $\prod_l z'_l$ is convergent to $1$. It follows that 
$\otimes_{z\cdot f}\not=[\prod_\alpha z_\alpha]\otimes_f=0$ but
$\otimes'_{z'\cdot f'}=[\prod_l z'_l]\otimes'_{f'}=\otimes'_{f'}$,
in particular, $[f]\not=[z\cdot f]$ but $[f']'=[z'\cdot f']'$.
\item[iii)] ~~~~~~~~\\
Our map is certainly not invertible : Consider, for the same index sets
as in i), the vector 
\be \label{4.13}
f_l':=\frac{1}{\sqrt{2}}[e_{2l-1}^1\otimes e_{2l}^1
+e_{2l-1}^2\otimes e_{2l}^2]
\ee
where we assume that ${\cal H}_\alpha$ is at least two-dimensional and 
we choose two orthonormal vectors $e_\alpha^j,\;j=1,2$ for each $\alpha$.
Then $||f'_l||_l=1$ and $f'$ is a $C_0$-sequence for ${\cal H}^{\otimes\prime}$.
However, we cannot write $f'$ as a finite linear combination of 
$C_0$-vectors of ${\cal H}^\otimes$ : Any attempt to use the distributive 
law and to write it as a linear combination of $C_0$-vectors 
for ${\cal H}^\otimes$ of the form 
$\otimes_l[e_{2l-1}^{j_l}\otimes e_{2l}^{j_l}]$ 
with $j_l\in\{1,2\}$ fails because all of these vectors are 
orthogonal (with respect to ${\cal H}^{\otimes\prime}$) to $\otimes_{f'}'$ :
\be \label{4.14a}
<\otimes_l[e_{2l-1}^{j_l}\otimes e_{2l}^{j_l}],\otimes_l f'_l>=
\prod_l \frac{1}{\sqrt{2}}=0
\ee
\end{itemize}
It is plausible and one can indeed show that these complications do not 
arise if $|{\cal L}|<\infty$.

\subsection{Von Neumann Algebras on the Infinite Tensor Product}
\label{s4.2}

The set of von Neumann algebras that one can define on the Infinite Tensor 
Product Hilbert space is of a surprisingly rich structure. In fact, every
possible type of von Neumann's factors (I$_\infty$, II$_1$, II$_\infty$,
III$_0$, III$_1$, III$_\lambda;\;\lambda\in(0,1)$) can be realized on the 
ITP. Physically, one will start from the local operators that ``come from 
the various ${\cal H}_\alpha$''. However, there are many more operators
which are not local and which are well-defined on the ITP. All the 
algebras that we consider are assumed to be unital.
\begin{Definition} \label{def4.9}
We denote by ${\cal B}({\cal H}_\alpha)$ the set of bounded operators on
${\cal H}_\alpha$ and by ${\cal B}^\otimes:={\cal B}({\cal H}^\otimes)$ the 
set of bounded operators on the ITP ${\cal H}^\otimes$.
\end{Definition}
The restriction to bounded operators is not a severe one since  
every unbounded operator can be written (up to domain questions) as a 
linear combination of self-adjoint ones and those are known if we know 
their spectral projections which are bounded operators. 
 
An operator on one of the tensor product factors is not a priori defined
on the ITP. The following lemma embeds ${\cal B}({\cal H}_\alpha)$ into
${\cal B}^\otimes$.
\begin{Lemma} \label{la4.11}
Let $\alpha_0\in{\cal I}$ and $A_{\alpha_0}\in {\cal B}({\cal 
H}_{\alpha_0})$. Then there exists a unique operator 
$\hat{A}_{\alpha_0}\in {\cal B}^\otimes$ such that for any $C$-sequence
$f$
\be \label{4.15}
\hat{A}_{\alpha_0}\otimes_f=\otimes_{f'} \mbox{ where }
f'_\alpha= 
\left\{ 
\begin{array}{cc}
f_\alpha & :\;\alpha\not=\alpha_0\\
A_{\alpha_0}f_{\alpha_0} & :\;\alpha=\alpha_0
\end{array}
\right.
\ee
We will use the notation
\be \label{4.16}
\hat{A}_{\alpha_0}\otimes_f=
[A_{\alpha_0}f_{\alpha_0}]\otimes[\otimes_{\alpha\not=\alpha_0} f_\alpha]
\ee
\end{Lemma}
This lemma gives rise to the following definition.
\begin{Definition} \label{def4.10}
We denote by ${\cal B}_\alpha$ the extension of 
${\cal B}({\cal H}_\alpha)$ to the ITP, that is, 
\be \label{4.17}
{\cal B}_\alpha=\{\hat{A}_\alpha;\;A_\alpha\in {\cal B}({\cal H}_\alpha)\}
\ee
Obviously ${\cal B}_\alpha\subset {\cal B}^\otimes$.
\end{Definition}
It is not difficult to prove that $A_\alpha\leftrightarrow\hat{A}_\alpha$
is in fact a $^\ast$ algebra isomorphism. The algebras ${\cal B}^\otimes,
{\cal B}({\cal H}_\alpha)$ are C$^\ast$ algebras by definition. 
Recall that, on the other hand, a von Neumann algebra over a Hilbert space
is a weakly (equivalently strongly) closed sub- $^\ast$ algebra of 
the algebra of bounded operators on that Hilbert space.
\begin{Lemma} \label{la4.12}
For all $\alpha\in {\cal I}$, the algebra ${\cal B}_\alpha$ is a von Neumann
algebra (v.N.a.) over ${\cal H}^\otimes$.
\end{Lemma}
The idea of proof is quite simple : One writes ${\cal B}^\otimes=
{\cal B}({\cal H}_\alpha\otimes{\cal H}_{\bar{\alpha}})$ where
$\bar{\alpha}={\cal I}-{\alpha}$. Next, it is 
almost obvious that ${\cal B}_\alpha$ coincides with 
${\cal B}_{\bar{\alpha}}'=\{\hat{B}\in {\cal B}^\otimes;\;[\hat{A},\hat{B}]=0
\;\forall\;\hat{A}\in {\cal B}_{\bar{\alpha}}\}$, the commutant of 
${\cal B}_{\bar{\alpha}}$. Then an appeal to the bicommutant (or von Neumann
density) theorem \cite{49} finishes the proof.

Actually, the correspondence of lemma \ref{la4.11} extends to von Neumann 
algebras ${\cal R}({\cal H}_\alpha)\subset{\cal B}({\cal H}_\alpha)$ as we
state in the subsequent theorem.
\begin{Theorem} \label{th4.5}
The one to one correspondence 
${\cal B}({\cal H}_\alpha)\ni A_\alpha\leftrightarrow
\hat{A}_\alpha\in {\cal B}_\alpha$ extends to a $^\ast$ 
isomorphism between von Neumann algebras
${\cal B}({\cal H}_\alpha)\supset{\cal R}({\cal H}_\alpha)
\leftrightarrow
{\cal R}_\alpha=\{ \hat{A}_\alpha;\;A_\alpha\in {\cal R}({\cal H}_\alpha)\}$
\end{Theorem}
The largest von Neumann algebra on ${\cal H}^\otimes$ that we can construct 
from the algebras ${\cal B}({\cal H}_\alpha)$ is the following one.
\begin{Definition} \label{def4.11a}
By ${\cal R}^\otimes$ we denote the smallest v.N.a. that contains all
the ${\cal B}_\alpha$, that is, the weak closure of the set
\be \label{4.18}
\cup_{\alpha\in{\cal I}} {\cal B}_\alpha
\ee
\end{Definition}
It turns out that not surprisingly ${\cal R}^\otimes$ is a proper subalgebra
of ${\cal B}^\otimes$ unless $|{\cal I}|<\infty$. Physically, the indices
$\alpha$ label local degrees of freedom and therefore the elements 
of ${\cal B}_\alpha$ correspond to local operators of a quantum field 
theory. Thus the algebra ${\cal R}^\otimes$ is the algebra of local 
observables represented on the ITP ${\cal H}^\otimes$. The remainder
${\cal B}^\otimes-{\cal R}^\otimes$ can therefore be identified with 
a set of non-local operators. Thus, while the algebra ${\cal R}^\otimes$
is rather important from the point of view of local (or algebraic) quantum 
field theory \cite{50} it is the remainder which offers challenging
possibilities in the sense that it could be the universal home for operators
that map a given physical system to a drastically different one. Examples
for this could be the change of energy by an infinite amount or {\it
topology change} of the underlying spacetime manifold. We will come back 
to this point in section \ref{s5}. These issues should be particularly
important for quantum general relativity since there all the (Dirac)
observables are supposed to be non-local.

In any case we should investigate the subalgebra ${\cal R}^\otimes$ in 
more detail. To that end, recall from lemma \ref{la4.6} that the equation
$\otimes_{z\cdot f}=[\prod_\alpha z_\alpha]\otimes_f$ is false only if
both $f,z\cdot f$ are $C_0$-vectors, $z_\alpha\not=0\;\forall \alpha$ but
$\prod_\alpha z_\alpha$ is only quasi-convergent. This fact gives rise to
the next definition.
\begin{Definition} \label{def4.11}
Two $C_0$-sequences $f,g$ are said to be weakly equivalent, denoted by
$f\sim g$, provided that there are complex numbers $z_\alpha$ such that
$z\cdot f$ and $g$ are strongly equivalent, that is, $z\cdot f\approx g$.
\end{Definition}
Important facts about weak equivalence are contained in the following 
lemma which also contains a necessary and suffient criterion.
\begin{Lemma} \label{la4.13}~~~~~\\
i) Definition \ref{def4.11} remains unchanged if we restrict to complex 
numbers with $|z_\alpha|=1$.\\
ii) Weak equivalence is an equivalence relation (reflexive, symmetric,
transitive).\\
iii) $f\sim g$ if and only if
\be \label{4.19}
\sum_\alpha |\;|<f_\alpha,g_\alpha>_\alpha|-1|
\ee
converges.
\end{Lemma}
Comparing with definition \ref{def4.4} we see that the ``only'' difference 
between strong and weak equivalence is the additional modulus for
$<f_\alpha,g_\alpha>_\alpha$ in (\ref{4.19}).
\begin{Definition} \label{def4.12}~~~~~~\\
i) For a $C_0$-sequence $f$ its weak equivalence class is denoted by 
$(f)$. The set of weak equivalence classes is denoted by $\cal W $.\\
ii) For given $(f)\in {\cal W}$ we denote by ${\cal H}_{(f)}$ the closure
of the set of finite linear combinations of $\otimes_{f'}$'s where
$f'\in (f)$. 
\end{Definition}
Obviously, weak equivalence is weaker than strong equivalence. Thus,
each $(f)\in{\cal W}$ decomposes into mutually disjoint $[f']\in{\cal S},\;
f'\in (f')$. It follows from this and the mutual orthogonality of the 
${\cal H}_{[f']}$'s (theorem \ref{th4.2}) that we may write 
\be \label{4.20}
{\cal H}_{(f)}=\overline{\oplus_{[f']\in{\cal S}\cap (f)} {\cal H}_{[f']}}
\ee
\begin{Lemma} \label{la4.14}
i) For every sequence of complex numbers $\{z_\alpha\}_\alpha$ such that
$|z_\alpha|=1\;\forall\;\alpha$ there exists a unique, unitary operator
$\hat{U}_z$, densely defined on (finite linear combinations of) 
$C_0$-vectors $f$ such that $\hat{U}_z\otimes_f=\otimes_{z\cdot f}$.\\
ii) Given $s\in {\cal S},\;w\in {\cal W}$ respectively, denote by
$\hat{P}_s,\;\hat{P}_w$ respectively the projection operators from
${\cal H}^\otimes$ onto the closed subspaces ${\cal H}_s,\;{\cal H}_w$ 
respectively. Then :\\
a) $[\hat{U}_z,\hat{P}_w]=0$,\\
b) $[\hat{U}_z,\hat{P}_s]=0$ if and only if $\prod_\alpha z_\alpha$ 
converges to $z,\;|z|=1$ in which case $\hat{U}_z=z 
\mbox{1}_{{\cal H}^\otimes}$ and\\
c) if $[\hat{U}_z,\hat{P}_s]\not=0$ then $\hat{U}_z{\cal H}_s={\cal H}_{s'}$
where $s\not=s'\in {\cal S}$, that is, $\hat{U}_z$ maps different 
$s$-adic ITP subspaces onto each other which are thus unitarily equivalent.
\end{Lemma}
The following theorem describes much of the structure of ${\cal R}^\otimes$.
\begin{Theorem} \label{th4.6}~~~~~~\\
i) An operator $\hat{A}\in{\cal B}^\otimes$ belongs actually to 
${\cal R}^\otimes$ if and only if it commutes with all the 
$\hat{U}_z,\hat{P}_s$
of lemma \ref{la4.14}. In particular, the elements of ${\cal R}^\otimes$
leave all the ${\cal H}_s,\;s\in {\cal S}$ invariant.\\
ii) For each $w\in {\cal W}$, fix once and for all an element
$s_w\in {\cal S}\cap w$. Suppose that we are given a family of 
bounded operators $\hat{A}_w$ on ${\cal H}_{s_w}$ for each $w\in {\cal W}$.
Then there exists an operator $\hat{A}\in {\cal R}^\otimes$ such that
its restriction $\hat{A}_{s_w}$ to ${\cal H}_{s_w}$ coincides with 
$\hat{A}_w$, provided
that the set of non-negative numbers $\{||\hat{A}_w||;\; w\in {\cal W}\}$ is 
bounded. In that case, $\hat{A}_w$ is actually unique.\\
iii) The norm of the operator $\hat{A}$ of ii) is given by
\be \label{4.20a} 
||\hat{A}||=\sup\; \{||\hat{A}_s||;\;s\in {\cal S}\}=
\sup\; \{||\hat{A}_w||;\;w\in {\cal W}\}
\ee
\end{Theorem}
This theorem tells us the following about ${\cal R}^\otimes$:\\
1)\\
As $\hat{P}_w=\oplus_{s\in{\cal S}\cap w}\hat{P}_s$, item i) reveals 
that each ${\cal H}_w,\;w\in {\cal W}$ is an invariant subspace for any 
element $\hat{A}\in{\cal R}^\otimes$, it is ``block diagonal'' with
respect to ${\cal H}^\otimes$ where the blocks correspond to the 
${\cal H}_w,\;w\in {\cal W}$. Within each of these blocks, $\hat{A}$
is further reduced by each ${\cal H}_s,\;s\in {\cal S}\cap w$. Moreover,
since $\hat{U}_z$ commutes with $\hat{A}$ and we obtain any ${\cal H}_s,\;
s\in {\cal S}\cap w$ by mapping ${\cal H}_{s_w}$ of theorem \ref{th4.6} 
with $\hat{U}_z$, knowlegde of $\hat{A}$ on ${\cal H}_{s_w}$ is sufficient
to determine it all over ${\cal H}_w$. 
2)\\
Item ii) tells us that certainly not every element of ${\cal B}^\otimes$
lies in ${\cal R}^\otimes$, actually it is easy to construct 
bounded operators, e.g. the $\hat{U}_z\in {\cal B}^\otimes$, which do not 
lie in ${\cal R}^\otimes$.

Finally we determine the cardinality of the set ${\cal S}\cap w$.
\begin{Lemma} \label{la4.15}~~~~~\\
i) If $|{\cal I}|<\infty=\aleph$ then ${\cal S}={\cal W},\;|{\cal S}|=1$ and 
${\cal H}^\otimes ={\cal H}_w={\cal H}_s$. \\
ii) If $|{\cal I}|\ge\aleph$ then $|{\cal S}\cap w|=2^{|{\cal I}|}$.\\ 
iii) If the number of $\alpha$'s such that $\dim({\cal H}_\alpha)\ge 2$ is
finite, then $|{\cal W}|=1$. Otherwise, $|{\cal W}|\ge
2^\aleph$.
\end{Lemma}
To investigate the structure of ${\cal R}^\otimes$ further we need 
to recall some of the notions from the theory of von Neumann algebras,
e.g. \cite{50}.
\begin{Definition} \label{def4.13}~~~~~\\
i) \\
Let ${\cal R}\subset{\cal B}({\cal H})$ be a v.N.a. over the Hilbert space
$\cal H$. The commutant of $\cal R$, denoted by ${\cal R}'$ is the set of
operators in ${\cal B}({\cal H})$ that commute with all elements of 
$\cal R$. For a v.N.a. we have ${\cal R}^{\prime\prime}={\cal R}$.
${\cal Z}({\cal R})={\cal R}\cap{\cal R}'$ is called the center of 
$\cal R$. The v.N.a. is called a factor if ${\cal Z}({\cal R})=
\{\lambda\mbox{1}_{{\cal H}};\;\lambda\in\Co\}$, that is, the center
consists only of the scalars.\\
ii) \\
Let $\hat{P},\hat{Q}\in {\cal R}$ be projections. We say that \\
a) $\hat{Q}$ is a subprojection of $\hat{P}$, denoted $\hat{Q}\le\hat{P}$,
iff $\hat{Q}{\cal H}\subset\hat{P}{\cal H}$.\\
b) $\hat{Q},\hat{P}$ are equivalent, denoted $\hat{Q}\sim\hat{P}$,
iff there exists a partial isometry \cite{51} with 
initial subspace $\hat{P}{\cal H}$ and final subspace $\hat{Q}{\cal H}$. \\
c) $\hat{P}\not=0$ is a minimal projection if there is no proper
subprojection $\hat{Q}\not=0$ of $\hat{P}$.\\
d) $\hat{P}\not=0$ is an infinite projection if there is a proper 
subprojection $\hat{Q}\not=0$ of $\hat{P}$ to which it is equivalent.\\
iii)\\ 
Let ${\cal R}$ be a factor. Then we call ${\cal R}$ of type \\
I : if $\cal R$ contains a minimal projection. If 1$_{{\cal H}}$ is an
infinite projection, then the type is $I_\infty$ otherwise it is $I_n$
where $n=\dim({\cal H})$.\\
III : every non-zero projection of ${\cal R}$ is infinite. A further
systematic classification of type III factors is due to Connes,
see e.g. \cite{52} and references therein. One distinguishes between
type III$_0$ (the Krieger factor \cite{53}), type III$_\lambda,\;\lambda\in
(0,1)$ (the Powers factor \cite{54}) and III$_1$ (the factor of Araki and 
Woods \cite{55}). \\
II : if {\cal R} is neither of type I nor of type III. If 1$_{{\cal H}}$
is an infinite projection, then ${\cal R}$ is called type II$_\infty$
otherwise type II$_1$.
\end{Definition}
One can show that factors of type I are isomorphic to algebras of bounded
operators on some Hilbert space. Factors of type II$_\infty$ are generated 
by operators of the form $A_1\otimes \mbox{1}_{{\cal H}_2},
\mbox{1}_{{\cal H}_1}\otimes A_2$ acting on the Hilbert space
${\cal H}_1\otimes{\cal H}_2$ where $A_1$ belongs to a factor of type
I$_\infty$ over ${\cal H}_1$ and $A_2$ to one of type II$_1$ over 
${\cal H}_2$. For factors of type I and II it is possible to introduce
a dimension function for projections, that is, a positive definite
function $\dim(\hat{P})\ge 0$ vanishing only if $\hat{P}=0$, uniquely 
determined by the two properties that  \\
1) $\dim(\hat{P}+\hat{Q})=\dim(\hat{P})+\dim(\hat{Q})$ if $\hat{P}\perp
\hat{Q}$ and \\
2) $\dim(\hat{P})=\dim(\hat{Q})$ if $\hat{P}\sim\hat{Q}$.\\
The range of that function is $0,1,2,..,n$ for type I$_n$,
$0,1,2,..,\infty$ for type I$_\infty$, $[0,1]$ for type II$_1$ and 
$[0,\infty]$ for type II$_\infty$. For type III a dimension function
can be introduced but it takes only the values $0,\infty$ and therefore 
cannot be used to obtain the finer subdivision of type III factors outlined
above for which the use of modular (or Tomita-Takesaki) theory and the 
Connes invariant is necessary (a self-contained exposition aimed at 
mathematical phyicists can be found in \cite{56}). 

We close this section by mentioning that the more unfamiliar
factors of type II and III are not only of academic interest. In fact,
they appear already in systems as simple as the infinite spin chain
(see the second reference of \cite{33}). If one represents the  
abstract CCR $C^\ast$algebra of spin $1/2$ operators 
$\hat{\sigma}^j_l;\; j=1,2,3;\;l=1,2,..$, via the GNS
theorem \cite{49}, for a state $\omega_s$ ($s\in[0,1]$) for which
we get GNS data $(\Omega_s^\infty,{\cal H}_s^\infty,\pi_s^\infty)$ where the 
cyclic vector is 
$$
\Omega^\infty_s=\otimes_{l=1}^\infty \Omega_s,\;
\Omega_s=[\sqrt{\frac{1+s}{2}}e_1\otimes e_1 
+\sqrt{\frac{1-s}{2}}e_2\otimes e_2], 
$$
the Hilbert space is the ITP 
${\cal H}_s^\infty= \otimes_{l=1}^\infty [\Co^2\otimes\Co^2]$ 
corresponding
to the index set $\cal I$ of pairs $\alpha=(l,\tau),\;\tau=1,2$,   
and the representation is
$\pi_s^\infty(\hat{\sigma}^j_l)$ acting only on the Hilbert $\Co^2$ 
(with standard orthonormal basis $e_1,e_2$) corresponding
to $\alpha=(l,1)$, then upon weak closure a factor of type I$_\infty$
or II$_1$ or III$_s$ results for $s=1$ or $s=0$ or $s\in (0,1)$.
The physical interpretation of the parameter $s$ is that 
$\Omega_s$ is the GNS datum for the mixed state 
$$
\omega_s(A)=
\frac{\mbox{tr}(A e^{\beta\sigma^3})}{\mbox{tr}(e^{\beta\sigma^3})}
$$
with $s=\mbox{th}(\beta)$ on ${\cal H}_s=\Co^4=\Co^2\otimes\Co^2$, thus 
type I$_\infty$, II$_1$ and III$_s$ respectively means zero, infinite
or finite temperture respectively.

Finally, the type of local algebras ${\cal R}({\cal O})$ appearing in 
algebraic quantum field theory is the unique 
hyperfinite factor of type III$_1$ for diamond regions $\cal O$ 
\cite{50} 
(intersections of past and future light cones in the obvious way; this 
result can be extended to arbitrary $\cal O$ in case that the theory has a 
scaling limit (short distance conformal invariance) \cite{50a}).
Here a v.N.a. is said to be hyperfinite if it is the inductive
limit of finite dimensional algebras. This brings to the next topic.

\subsection{Inductive Limits of Hilbert Spaces and von Neuman Algebras}
\label{s4.3}

For the applications that we have in mind, specifically quantum gravity
and quantum gauge theory coupled to gravity, the framework of the 
Infinite Tensor Product is not general enough for the following reason.
Recall from section \ref{s2} that the degrees of freedom of these 
field theories are labelled by graphs. Moreover, given 
a graph $\gamma$ the degrees of freedom associated with it are
labelled by the edges of that graph. Thus, it seems that we are in position
to apply the theory outlined in sections \ref{s4.1} and \ref{s4.2}
by choosing ${\cal I}=E(\gamma)$. While that is indeed true for the given
graph $\gamma$, rather than working with a fixed, infinite graph $\gamma$ 
we are working with all of them because we do not have a lattice gauge field 
theory but a continuum one. So we actually get an uncountably infinite
family of ITP's. That would not pose any problems if we could treat each of 
them independently, however, this is not the case, e.g. not if a graph is
contained in a bigger one. The inductive limit construction is well
suited to handle this problem.
\begin{Definition} \label{def4.14}~~~~\\
0)\\ 
Let $\prec$ be a partial order (that is, a reflexive, antisymmetric 
and transitive relation) on the index set $\cal L$. The index set is said 
to be directed if for any $l,l'$ there exists 
$l^{\prime\prime}$ such that $l\prec l^{\prime\prime}$ 
and $l'\prec l^{\prime\prime}$.\\ 
1i) \\
Let $\{{\cal R}_l\}_{l\in{\cal L}}$ be a family of 
$C^\ast$ algebras (v.N.a.'s) labelled by a directed index set $\cal L$. 
Suppose that for all $l,l'$ with $l\prec l'$ there is
a $^\ast$ monomorphism (injective homomorphism) 
$F_{ll'}:\;{\cal R}_l\mapsto {\cal R}_{l'}$
satisfying \\
1) $F_{ll'}(\mbox{1}_{{\cal R}_l})
=\mbox{1}_{{\cal R}_{l'}}$ and \\
2) $F_{ll'}\circ F_{l' l^{\prime\prime}}=
F_{l l^{\prime\prime}}$ for any 
$l\prec l'\prec l^{\prime\prime}$.\\
Then the pair of families $\{{\cal R}_l,F_{l l'}\}$ is called
a directed system of $C^\ast$ algebras (v.N.a.'s).\\
1ii)\\
Let $\{{\cal R}_l\}_{l\in{\cal L}}$ be a family of 
$C^\ast$ algebras (v.N.a.'s) labelled by a directed index set $\cal L$. 
A $C^\ast$ algebra (v.N.a.) $\cal R$ is said to be the $C^\ast$ 
($W^\ast$) inductive limit of the ${\cal R}_l$ provided there exist
$^\ast$ monomorphisms $F_l:\;{\cal R}_l\mapsto {\cal R}$ such 
that \\
1) $F_l(\mbox{1}_{{\cal R}_l})=\mbox{1}_{{\cal R}}$ and \\
2) $\cup_l F_l({\cal R}_l)$ is uniformly (weakly) dense
in $\cal R$.\\
2i)\\
Let $\{{\cal H}_l\}_{l\in{\cal L}}$ be a family of 
Hilbert spaces labelled by a directed index set $\cal L$. 
Suppose that for all $l,l'$ with $l\prec l'$ there is
an isometric monomorphism 
$\hat{U}_{ll'}:\;{\cal H}_l\mapsto {\cal H}_{l'}$
such that $\hat{U}_{ll'}\hat{U}_{l' l^{\prime\prime}}=
\hat{U}_{l l^{\prime\prime}}$ for any 
$l\prec l'\prec l^{\prime\prime}$.\\
Then the pair of families $\{{\cal H}_l,\hat{U}_{l l'}\}$ is called
a directed system of Hilbert spaces.\\
2ii)\\
Let $\{{\cal H}_l\}_{l\in{\cal L}}$ be a family of 
Hilbert spaces labelled by a directed index set $\cal L$. 
A Hilbert space $\cal H$ is said to be the
inductive limit of the ${\cal H}_l$ provided there exist
isometric monomorphisms $\hat{U}_l:\;{\cal H}_l\mapsto {\cal H}$ such 
that $\cup_l \hat{U}_l {\cal H}_l$ is dense in $\cal H$.\\
3i) \\
Given a directed system of Hilbert spaces ${\cal H}_l$, suppose 
that we are given a family of operators 
$\hat{A}_l\in {\cal R}_l\subset {\cal B}({\cal H}_l)$ such that\\
1) $\sup \{||\hat{A}_l||_l;\; l\in {\cal L}\}<\infty$ and \\
2) there exists $l_0\in {\cal L}$ so that 
$\hat{U}_{ll'}\hat{A}_l=\hat{A}_{l'}\hat{U}_{ll'}$ for any 
$l_0\prec l\prec l'$.\\
Then the family is called a directed system of operators.\\
3ii) \\
Given an inductive limit $\cal H$ of Hilbert spaces ${\cal H}_l$ together 
with a family of operators 
$\hat{A}_l\in {\cal R}_l\subset {\cal B}({\cal H}_l)$, an operator
$\hat{A}\in {\cal R}\subset {\cal B}({\cal H})$ is called the inductive limit
of the $\hat{A}_l$ provided that there exists $l_0\in {\cal L}$ so that 
$\hat{U}_l\hat{A}_l=\hat{A}\hat{U}_l$ for any $l_0\prec l$.
\end{Definition}
The connection between ni) and nii), $n=1,2,3$ is made through the following
theorem.
\begin{Theorem} \label{th4.7}~~~~\\
1)\\
Given a directed system of $C^\ast$ algebras (v.N.a.'s) 
$\{{\cal R}_l,F_{l l'}\}$ there exists a unique (up to $^\ast$ isomorphisms) 
$C^\ast$ ($W^\ast$) inductive
limit ${\cal R}$ of the ${\cal R}_l$ where the corresponding $^\ast$ 
monomorphisms $F_l$ satisfy the compatibility condition
$F_{l'}\circ F_{ll'}=F_l$.\\
2)\\
Given a directed system of Hilbert spaces
$\{{\cal H}_l,\hat{U}_{l l'}\}$ there exists a unique (up to unitarity) 
inductive limit ${\cal H}$ of the ${\cal H}_l$ where the corresponding 
isometric monomorphisms $\hat{U}_l$ satisfy the compatibility condition
$\hat{U}_{l'} \hat{U}_{ll'}=\hat{U}_l$.\\
3)\\
Given a directed system of operators
$\{\hat{A}_l\}$ on a directed system of Hilbert spaces ${\cal H}_l$, there 
exists a unique (up to unitarity) 
inductive limit operator $\hat{A}$ on the inductive limit Hilbert space 
$\cal H$. 
\end{Theorem}
The proof of this theorem can be found in the second volume of the 
first reference of \cite{49}. Notice that inductive and projective limits
(as used, e.g. in \cite{5,6}) are essentially identical, just that the 
projective
limit employs ``projections downwards'' a chain in the directed system
while the inductive limit employs ``embeddings upwards'' the chain.

The importance of the inductive limit for our purposes lies 
in the following construction. Suppose we are given an index set $\cal I$
and consider the set $\cal L$ of all possible subsets of $\cal I$
(notice that we allow the cardinality of $l\in{\cal L}$ to be infinite). 
Then $\cal L$ is a directed set where the partial order $\prec$ 
is given by the inclusion relation $\subset$. For each $l\in {\cal L}$
we can form the Infinite Tensor Product ${\cal H}_l^\otimes$ of the 
${\cal H}_\alpha,\;\alpha\in l$ and the corresponding von Neumann
algebra ${\cal R}_l^\otimes$. Moreover, we have for $l\prec l'$ the 
obvious $^\ast$ monomorphism $F_{ll'}$ assigning to $\hat{A}_l\in
{\cal R}_l$ the operator $\hat{A}_l\otimes[\otimes_{\alpha\in l'-l}
\mbox{1}_{{\cal H}_\alpha}]\in{\cal R}_{l'}$. Finally, choose for each  
$\alpha\in {\cal I}$ a fixed standard unit vector 
$\Omega_\alpha\in {\cal H}_\alpha$, then for $l\prec l'$ we have 
isometric monomorphisms $\hat{U}_{ll'}$ mapping $\xi_l\in{\cal H}_l$
to $\xi_l\otimes [\otimes_{\alpha\in l'-l}\Omega_\alpha]$. It is easy to
see that $F_{ll'},\hat{U}_{ll'}$ satisfy the requirements of 
definition \ref{4.14} and so we can form the inductive limit
von Neumann algebra ${\cal R}^\otimes_\infty$ and inductive limit
Hilbert space ${\cal H}^\otimes_\infty$ respectively which are the 
universal objects from which our various ``lattice'' algebras 
${\cal R}_l$ and Hilbert spaces ${\cal H}_l$ respectively can be obtained
by theorem \ref{th4.7}.

\section{Infinite Tensor Products and Continuum Quantum Gauge Theories}
\label{s5}

We will now apply the machinery of section \ref{s4} to quantum gauge field
theories on globally hyperbolic, spatially non-compact manifolds along
the lines suggested by the exposition of section \ref{s2} and make contact
with the semi-classical analysis machinery in connection with the coherent 
states as outlined in section \ref{s3}. We proceed in several steps.

\subsection{Kinematical Framework}
\label{s5.1}

In this subsection we carefully carry over the Ashtekar-Isham-Lewandowski
kinematical framework developed for the finite analytical category to the 
infinite analytical one.

\subsubsection{Properties of Infinite Graphs}
\label{s5.1.1}

Notice that in order that $\gamma\in\Gamma^\omega_\sigma$ has an infinite 
number of edges, $\Sigma$ must not be compact by the very definition of  
compactness.

Next, while $\Gamma^\omega_\sigma$ contains graphs with an infinite 
number of edges, the number of these edges is at most countably infinite
if $\Sigma$ is paracompact as we assume here as otherwise integration
theory cannot be employed. To see this, notice that since a finite 
dimensional manifold $\Sigma$ is locally compact we can apply 
the theorem in \cite{57} chapter I, paragraph 9 which says that a 
(connected) 
locally compact space is paracompact if and only if it is the countable union 
of compact sets. Assume now that $\gamma$ has an uncountably  
infinite number of edges and let $U_n,\;n=1,2,..$ be a countable compact 
cover of $\Sigma$. We conclude that at least one of the $U_n$ must contain  
an uncountably infinite number of edges of $\gamma$ because 
$\gamma$ has an uncountable number of edges and the countable union
of countable sets is countable. But this cannot happen if $\gamma$ is 
piecewise analytic and $\sigma$-finite by definition. 

We conclude that each element $\gamma\in\Gamma^\omega_\sigma$ is of a rather 
controllable form with at most a countable number of edges and vertices and
no accumulation points as it would happen for webs. They thus resemble 
maximally the lattices that one is used to from lattice gauge theory
and this is the form of graphs which are clearly most suitable for 
semiclassical analysis and the continuum limit. (The typical element
of $\Gamma^\infty_0$ has at least one accumulation point of vertices
and on such graphs one will certainly not approximate actions, 
Hamiltonians and the like).

Moreover, we have the following basic lemma
and this is where analyticity comes in.
\begin{Lemma} \label{la5.1}
The set $\Gamma^\omega_\sigma$ is a directed set by inclusion.
\end{Lemma}
Proof of Lemma \ref{5.1} :\\
Notice that if $\gamma,\gamma'$ are two piecewise analytical, $\sigma$-finite
graphs then $\gamma^{\prime\prime}:=\gamma\cup\gamma'$ is also
piecewise analytic. We claim that it is also $\sigma$-finite. 
Suppose this was not the case. Then, 
either a) there exists a compact subset $U\subset\Sigma$ such that 
$\gamma^{\prime\prime}\cap U$ is an infinite graph or b) there 
exists a compact cover $\cal U$ such that the set 
$\{|E(\gamma^{\prime\prime}\cap U)|;\; U\in{\cal U}\}$ is unbounded.

As for case a), we know that $\gamma\cap U,\gamma'\cap U$ are both
finite graphs with finite number of edges $e,e'$ respectively.
Since $\gamma^{\prime\prime}\cap U=[\gamma\cap U]\cup[\gamma'\cap U]$
the only way that $\gamma^{\prime\prime}\cap U$ can possibly be infinite
is that there is at least one edge $e$ of a $\gamma$ and one edge
$e'$ of $\gamma'$ such that $e\cup e'$ intersect each other in an
infinite number of isolated points. (The possibility that they overlap 
in a finite segment is excluded by analyticity as they would be 
analytic extensions of each other in this case and thus would make
up a single analytical curve). However, two analytical curves that
coincide in an infinite number of points are analytical extensions of 
each other. Thus, case a) cannot occur.

As for case b), we find compact sets $U_n$ labelled by natural numbers
$n$ such that $\gamma^{\prime\prime}\cap U_n$ has at least $n$ edges.
However, we know that there are natural numbers $N,N'$ such that
$|E(\gamma\cap U_n)|<N,|E(\gamma'\cap U_n)|<N'$ for all $n$.
It follows that $U_\infty$ has the property of case a) which we excluded 
already. Thus, case b) can also not occur.\\
$\Box$\\
That $\Gamma^\omega_\sigma$ is a directed set is of paramount importance
for inductive limit constructions.

\subsubsection{Quantum Configuration Space}
\label{s5.1.2}

Recall that in section \ref{s2.2}
the quantum configuration space $\ab$ arose as the Gel'fand spectrum
of the Abelian $C^\ast$ algebra generated by finite linear combinations of 
functions of smooth connections, restricted to {\it finite} graphs 
(cylindrical functions)
and completed in the supremum norm. It is natural to ask whether we can 
extend this construction to functions of smooth connections restricted to
{\it infinite} graphs and to see if the size of the quantum configuration 
space is changed. The following simple example reveals that a naive 
transcription of this method is problematic :\\
Take $\Sigma=\Rl^3,\;G=SU(2)$ and let $\gamma$ be the $x$-axis split into the 
countably infinite number of intervals $e$ of equal unit length. 
Thus, $\gamma$ is a piecewise analytic, $\sigma$-finite graph. Let us 
consider the following function of smooth connections 
\be \label{5.1}
A\mapsto f(A):=\prod_e [k \chi_j(h_e(A))] 
\ee
where $k$ is a constant,
$\chi_j(h)=\mbox{tr}(\pi_j(h))$ is the character of the spin $j$ 
representation and the convergence of (\ref{5.1}) is meant in the sense of
definition \ref{def4.1}. By definition, the sup-norm of that function is
$||f||=\sup_{A\in \a} \prod_e |k\chi_j(h_e(A))|$. Now the zero
connection is certainly an element of $\a$, so
$||f||\ge \sup_{A\in \a} \prod_e |k(2j+1)|$ and this infinite product
converges to $0$ if $|k|<1/(2j+1)$, to $1$ if $|k|=1/(2j+1)$ and diverges 
otherwise. Now it is easy to see that for any $h\in SU(2)$ we have in fact
$|\chi_j(h)|\le 2j+1$ and equality is reached for $h=1$ so that indeed
$||f||=\sup_{A\in \a} \prod_e |k(2j+1)|$. It follows that in the only case 
that the norm is finite and non-vanishing, we have that $f(A)$ is 
non-vanishing iff $\sum_e |k\chi_j(h_e(A))-1|$ converges which means 
that for each $\epsilon>0$ the set of $e's$ such that 
$|k\chi_j(h_e(A))-1|\ge\epsilon$ is finite. In other words, $f(A)$ is  
almost given by $\prod_e \delta_{h_e(A),1}$, an infinite product over 
Kronecker $\delta$'s rather than $\delta$-distributions and it is 
almost granted that its support is of measure zero for every reasonable 
measure even if we extend $\a$ to $\ab$. We will prove shortly that this 
is indeed the case with respect to the Ashtekar-Lewandowski measure
which turns out to be extendible to our context. Thus, we face the problem
that the (Gel'fand transforms of) functions of finite sup-norm (and thus
all the elements of the Abelian $C^\ast$ algebra) seem to be supported 
on measure zero subsets of interesting measures on the resulting spectrum.

On the other hand, it is physically plausible that the quantum
configuration space as obtained from $\Gamma^\omega_0$ should not change
when we extend to $\Gamma^\omega_\sigma$. The reason is that, by the very 
definition of $\sigma$-finiteness, if we consider a function depending
on the infinite number of degrees of freedom labelled by the edges of 
$\gamma\in \Gamma^\omega_\sigma$ but restrict its dependence to a finite 
number of degrees of freedom by ``freezing'' all degrees of freedom 
labelled by $\gamma-[\gamma\cap U]$ for any compact set $U\subset\Sigma$
then we get a function cylindrical over 
$\gamma\cap U\subset \Gamma^\omega_0$ whose behaviour is certainly not 
different from the ones considered in section \ref{s2.2}. In other
words, functions over $\gamma$ satisfy a locality property.

Thus, rather than deriving the spectrum $\ab$ as the Gel'fand spectrum
arising from an Abelian C$^\ast$ algebra of cylindrical functions over
truly infinite graphs it is the characterization of the 
Ashtekar-Isham spectrum derived in \cite{4,5} for finite graphs
which we {\it simply extend} to the infinite category ! This works as 
follows :\\
We need the set $W^\omega_0$ that one obtains as the 
union of a {\it finite} number of, not necessarily compactly supported  
{\it anlytical} paths. Since analytical paths of non-compact range can 
intersect each other in an infinite number of isolated points and since
generic elements of $\Gamma^\omega_\sigma$ cannot be otained as  
the union of a finite number of paths we see
that we have the proper inclusions $\Gamma^\omega_0\subset
W^\omega_0\subset\Gamma^\omega_\sigma$. The set 
$W^\omega_0$ is trivially directed by inclusion, in a sense 
it is very similar to the set $\Gamma^\infty_0$. In fact, if one would  
blow up the neighbourhood of the source of a tassel \cite{6a} by an 
infinite amount then one gets, apart from the difference between 
smooth and analytic paths, precisely the kind of objects that lie
in $W^\omega_0$. For this reason, we will call them
{\it analytical webs}. Notice that in contrast to smooth webs the 
paths that determine an analytical web are obviously holonomically 
independent
because 1) they cannot overlap each other in a finite segment due to 
analyticity, they can only intersect each other in a possibly infinite
number of isolated points and 2) because they always have a 
(non-overlapped) segment in the bulk of $\Sigma$ where no fall-off conditions 
on $A$ restrict the range of the holonomy along that segment.\\ 
Let now $\a$ be the classical configuration space of section \ref{s2.1} 
where appropriate fall-off conditions at spatial infiniy are obeyed.
Then the holonomy of $A\in\a$ along an analytic path of infinite
range is in fact well-defined precisely  
due to the fall-off conditions on $A$ at spatial infinity. As in the context
of $\Gamma^\omega_0$ we can now consider the $^\ast$algebra of cylindrical
functions of $A\in\a$ which are simply finite linear combinations of 
functions of the form $f(A)=f_w(\{h_e(A)\}_{e\in w})$ where $w\in W^\omega_0$ 
denotes the analytical web and $f_w$ is a complex valued function on
$G^{|w|},\; |w|$ the number of paths that determine $w$. Now the 
complications that we observed above in the context of cylindrical
functions over $\gamma\in\Gamma^\omega_\sigma$ are out of the way because 
the cylindrical functions for webs depend on a finite number
of arguments only. We can therefore complete the $^\ast$ algebra in the  
sup-norm just as in section \ref{s2.2}, obtain a $C^\ast$ algebra and 
can follow exactly the same steps reviewed there for $\Gamma^\omega_0$
to arrive at the Ashtekar-Isham spectrum $\ab$ as the Gel-fand spectrum of 
that algebra. Finally, by following 
exactly the same proofs as in \cite{4,5,39} we find $\ab$ to be 
in one to one correspondence with
the set of {\it all} homomorphisms from the groupoid $\cal X$ of 
(composable) analytic paths in $\Sigma$ into the gauge group $G$. 
The isomorphism is the same as the one from \cite{4,5}, that is, 
\be \label{5.1a}
\ab\ni \bar{A}\mapsto H_{\bar{A}}\in \mbox{Hom}({\cal X},G);\;
(H_{\bar{A}}(e))_{mn}:=\bar{A}((h_e)_{mn})
=(\hat{h}_e)_{mn}(\bar{A}) 
\ee
Here $m,n$ are the indices of the matrix elements of the defining 
represenation of $G$ and $\wedge$ denotes the Gel'fand transform.
Notice that in contrast to \cite{57a} it was not necessary to consider the 
one point compactification $\hat{\Sigma}$ of $\Sigma$. In fact, we refrained
from doing that because we now can consider the paths $e\in{\cal X}$
that determine an analytical web $w$ also as possible edges of a truly 
infinite graph $\gamma\in\Gamma^\omega_\sigma$. Clearly, considering 
the one point compactification $\hat{\Sigma}$ with an embedded generic
element of $\Gamma^\omega_\sigma$ results in a highly singular object 
and therefore we do not have the luxury to do this.

In summary, essentially we do not change the spectrum $\ab$ as compared 
to \cite{4,5} except that the correspondence (\ref{5.1a}) is now
extended to paths with non-compact range and therefore all the properties of 
$\ab$ derived in the literature are preserved.

One could ask whether there is a more fundamental reason for this choice,
trying to define, as in the finite category, an Abelian $C^\ast$ algebra
of cylindrical functions depending on an infinite graph. This meets 
mathematical difficulties which are once more related to the fact that 
the associative law does not hold in general for the ITP and boils
down to saying that one cannot really build an algebra of cylindrical
functions over infinite graphs, only a vector space. We thus 
just adopt the above point of view with respect to definition of 
$\ab$. However, an outline
of these difficulties will be given in the subsequent digression 
since it is instructive and gives rise to some natural definitions.\\
\\
A natural way to proceed with infinite graphs comes from the observation that
the set $\Gamma^\omega_0$ 
is a subset of $\Gamma^\omega_\sigma$ which 
consists of compactly supported graphs. This observation motivates the 
following definition.
\begin{Definition} \label{def5.1}
Let $\gamma\in\Gamma^\omega_\sigma$.\\ 
i)\\
A function $f$ on $\a$ 
is said to be a $C$ function (not to be confused with the $C$ vectors
of section \ref{s4.1}) over 
$\gamma$ with values in 
$\Co\cup\{\infty\}$ provided that for each $e\in E(\gamma)$ there exist 
functions $f_e$ on $\a$ of the form $f_e(A)=F_e(h_e(A))$, where $F_e$ is a 
complex valued function on $G$, such that
\be \label{5.2}
f(A)=\prod_{e\in E(\gamma)} f_e(A)
\ee
and convergence is defined as in definition \ref{def4.1} where we set
$f(A)=\infty$ if $\prod_e |f_e(A)|=\infty$ irrespective of the 
phases of the $f_e(A)$. \\
ii) \\
A function $f$ on $\a$ is said to be cylindrical over $\gamma$ if it is
a finite linear combination of $C$ functions over $\gamma$. The set
of cylindrical functions over $\gamma$ is denoted by {\rm Cyl}$_\gamma$.\\
iii)\\
A function $f$ on $\a$ is said to be cylindrical if it is a finite 
linear combination of cylindrical functions over some graphs $\gamma$.
The set of cylindrical functions is denoted {\rm Cyl}.\\
iv)\\
An element $0\not=f=\sum_{n=1}^N z_n \prod_{e\in E(\gamma)} f^{(n)}_e
\in \mbox{Cyl}_\gamma,\; z_n\in\Co$ is said to be $\sigma$-bounded
if and only if 
\be \label{5.3}
||f||_\gamma:=\sup_{U\subset\Sigma}\sup_{A\in\a}
|\sum_{n=1}^N z_n \prod_{e\cap U\not=\emptyset} f^{(n)}_e(A)|
\ee
is finite where $U$ runs over all compact subsets of $\Sigma$. For $f=0$ 
we set $||f||=0$. Notice that the argument of the modulus in (\ref{5.3}) is a
cylindrical function in the sense of section \ref{s2.2}.
We will denote the set of $\sigma$-bounded, cylindrical functions by 
{\rm Cyl}$^b_\gamma$. Notice that {\rm Cyl}$^b_\gamma$ is not empty precisely due 
to the usual boundary conditions on smooth connections $\a$ for non-compact
$\Sigma$.
\end{Definition}
The norm (\ref{5.3}) assigns a finite value to functions $f$ even if there
is $A\in\a$ such that $f(A)=\infty$ which corresponds to our motivation
to take over the structure from $\Gamma^\omega_0$.
\begin{Lemma} \label{la5.2}
The space of $\sigma$-bounded cylindrical functions over $\gamma$ forms a 
$^\ast$ algebra with the $C^\ast$ property.
\end{Lemma}
Proof of Lemma \ref{la5.1} :\\
That Cyl$^b_\gamma$ is closed under linear combination, multiplication 
by scalars and factor-wise complex conjugation is obvious.
Suppose now that $f=\sum_{m=1}^M u_m \prod_{e\in E(\gamma)} f^{(m)}_e$,
$g=\sum_{n=1}^N v_n \prod_{e\in E(\gamma)} g^{(n)}_e$ are given and we 
define
\be \label{5.4}
fg:=\sum_{m,n} u_m v_n \prod_{e\in E(\gamma)} f_e^{(m)} g_e^{(n)}
\ee
simply by factorwise multiplication. Then 
\ba \label{5.5}
||fg||_\gamma 
&=&\sup_{A,U} |\sum_{m,n} u_m v_n \prod_{e\cap U\not=\emptyset}
f_e^{(m)}(A) g^{(n)}_e(A)| \nonumber\\
&=&\sup_{A,U} |
[\sum_{m=1}^M u_m \prod_{e\cap U\not=\emptyset} f_e^{(m)}(A)]
[\sum_{n=1}^N v_n \prod_{e\cap U\not=\emptyset} g_e^{(n)}(A)] |
\nonumber\\
&\le& 
[\sup_{A,U}|\sum_{m=1}^M u_m \prod_{e\cap U\not=\emptyset} f_e^{(m)}(A)|]
[\sup_{A,U}|\sum_{n=1}^N v_n \prod_{e\cap U\not=\emptyset} g_e^{(n)}(A)|]
\nonumber\\
&=& ||f||_\gamma \; ||g||_\gamma
\ea
is also bounded. The $C^\ast$ property follows from 
$|\overline{f_U(A)}|=|f_U(A)|$ and 
$\sup_{U,A}|f_U(A)|^2=(\sup_{U,A} |f_U(A)|)^2$.\\
$\Box$\\
So far we have considered only one cylindrical algebra Cyl$^b_\gamma$.
Can we consider the algebra Cyl$^b$ of finite linear combinations of 
elements of Cyl$^b_\gamma$ for some $\gamma$'s ? As we have shown in  
lemma \ref{la5.1}, $\Gamma^\omega_\sigma$ is a directed set so that
for any finite collection $\gamma_1,..,\gamma_n$ there exists a 
$\gamma$ containing each of them. However, it may no longer be true 
that a given $f_k\in Cyl^b_{\gamma_k},\;k=1,..,n$ can be written as 
a finite linear combination of $C$ functions over $\gamma$, in fact, this 
will almost never be the case. Thus, while linear combinations pose no 
problem, products do as we then can no longer multiply factor-wise
without having to consider infinite linear combinations of $C$ functions.
In other words, as soon as we allow linear combinations of functions
cylindrical over different infinite graphs, we end up
having no algebra any more, products are ill-defined.
The only exception is that for each of $\gamma_1,..,\gamma_n$ only a finite
number of edges have to be decomposed into a finite number of segments 
each of which is an edge of $\gamma$. In that case, each of $f_k$
can be considered already as a function in Cyl$^b_\gamma$ so that
nothing new is gained. Thus, the only way to proceed along the lines
of \cite{4,5} is  
therefore to consider all the Cyl$^b_\gamma$ separately.\\
\\
Once this is agreed on, the remainder is now standard. We complete the 
$^\ast$ 
algebra Cyl$^b_\gamma$ in the norm (\ref{5.3}) and obtain an Abelian
$C^\ast$ algebra ${\cal B}_\gamma$ which now depends on $\gamma$,
in contrast to section \ref{s2.2}. By the Gel'fand theorem we obtain
the spectrum $\ab_\gamma$ of this algebra and ${\cal B}_\gamma$ is,
via the Gel'fand transform $f\mapsto\hat{f}$, isometrically
isomorphic to the algebra of continuous functions $C^0(\ab_\gamma)$
over the compact Hausdorff space $\ab_\gamma$. But now we meet 
the next difficulty and this finishes our trial to proceed this way :
Namely, the set $\a$ is now no longer a subset of $\ab_\gamma$. Namely,
let $A_0\in\a\cap\ab_\gamma$ then we have from isometricity
\be \label{5.6}
||f||=||\hat{f}||=\sup_{\bar{A}\in\ab_\gamma}|\hat{f}(\bar{A})|
=\sup_{\bar{A}\in\ab_\gamma}|\bar{A}(f)|
\ge|A_0(f)|=|f(A_0)|
\ee
which from the definition (\ref{5.3}) can be true only if $A_0$ has compact
support. However, we are precisely interested in (distributional) 
connections which are supported everywhere in $\Sigma$ as this corresponds
to the intended physical application in connection with the clasical limit 
for non-compact $\Sigma$. There is no claim that one could not introduce 
a different $C^\ast$ norm on cylindrical functions which would lead to 
the desired 
distributional extension of $\a$ but there seems to be no obvious, natural
candidate as the above discussion reveals. We leave the question on the 
existence of such a norm for future research. This terminates our
digression.\\
\\
We thus will not use the norm (\ref{5.3}) any further but simply
consider the vector space Cyl of arbitrary cylindrical functions of $\a$
without any convergence requirements, to 
begin with. As we will see, a subset of this space, extended to distributional 
connections, is dense in the Hilbert space 
which we are going to construct and although it is not an algebra, inner 
products {\it can be computed} even if we have linear combinations of 
functions over different infinite graphs. 

This extension works as follows : Since every cylindrical
function is a finite linear combination of $C$ functions over some 
$\gamma$ we can also extend any $f\in$Cyl to a function $\hat{f}$ on 
$\ab$ simply by the pull-back of the Gel'fand transform on $C$ functions
\be \label{5.7}
\hat{f}:=\prod_{e\in E(\gamma)} \hat{f}_e \mbox{ where }
\hat{f}_e(\bar{A})=F_e(\bar{A}(h_e))=F_e(\hat{h}_e(\bar{A}))=
(\wedge^\ast f_e)(\bar{A})
\ee
extended by linearity. The notation means that $\hat{f}$ is the Gel'fand
transform of $f=\prod_e f_e,\;f_e=F_e\circ h_e$ extended from finite to 
infinite graphs. 
We will continue to call the extensions $\hat{f}$ cylindrical functions.

Although a general $\hat{f}\in$Cyl will take an infinite value 
on almost every point $\bar{A}\in\ab$ it is still possible to equip Cyl
with a topology which is weaker than the Hilbert space topology
that we are going to construct, moreover, the Hilbert space measure 
is such that these infinite values are integrable. This is important 
in order to have a framework
for solving quantum constraints via (analogs of) Gel'fand triples \cite{7}.
we postpone the definition of this topology to subsection \ref{s5.1.4}.

\subsubsection{Measure and Hilbert Space}
\label{s5.1.3}

Consider for a moment the set ${\cal C}^\omega$ {\it of all possible 
oriented, 
analytic curves in $\Sigma$}. Clearly, at most countable collections of 
elements $e\in{\cal C}^\omega_\sigma$ constitute an element 
$\gamma\in {\cal G}^\omega_\sigma$ through their union if that 
union is $\sigma$-finite. The idea is now to 
construct the Infinite Tensor Product Hilbert spaces 
${\cal H}^\otimes_\gamma$ associated 
with the Hilbert spaces ${\cal H}_e,\;e\in E(\gamma)$ of section
\ref{3.2}, that is,
\be \label{5.8}
{\cal H}^\otimes_\gamma:=\otimes_{e\in E(\gamma)} {\cal H}_e
\ee
Using the notation of section \ref{s4.3} we would have index sets
${\cal I}={\cal C}^\omega$ and the set of {\it arbitrary} index subsets 
$\cal L$ (or power set) of $\cal I$ of which 
${\Gamma}^\omega_\sigma\subset{\cal L}$ is a proper subset.

The reader may now wonder why we do not use the full power of the
Infinite Tensor Product of being able to deal with index sets of
arbitrary cardinality and rather stick with $\Gamma^\omega_\sigma$. Indeed,
an interesting observation is now the following : Consider instead 
of $\cal L$ the slightly smaller set $\cal P$ of 
{\it arbitrary} subsets $C$ of ${\cal C}^\omega$ (not necessarily elements
of $\Gamma^\omega_\sigma$) such that no element $e\in C$ can be written 
as a composition of elements of $C-\{e\}$ and their inverses. Then we say 
$C\prec C'$ if every element $e\in C$ 
can be written as a composition of elements $e'\in C'$ and their inverses
which gives also $\cal P$ a partial order. For $C\prec C'$ we define 
$C\cup C'=C'$.
Recall that a subset $P\subset {\cal P}$ is called a chain if all elements
$C\in P$ are in relation $\prec$. Given a chain $P$, consider the element 
$C_P:=\cup_{C\in P} C$ which is an element of ${\cal P}$ (not necessarily
of $P$), moreover,
$C\prec C_P\;\forall C\in P$. In other words, every chain in ${\cal P}$
has an upper bound in ${\cal P}$ and by the lemma of Zorn we obtain that
$\cal P$ has a maximal element $C_\infty$, that is, $C\prec C_\infty$
for all $C\in {\cal P}$. Certainly, there are infinitely many such
maximal elements each of which we will call a ``supergraph''. By construction,
every element $e\in C_\infty$ is not composition of elements of 
$C_\infty-\{e\}$ and thus they are holonomically independent. 
(This construction can obviously be repeated for the smooth category of 
curves as well). Notice that the existence
of $C_\infty$, while of theoretical interest since it allows us to
construct the universal ITP ${\cal H}^\otimes_\infty:=
\otimes_{e\in C_\infty} {\cal H}_e$, universal in the sense  
that every possible piecewise analytic graph $\gamma$ can be written as 
composition of elements of $C_\infty$, it is practically so far of modest
interest only because 1) no one knows how to describe $C_\infty$ 
explicitly and 2) even if one knew $C_\infty$ explicitly, given 
$\gamma\in {\Gamma^\omega}$, every edge $e$ of $\gamma$ would generically 
decompose into an infinite number of segments each of which is an element of
$E(C_\infty)$. Thus, even a very simple function from the point of view 
of $\gamma$ would look very complicated from the point of view 
of $C_\infty$. In particular, as we have seen already in section 
\ref{s4.1}, the associative law fails for the ITP and it will in general 
happen that a function on an incomplete ITP associated with some $\gamma$
cannot be written as an element of the universal ITP. We are therefore 
forced to work with all the ${\cal H}^\otimes_\gamma$ simultaneously
rather than with the single universal object ${\cal H}^\otimes_\infty$
only.

However, the supergraph $\gamma_\infty$ allows us to give a simple proof 
of the existence of a $\sigma$-additive,
faithful, Borel measure on $\ab$ with respect to which we can compute
arbitrary inner products of cylindrical functions. 
This is a simple corollary of the Kolmogorov theorem for the case of 
an uncountably infinite tensor product of probability measures 
\cite{58} and works as follows in the present context :\\
The supergraph $C_\infty\in {\cal P}$
is a generating system of holonomically independent analytic curves
for every element $P\in{\cal P}$, 
in particular, for every element $\gamma\in\Gamma^\omega_\sigma$. 
Each element $\bar{A}$ of the Ashtekar-Isham
space $\ab$ of generalized connections assigns to each 
curve $e\in C_\infty$ an element $\bar{A}(h_e)=\hat{h}_e(\bar{A})$
of $G$ and as $\bar{A}$ varies, this map is onto (except if $e$ is 
just a point in which case $\bar{A}(h_e)=1_G$). Given $P\in
{\cal P}$ we consider the $\sigma$-algebra ${\cal M}_P$ generated by 
preimages of Borel subsets of $G^{|P|}$
under the map $p_P:\;\ab\mapsto G^{|P|};\bar{A}\mapsto 
\{\bar{A}(h_e)\}_{e\in P}$ where $|P|$ denotes the cardinality of the set
$P$. Consider the $\sigma$-algebra $\cal M$ generated by all the 
${\cal M}_P$ displaying $(\ab,{\cal M})$ as a measurable space.
We say that a function $f$ is measurable if it is of the form  
$f=F\circ p_P$ for some $P\in {\cal P}$ and some function $F$ on 
$G^{|P|}$. A measure on 
$\ab$ can now be defined on measurable functions by 
\be \label{5.8a}
\mu_0(f):=\int_{G^{|P|}} \otimes_{e\in P} d\mu_H(h_e) F(\{h_e\}_{e\in P})
\ee
where $\mu_H$ is the Haar measure on $G$. The normalization, right -- and
left invariance and the invariance under inversion display this 
measure as a consistently defined measure on measurable functions, 
the proof is completely analogous to the one displayed in \cite{5}
so that we can omit it here. Notice, however, that in contrast to 
\cite{5} we allow $|P|=\infty$. Being consistently defined, the measure
qualifies as one to have a $\sigma$-additive extension to $\cal M$.
Rather than using the existence theorem of \cite{6} we simply write 
it down :
\be \label{5.8b}
\mu_0(.):=\int_{G^{|C_\infty|}} \otimes_{e\in C_\infty} d\mu_H(h_e) (.)
\ee
We will call it the {\it extended Ashtekar-Lewandowski measure}.
Again, the remarkable properties of the Haar measure on $G$ reveal
that (\ref{5.8b}) reduces to (\ref{5.8a}) for $f=F\circ p_P$ and 
theorem 12.1 in \cite{58} guarantees that (\ref{5.8b}) has the required 
properties. Thus, although we do not know the object $C_\infty$,
its mere existence can be used to define $\mu_0$.
Henceforth we will denote ${\cal H}:=L_2(\ab,d\mu_0)$.\\
\\
A more explicit construction of that Hilbert space is as follows
and this provides a simple way to embed the kinematical
framework of section \ref{s2.2} for finite piecewise analytical graphs 
into the context of $\Gamma^\omega_\sigma$ of infinite piecewise analytical 
$\sigma$-finite graphs.

Given $\gamma\in\Gamma^\omega_\sigma$ we can use the 
inductive limit construction of section \ref{s4.3} to obtain 
${\cal H}^\otimes_\gamma$ for 
infinite $\gamma$ from the Hilbert spaces ${\cal H}_\gamma$ constructed in
\cite{5} for finite $\gamma$. Notice that we get this way a genuine 
extension of the so-called Ashtekar-Lewandowski Hilbert space 
\be \label{5.9}
{\cal H}_{AL}:=\overline{\cup_{\gamma\in\Gamma^\omega_0} {\cal H}_\gamma}
\ee
as we will see in a moment. In fact, the Hilbert space that we will 
construct is defined by 
\be \label{5.10}
{\cal H}^\otimes:=\overline{\cup_{\gamma\in\Gamma^\omega_\sigma} 
{\cal H}^\otimes_\gamma} 
\ee
Let us then proceed to the explicit construction. Recall that ${\cal 
H}_\gamma$ is the completion, with respect to the Ashtekar-Lewandowski 
measure $\mu_0$ of section \ref{s2.2}, of the space of cylindrical 
functions $\mbox{Cyl}_\gamma$ over $\gamma\in\Gamma^\omega_0$.
Since $\mbox{Cyl}_\gamma$ can be replaced by the finite linear 
combinations of (non-coloured) spin-network functions over $\gamma$
we see that ${\cal H}_\gamma$ can be equivalently described as the 
closure of the finite 
linear combinations of $C_0$-vectors of the {\it finite} tensor product
\be \label{5.11}
{\cal H}^\otimes_\gamma=\otimes_{e\in E(\gamma)} {\cal H}_e
\ee
where each ${\cal H}_e$ is isometric isomorphic with $L_2(G,d\mu_H)$
where $\mu_H$ is the Haar measure. Thus, ${\cal H}_\gamma=
{\cal H}^\otimes_\gamma$ for $\gamma\in\Gamma^\omega_0$.

Indeed, as it is immediately obvious from
the cylindrical consistency of the measure $\mu_0$, it reduces on 
${\cal H}_\gamma$ precisely to the tensor product Haar measure, 
corresponding to one copy of $G$ for each $e\in \gamma$ and this is  
precisely the original definition of the Ashtekar-Lewandowski
measure in terms of its cylindrical projections given in \cite{5}.

Let now $\gamma\in\Gamma^\omega_\sigma$ be given, then we find a sequence of 
elements $\gamma_n\in\Gamma^\omega_0$ such that $\gamma_n\subset\gamma$
and $\gamma_{n}\subset\gamma_{n+1}$ for each $n=1,2..$, moreover
$\cup_{n=1}^\infty \gamma_n=\gamma$. By means of the isometric 
monomorphisms defined on $C_0$-vectors for $m\le n$
\be \label{5.12}
\hat{U}_{\gamma_m\gamma_n}:\;{\cal H}^\otimes_{\gamma_m}\mapsto 
{\cal H}^\otimes_{\gamma_n};
\;\otimes_{e\in E(\gamma_m)} f_e\mapsto [\otimes_{e\in E(\gamma_m)} f_e]
\;\otimes\;[\otimes_{e\in E(\gamma_n-\gamma_m)} 1]
\ee
and extended by linearity,
where $1(A)=1$ is the unit function, we display the system of Hilbert 
spaces ${\cal H}_\gamma\simeq {\cal H}^\otimes_\gamma$ as a directed
system. By theorem \ref{th4.7} the unique inductive limit of the ${\cal 
H}_{\gamma_n}$ exists and can be identified with the Infinite Tensor Product
for each $\gamma\in \Gamma^\omega_\sigma$
\be \label{5.13}
{\cal H}^\otimes_\gamma=\otimes_{e\in E(\gamma)} {\cal H}_e
\ee
and indeed the required isometric isomorphisms are given by
\be \label{5.14}
\hat{U}_{\gamma_n}:\;{\cal H}^\otimes_{\gamma_n}\mapsto 
{\cal H}^\otimes_\gamma;
\;\otimes_{e\in E(\gamma_n)} f_e\mapsto 
[\otimes_{e\in E(\gamma_n)}f_e]\;\otimes\;
[\otimes_{e\in E(\gamma-\gamma_n)} 1]
\ee
Thus, for a truly infinite graph $\gamma$ the Hilbert space 
${\cal H}^\otimes_\gamma$ is hyperfinite, that is, it is the inductive limit
of the finite dimensional Hilbert spaces ${\cal H}^\otimes_{\gamma_n}$.
 
Several remarks are in order : 
\begin{itemize}
\item[A)] From the point of view of ${\cal H}^\otimes_\gamma$ the vectors of
${\cal H}_{\gamma_n}\simeq{\cal H}^\otimes_{\gamma_n}$ lie in the strong
equivalence class of the $C_0$-sequence $f^0=\{f^0_e\}_{e\in E(\gamma)}$
where $f^0_e=1$ for each $e$. This is an immediate consequence of lemma 
\ref{la4.9}. It follows that the Hilbert space (\ref{5.9}) is just a tiny 
subspace of the Hilbert space (\ref{5.10}) since every vector over 
$\gamma$ which is not
in the strong equivalence class of $f^0$ is orthogonal to all of the 
${\cal H}_{\gamma_n}$ and there are uncountably infinitely many
different strong eqivalence classes even for fixed $\gamma$ as follows 
from lemma \ref{la4.15} since $|E(\gamma)|=\aleph$. To see this in more 
detail, notice that if a generic element $\xi\in{\cal H}^\otimes_\gamma$
would be a Cauchy sequence of elements $\xi_n\in{\cal H}_{AL}$ then 
for any $\epsilon>0$ we would find $n_0(\epsilon)$ such that 
$||\xi-\xi_n||<\epsilon\;\forall\;n>n_0(\epsilon)$. Now each $\xi_n$
can be chosen to be in some ${\cal H}_{\gamma_n}$ with 
$\Gamma^\omega_0\ni\gamma_n\subset\gamma$
since any vector in the completion of ${\cal H}_{AL}$ can be approximated 
by vectors of that form and since any vector depending on a coloured 
graph which is not contained in $\gamma$ is automatically orthogonal to
$\xi$. However, if we choose, e.g., 
$\xi$ to be a linear combination of $C_0$-vectors each of which lies
in a different strong equivalence class Hilbert space than the vector
$f^0$ above then we get the contradiction 
$||\xi||^2<||\xi_n||^2+||\xi||^2<\epsilon\;\forall\;n>n_0(\epsilon)$. 
\item[B)] While the Ashtekar-Lewandowski Hilbert space is just a tiny 
subspace of ${\cal H}^\otimes$ in (\ref{5.4}), the Ashtekar-Lewandowski 
measure is still the appropriate measure to use in our extended context.

Indeed, it has been identified already as the $\sigma$-additive extension
of the cylindrically defined measure of \cite{5} to the projective 
(or inductive) limit of arbitrarily large and complicated, but 
finite piecewise analytic graphs in \cite{6}. 
Therefore, it could be used to date only in order to integrate special 
functions depending on an infinite number degrees of freedom (i.e.
depending on infinite graphs) : Namely those which can be written as 
infinite {\it sums} of functions each of which depends only on a 
finite graph, an exception being \cite{59} where some sort of
infinite volume limit has been taken. One contribution of the present paper 
is to show that the measure
can be used to integrate more general functions depending on an infinite
number of degrees of freedom : namely those which are infinite {\it products}
of functions each of which depends only on a finite graph.
\item[C)] In the context of finite graphs we can (even for 
non-gauge-invariant states) write down a complete orthonormal (with 
respect to the Ashtekar-Lewandowski measure $\mu_0$) basis, the 
so-called spin-network basis of section \ref{s2.2}. It is 
frequently stressed that the Ashtekar-Lewandowski measure can then be 
dispensed with by just requiring
these functions to be orthogonal and to check that a positive definite 
sesquilinear form results in this way \cite{8,10}. Adopting this point
of view, given arbitrary functions in ${\cal H}_{AL}$ one can explicitly
compute their inner products by writing them in terms of spin-network
functions and using sesquilinearity. {\it This is no longer possible in
the context of the Infinite Tensor Product (spatially non-compact $\Sigma$), 
here the Ashtekar-Lewandowski
measure is the only way to calculate inner products !} To see this
we just need to display {\it one} simple example :\\
Consider an infinite graph $\gamma$ with a countable number of analytic edges
$e$ (say a cubic lattice in $\Sigma=\Rl^3$). 
Consider the $C_0$-sequence $f:=\{f_e\}$ where (from now on we drop
the bar in $\bar{A}$ for a distributonal connection and we write
$h_e$ instead of the Gel'fand transform $\hat{h}_e$)
\be \label{5.15}
f_e(A):=f^0(h_e(A)):=\frac{1+\chi_{j}(h_e(A))}{\sqrt{2}}
\ee
where $\chi_j$ is again the character in the spin $j>0$ representation
of $SU(2)$.
Using the extended Ashtekar Lewandowski measure 
(\ref{5.8b}), which on this infinite graph
just reduces to $d\mu_{0\gamma}=\otimes_{e\in E(\gamma)} d\mu_H(h_e)$ we 
immediately 
verify that the norm of the $C_0$-vector $\otimes_f$ equals unity.
Suppose now we wanted to use only the knowledge that the set of functions
\be \label{5.16}
\prod_{k=1}^n \chi_j(h_{e_k})
\ee
for finite $n$ and mutually distinct
$e_1,..,e_n\in E(\gamma)$ are mutually orthogonal spin-network functions.
Then, in order to compute the norm of $\otimes_f$ we would need to 
decompose this vector into the latter set of functions which
at least formally can be done using the distributive law over and over 
again. However, it is easy to see that each of these infinite number of 
terms comes with the coefficient $(1/\sqrt{2})^\infty$ and so our attempt 
to compute the norm would result in the ill-defined expression $0\cdot\infty$.
More precisely, this ill-defined result is due to the fact that the inner
product between the vectors (\ref{5.15}) and (\ref{5.16}) for  
$n\to\infty$ equals zero.

Concluding, in the ITP or $\Gamma^\omega_\sigma$ category the spin-network 
functions no longer provide a basis 
(a related observation has been made independently already in \cite{6a,39} 
in the context of webs or $\Gamma^\infty_0$), simply because, even for a 
single 
$\gamma\in\Gamma^\omega_\sigma$, the orthonormal set of functions given by 
\be \label{5.17}
A\mapsto \prod_e [\sqrt{2j_e+1}\pi_{j_e}(h_e(A))_{m_e n_e}]
\ee
where $e\in E(\gamma);\;2j_e=0,1,2,..;m_e,n_e=-j_e,-j_e+1,..,j_e$
and which from experience with spin-network functions one might think
to provide a basis, {\it is not complete} ! For instance, the unit 
$C_0$-vector
\be \label{5.18} 
A\mapsto \prod_e \chi_j(h_e(A))
\ee
is orthogonal to all of them for any $j>0$, even if we choose 
$j_e=j$ for all $e$ since 
$|<\sqrt{2j+1}\pi_{jmn},\chi_j>_{L_2(SU(2),d\mu_H)}|\le 1/\sqrt{2j+1}<1$.
The ITP Hilbert space has many more orthogonal directions than one is 
used to due to its non-separability. A complete orthonormal bases 
on a single $\gamma\in \Gamma^\omega_\sigma$ is not given by
spin-network functions but rather by a von Neumann basis 
defined in lemma \ref{la4.10} and corollary \ref{col4.2}, one for each
$[f]$-adic Infinite Tensor Product subspace of ${\cal H}_\gamma$.
The only $[f]$-adic ITP that has indeed a spin-network basis is 
the one given by the trivial strong equivalence class $[f^0]$
where for any given $\gamma\in\Gamma^\omega_\sigma$ we have $f^0_e=1$
for each $e\in E(\gamma)$.
\end{itemize}
Our treatment is still incomplete because, while we can compute
inner products between finite linear combinations of $C_0$
vectors over a single $\gamma$,
nothing has so far been said about inner products between finite linear 
combinations of $C_0$ vectors over different $\gamma$'s and this is what 
we need if we wish to glue together the ${\cal H}^\otimes_\gamma$ as 
displayed in (\ref{5.10}). The idea is, of course, to use the inductive 
limit construction once again, however, as far as inner products are 
concerned we 
have to go somewhat beyond von Neumann's theory
which tells us only how to compute inner products between finite linear
combinations of $C_0$ vectors over the same $\gamma$.

A concrete and natural definition can be given employing the 
extended Ashtekar-Lewandowski measure. Let us derive it, proceeding
formally to begin with :\\
\\
Let $\gamma,\gamma'\in\Gamma^\omega_\sigma$ and let $f_\gamma,g_{\gamma'}$
respectively be $C_0$-sequences over $\gamma,\gamma'$ respectively.
Consider the graph $\gamma^\dprime:=\gamma\cup\gamma'$. The idea
is to define the inner product between the corresponding $C_0$-vectors 
$\otimes_{f_\gamma},\otimes_{g_{\gamma'}}$ by
\ba \label{5.19}
<\otimes_{f_\gamma},\otimes_{g_{\gamma'}}> &:=&
\int_{G^{|C_\infty|}} [\otimes_{e^\dprime\in C_\infty} 
d\mu_H(h_{e^\dprime})]\;
\overline{[\otimes_{f_\gamma}]}\;[\otimes_{g_{\gamma'}}]
\nonumber\\
& = &\int_{G^{|E(\gamma^\dprime)|}} 
[\otimes_{e^\dprime\in E(\gamma^\dprime)} d\mu_H(h_{e^\dprime})]\;
\overline{[\otimes_{f_\gamma}]}\;[\otimes_{g_{\gamma'}}]
\ea
where the second equality follows from cylindrical consistency. 
The problem, that by now we are already used to
with these infinite tensor products, is that the associative law does
not hold. In other words, the ITP
\be \label{5.20}
{\cal H}^\otimes_{\gamma^\dprime}:=
\otimes_{e^\dprime \in E(\gamma^\dprime)} {\cal H}_{e^\dprime}
\ee
is in general quite different from the subdivisions
\ba \label{5.21}
&& (\otimes_{e\in E(\gamma)}[\otimes_{e^\dprime\in E(\gamma^\dprime)\cap e} 
{\cal H}_{e^\dprime}])\;\otimes\; 
(\otimes_{e^\dprime \in E(\gamma^\dprime)-\gamma}
{\cal H}_{e^\dprime}) \mbox{ and}\nonumber\\ 
&& (\otimes_{e'\in E(\gamma')}
[\otimes_{e^\dprime\in E(\gamma^\dprime)\cap e'} {\cal H}_{e^\dprime}])\;
\otimes \;
(\otimes_{e^\dprime \in E(\gamma^\dprime)-\gamma'} {\cal H}_{e^\dprime})
\ea
to which $f_\gamma,g_{\gamma'}$ belong respectively. This is precisely the 
problem outlined at the end of section \ref{s4.1} : The correspondence with 
the notation there is that ${\cal I}=E(\gamma^\dprime),\;{\cal L}=
E(\gamma)\cap \{\gamma^\dprime-\gamma\},\;{\cal I}_l=
E(\gamma^\dprime)\cap e$ for $l=e$ and ${\cal I}_l=E(\gamma^\dprime)-\gamma$ 
for $l=\gamma^\dprime-\gamma$ and similar for $\gamma'$. 
Here we have identified ${\cal H}_e$ with 
$\otimes_{e^\dprime\in E(\gamma^\dprime)\cap e} {\cal H}_{e^\dprime}$
and similar for $e'$. Notice that in general 
$|E(\gamma^\dprime)\cap e|,|E(\gamma^\dprime)\cap e'|=\infty$,
an example being given by two graphs consisisting of a single
edge only, $\gamma=e,\gamma'=e'$, which however both have non-compact range
and intersect each other an infinite number of times in isolated
points. This is not excluded by piecewise analyticity since there is
no accumulation point of intersection points (take, e.g. $e=(x,0)$
and $e'=(x,\sin(x))$ in $\Rl^2$).\\ 
Step I)\\
In order to proceed, we subdivide $\gamma^\dprime$ into the mutually
disjoint sets $\gamma^*:=\gamma\cap\gamma',
\bar{\gamma}=\gamma^\dprime-\gamma,\bar{\gamma}'=\gamma^\dprime-\gamma'$.
Then we formally embed $\otimes_{f_\gamma}$ into
${\cal H}_{\gamma^\dprime}$ by identifying it with 
$(\otimes_{f_\gamma})\;\otimes\; 
(\otimes_{e^\dprime \in E(\gamma^\dprime)\cap\bar{\gamma}} 1)$ and similarly
we identify $\otimes_{g_{\gamma'}}$ with 
$(\otimes_{g_{\gamma'}})\;\otimes\; 
(\otimes_{e^\dprime \in E(\gamma^\dprime)\cap\bar{\gamma}'} 1)$. 
Clearly, we will now perform first the easy integrals corresponding to
the tensor products factors of the unit function. In order to do this,  
for given $e\in E(\gamma)$ we recall that we can write 
$f_e(h_e)=\sum_{\pi} f^{mn}_{e\pi} \pi_{mn}(h_e)$ by the Peter\& Weyl theorem
where the sum is over a complete set of equivalence classes of irreducible 
representations of 
$G$, $\pi_{mn}$ denotes the matrix elements of a group element in the 
representation 
$\pi$ and $f^{mn}_{e\pi}$ are the Fourier coefficients of $f_e$. Now suppose 
that $e^\dprime\subset e$
and that $e^\dprime\not\subset\gamma'$, that is, $e^\dprime\in
E(\gamma^\dprime)\cap (e\cap \bar{\gamma}')$. Then we can write
$f_e(h_e)=\sum_\pi f^{mn}_{e\pi} \pi_{mn}(h^{(1)}_e h_{e\dprime} h^{(2)}_e)$ 
where $h{(1)}_e,h^{(2)}_e$ depend on $e-e^\dprime$.
In other words, we can consider it as a function $F$ of $h_{e^\dprime}$ only
and as far as the integral over $h_{e^\dprime}$ is concerned it reduces 
to evaluating $<F,1>_{L_2(G,d\mu_H)}=\overline{f_{e\pi_0}}=
<f_e,1>_{{\cal H}_e}$. It follows that for any $e\in E(\gamma)$ which  
is not fully overlapped by edges of $E(\gamma')$ we can replace 
$f_e$ by $<f_e,1>$ (we drop the index at the inner product). Likewise,
for any $e'\in E(\gamma)$ which  
is not fully overlapped by edges of $E(\gamma)$ we can replace 
$g_{e'}$ by $<1,g_{e'}>$. This is the result of performing the integral
over all $h_{e^\dprime}$ with $e^\dprime\in 
E(\gamma^\dprime)\cap[\bar{\gamma}\cup\bar{\gamma}']
=E(\gamma^\dprime)\cap[\gamma^\dprime-\gamma^*]$. It remains to perform the 
integral over the edges of $E(\gamma^\dprime)\cap \gamma^*$. \\
Step II)\\
Now notice
that from $\otimes_{e\in E(\gamma)} f_e$ only those factors are left
corresponding to edges $e$ fully overlapped by edges of $E(\gamma')$ and 
from $\otimes_{e'\in E(\gamma')} g_{e'}$ only those factors are left
corresponding to edges $e'$ fully overlapped by edges of $E(\gamma)$. 
Let us denote the corresponding subsets by 
$E(\gamma)_{|\gamma^*}\subset E(\gamma), 
E(\gamma')_{|\gamma^*}\subset E(\gamma')$.
The union of both sets of edges is contained in $\gamma^*$.
Suppose now that $e\in E(\gamma)_{|\gamma^*}$ is overlapped by a collection 
of edges $e'$ of $E(\gamma')$, that is, there is a countable
number of edges $e'_{10},..,e'_{11}$
of $E(\gamma')$ so that the endpoint of one is the starting point of the
next, such that $e$ is contained in their union and such that
it is not any more contained if we remove $e'_{10}$ or $e'_{11}$ from the
collection. It follows that $e'_{10},..,e'_{11}$ are analytical continuations 
of each other. \\
Step III)\\
Let us first focus on $e'_{10}$. Now either, A) $e'_{10}$ is also contained 
in $e$ or, B) it is not.
In case B), if there are no other edges of $E(\gamma)$ overlapping the 
remaining segment of $e'_{10}$ not contained in $e$ then the edge $e'_{10}$
does not appear any more in $E(\gamma')_{|\gamma^*}$. By the same argument 
as in Step I), if we now perform the integral over any $h_{e^\dprime}$
with $e^\dprime$
contained in $e-e'_{10}$ then we can replace $\overline{f_e}$ by $<f_e,1>$ and 
that factor also drops out of the integral. Thus, we can focus on the case 
that
$e'_{10}\in E(\gamma')_{|\gamma^*}$ , that is, there are such other edges 
$e_0,..,e_1$ of $E(\gamma)$ where 
$e_0$ is adjacent to $e$, an endpoint of one is the starting point of the 
next and if $e_1$ is removed, the collection
$e,e_0,..,e_1$ no longer overlaps $e'_{10}$. We see that $e,e_0,..,e_1$  
are analytical continuations of each other. 
Now either, A) $e_1$ is also contained in 
$e'_{10}$ or B), it is not. In case B), if there are no other edges of 
$E(\gamma')$ overlapping the remaining segment of $e_1$ not contained in 
$e'_{10}$ then $e_1$ does not belong to $E(\gamma)_{|\gamma^*}$ and so 
as in Step I) we can replace $g_{e'_{10}}$ by $<1,g_{e'_{10}}>$ and so that
factor drops out of the integral. However, then as just explained also
$\overline{f_e}$ drops out of the integral. Thus, we may assume that 
$e_1\in E(\gamma)_{|\gamma^*}$ and there are new edges $e'_{20},..
e'_{21}$ with $e'_{21}$ adjacent to $e'_{01}$, the endpoint of one 
is the starting point of the next and such that $e_1$ is no longer 
overlapped if we remove $e'_{20}$. 

Let us now rename $e\circ e_0\circ .. \circ e_1$ by $e$ and the collection 
$e'_{20},..,e'_{21},e'_{10},..,e'_{11}$ by $e'_{10},..,e'_{11}$. Then
we are in the same situation as in the beginning of Step III). Iterating,
we conclude that either we end up with case A) or with case B) but that
$e'_{10}$ is no longer overlapped. In case B) the whole chain collapses 
like a cardhouse and we can replace $\overline{f_e}$ by $<f_e,1>$.
In case A) we see that we found a maximal analytical continuation of 
the original $e$, into the direction of its starting point, by other edges of 
$E(\gamma)$ all of which are overlapped by edges of $E(\gamma')$ and those 
edges are also contained in that maximal analytical continuation
contained in $\gamma$.\\
Step IV)\\
Now we focus on $e'_{11}$ and proceed completely analogously.
The end result is that $\overline{f_e}$ can be replaced by $<f_e,1>$
unless there exists a maximal bothsided maximal analytic continuation
$\tilde{e}$ of $e$ by edges of $E(\gamma)$ completely overlapped by edges
of $E(\gamma')$ and those edges of $E(\gamma')$ are also completely 
overlapped by $\tilde{e}$.\\
Step V)\\
As the argument is completely symmetric with respect to $\gamma,\gamma'$
we conclude that the remaining integral depends only on the graph 
$\tilde{\gamma}$ consisting of analytical edges $\tilde{e}$ which
can be written simultaneously as compositions of edges of $E(\gamma)$ alone
and edges of $E(\gamma')$ alone. 
For all other edges $e\in E(\gamma)-\tilde{\gamma}$ we can replace 
$\overline{f_e}$ by $<f_e,1>$ and for all other edges 
$e'\in E(\gamma')-\tilde{\gamma}$ we can replace $g_{e'}$ by
$<1,g_{e'}>$. We are thus left with
\ba \label{5.22}
&& <\otimes_{f_\gamma},\otimes_{g_{\gamma'}}>\\
&=& [\prod_{e\in E(\gamma)-\tilde{\gamma}} <f_e,1>]\;
[\prod_{e'\in E(\gamma')-\tilde{\gamma}} <1,g_{e'}>]\;
[\prod_{\tilde{e}\in E(\tilde{\gamma})} 
<[\otimes_{e\in E(\gamma)\cap \tilde{e}}\; f_e],
[\otimes_{e'\in E(\gamma')\cap \tilde{e}} \;g_{e'}]>]
\nonumber
\ea
Step VI)\\
It remains to compute the inner product labelled by $\tilde{e}$
in (\ref{5.22}). For each $\tilde{e}\in \tilde{\gamma}$
consider its unique breakup into segments $e^\dprime\in E(\gamma^\dprime)$ 
defined by
the breakpoints given by the union of the endpoints of the 
$E(\gamma)\ni e\subset\tilde{e}$ and 
$E(\gamma')\ni e'\subset\tilde{e}$ respectively. Then the last 
inner product in (\ref{5.22}) is defined by
\ba \label{5.23}
&& <\otimes_{e\in E(\gamma)\cap \tilde{e}} f_e,
\otimes_{e'\in E(\gamma')\cap \tilde{e}} g_{e'}>
\\
&:=&\int_{G^{|\tilde{e}|}} [\otimes_{e^\dprime\subset \tilde{e}} \;
d\mu_H(h_{e^\dprime})]\; 
\overline{[\prod_{e\in E(\gamma)\cap \tilde{e}} 
f_e(\prod_{e^\dprime\subset e} h_{e^\dprime})]}\;
[\prod_{e'\in E(\gamma')\cap \tilde{e}} 
g_{e'}(\prod_{e^\dprime\subset e'} h_{e^\dprime})]
\nonumber
\ea
where we have symbolically written the holonomies along the 
edges $e,e'$ respectively as products of holonomies 
along the $e^\dprime$. The integral (\ref{5.23}) is already well-defined 
if the number of $e^\dprime\subset\tilde{e}$ is finite, if not, then we 
proceed as follows :\\ 
Since $\gamma,\gamma'$ are both $\sigma$-finite graphs, 
$\tilde{e}$ must be an infinite curve in $\Sigma$ with either A) one 
or B) both ends at infinity, otherwise there would be 
an accumulation point. If only one endpoint is at infinity, choose  
the other point as the starting point of $\tilde{e}$. If both endpoints 
are at infinity, choose an arbitrary breakpoint $p$ on $\tilde{e}$ and 
choose it as the startpoint of the the resulting semi-infinite curves,
that is, 
$\tilde{e}=([\tilde{e}^{(1)}]^{-1}\circ [\tilde{e}^{(2)}])^{\pm 1}$ is a 
choice of orientation of $\tilde{e}$. 
Since $\gamma,\gamma'$ are both $\sigma$-finite graphs, the number 
of $e^\dprime\subset \tilde{e}$ is at most countable and we can label
them by integers which are increasing into the direction of the orientation
of $\tilde{e}$ in case A) and of $\tilde{e}^{(1,2)}$ respectively in case 
B), that is, $\tilde{e}=e^\dprime_1\circ e^\dprime_2\circ...$
and $\tilde{e}^{(1,2)}=e^{(1,2)\dprime}_1\circ e^{(1,2)\dprime}_2\circ...$
respectively.
The integral is then defined by performing the integrals over 
the $h_{e^\dprime_n}$ in case A) and over the pairs 
$h_{e^{1\dprime}_n},h_{e^{2\dprime}_n}$ in case B) in both cases in the order 
of increasing $n$. It is easy to see that 
the prescription in case B) is independent of the choice of breakpoint
$p$ because the two integrals differ by a change of the order of a finite 
number of integrations which is irrelevant by properties of the measure
$\mu_0$ and the compactness of $G$. Namely, all appearing functions are 
certainly absolutly integrable
in any order and the assertion follows from Fubini's theorem. \\
\\
Steps I)-VI) provide the motivation for the following definition.
\begin{Definition} \label{def5.2}
Let $\gamma,\gamma'\in\Gamma^\omega_\sigma$ and let $f_\gamma,g_{\gamma'}$
respectively be $C_0$-sequences over $\gamma,\gamma'$ respectively.
Let $\gamma^\dprime=\gamma\cup\gamma'$ and $\tilde{\gamma}\subset 
\gamma\cap\gamma'$ be the piecewise
analytic, $\sigma$-finite graph consisting of analytic edges $\tilde{e}$
which can be written simultaneously as the (countable) composition of 
edges of $E(\gamma)$ alone and of edges $E(\gamma')$ alone. For 
$\tilde{e}\in\tilde{\gamma}$ we define 
\ba \label{5.24}
&& <\otimes_{f_\gamma},\otimes_{g_{\gamma'}}>_{\tilde{e}}
\\
&:=&\int_{G^{|\tilde{e}|}} 
[\otimes_{e^\dprime\in E(\gamma^\dprime)\cap\tilde{e}} 
d\mu_H(h_{e^\dprime}) ]\;
\overline{[\prod_{e\in E(\gamma)\cap \tilde{e}} 
f_e(\prod_{e^\dprime\subset e} h_{e^\dprime})]}\;
[\prod_{e'\in E(\gamma')\cap \tilde{e}} 
g_{e'}(\prod_{e^\dprime\subset e'} h_{e^\dprime})]
\nonumber
\ea
where $e,e'$ have been written as their decompositions over
$\gamma^\dprime$ and the order of integrations is defined in step VI)
above. Then the scalar product between the $C_0$ vectors over 
$\gamma,\gamma'$ is defined by
\ba \label{5.25}
<\otimes_{f_\gamma},\otimes_{g_{\gamma'}}>
\\
&:=&
[\prod_{e\in E(\gamma)-\tilde{\gamma}} <f_e,1>]\;
[\prod_{e'\in E(\gamma')-\tilde{\gamma}} <1,g_{e'}>]\;
[\prod_{\tilde{e}\in E(\tilde{\gamma})} 
<\otimes_{f_\gamma},\otimes_{g_{\gamma'}}>_{\tilde{e}}]
\nonumber
\ea
where the separate convergence of the infinite products in the square 
brackets is in the sense of definition \ref{def4.1}.
\end{Definition}
In order to define a scalar product on finite linear combinations 
of $C_0$ vectors over different $\gamma$'s we extend definition
\ref{def5.2} by sesquilinearity. Notice that the definition 
reduces to the scalar product on ${\cal H}_\gamma$ if both vectors
are finite linear combinations of $C_0$ vectors over $\gamma$.

Of course, in order to serve as a scalar
product we must check that the scalar product is positive definite.
However, this is obvious from the explicit measure theoretic expression 
(\ref{5.19}) and can be verified by direct means as well. 
\begin{Definition} \label{def5.3}
The pre-Hilbert space of finite linear combinations of $C_0$ vectors over
graphs $\gamma\in \Gamma^\omega_\sigma$ completed in the scalar
product (\ref{5.25}) defines the Hilbert space ${\cal H}^\otimes$\
of (\ref{5.10}).
\end{Definition}
Why do we choose the Hilbert space $\cal H$ of definition \ref{def5.3}
as our quantum mechanical starting point ? The reason is the same as in the 
case of the original Hilbert space ${\cal H}_{AL}$ in which finite 
linear combinations 
of cylindrical functions over finite graphs were dense : the basic
operators of the theory are still the same local operators as in section 
\ref{s2.2}. They can be realized as operators on the infinite tensor product
following the operator extension procedure of lemma \ref{la4.11}
in section \ref{s4.2}. 
Therefore, canonical commutation relations and 
adjointness relations are completely unchanged compared to the finite 
category.

\subsection{Inductive Limit Structure}
\label{s5.1.3a}

In the previous subsection we showed that any ${\cal H}^\otimes_\gamma$
for $\gamma\in\Gamma^\omega_\sigma$ can be obtained as the inductive 
limit of a sequence of Hilbert spaces ${\cal H}^\otimes_{\gamma_n}$ where 
$\gamma_n\in\Gamma^\omega_0$. It is therefore natural to ask 
whether not all of ${\cal H}^\otimes$ arises in turn as the inductive limit
of the ${\cal H}^\otimes_\gamma$ for $\gamma\in\Gamma^\omega_\sigma$.

The answer turns out to be negative, however, there is an inductive
substructure which we now describe.

Notice that if either 1) any of 
$\gamma-\tilde{\gamma},\gamma'-\tilde{\gamma}$ is an infinite graph or 2) 
any $\tilde{e}\in\tilde{\gamma}$ is a composition of an infinite number
of edges of $\gamma$ or $\gamma'$, or 3) the number of those $\tilde{e}$,
which are compositions of more than one edge of $\gamma$ or $\gamma'$ 
respectively, is infinite then almost always 
the expression (\ref{5.25}) will vanish, simply again because the 
associative law does not hold on the ITP. It follows that if 
$\gamma\subset\gamma'$ but one of the three cases
1), 2), 3) just listed applies, a generic function $f_\gamma\in {\cal 
H}^\otimes_\gamma$ cannot be written as a linear combination of functions
$f_{\gamma'}\in {\cal H}^\otimes_{\gamma'}$. 

This implies that, although $\Gamma^\omega_\sigma$
is a set directed by inclusion, we {\it cannot} simply define 
${\cal H}^\otimes$ as the inductive limit of the ${\cal H}^\otimes_\gamma$.
To see this, notice that given
$\gamma,\gamma'\in\Gamma^\omega_\sigma$ 
with $\gamma\subset\gamma'$ there is only one natural candidate for a 
unitary map $\hat{U}_{\gamma\gamma'}:\;{\cal H}_\gamma\mapsto
{\cal H}_{\gamma'}$ :\\
For any $e\in E(\gamma)$, find its breakup $e=e_1^{\prime n_1}\circ..\circ 
e_N^{\prime n_N},\;N\le \infty$ into edges of $\gamma'$ where $n_k=\pm 1$.
We then consider the functon $p_{\gamma\gamma'}:\;\ab_\gamma\mapsto
\ab_{\gamma'};\;h_e\mapsto h_{e_1^{\prime n_1}}..h_{e_N^{\prime n_N}}$
and then define 
\be \label{5.25c}
\hat{U}_{\gamma\gamma'} f_\gamma:=[p_{\gamma\gamma'}^\ast f_\gamma]\;
\otimes\; [\otimes_{e'\in E(\gamma')-\gamma} 1]
\ee
This map is unitary when considered as a map from ${\cal H}^\otimes_\gamma$
into ${\cal H}^\otimes$, with ${\cal H}^\otimes$ as defined in the previous
section, since the extended Ashtekar
Lewandowski measure is consistently defined. However, the right hand side of
(\ref{5.25c}) will for a generic element $f_\gamma\in 
{\cal H}^\otimes_\gamma$
simply not define an element of ${\cal H}^\otimes_{\gamma'}$ for the reason
already explained.

This state of affairs is in sharp contrast with the situation for the 
category
$\Gamma^\omega_0$ where the Hilbert space could indeed be written as 
the inductive limit of the various ${\cal H}_\gamma\equiv
{\cal H}^\otimes_\gamma$. For the category $\Gamma^\omega_\sigma$ 
the only way to define the Hilbert space structure is through
(\ref{5.25}).

The inductive limit still has a limited application in the following 
sense :\\ 
First, we define a new partial order $\sqsubset$ on 
$\Gamma^\omega_\sigma$, motivated by the conditions 1), 2) and 3)
at the beginning of this subsection.
\begin{Definition} \label{def5.3b}
For $\gamma,\gamma'\in\Gamma^\omega_\sigma$ we define 
$\gamma\sqsubset\gamma'$ if and only if 1) $\gamma\subset\gamma'$ 
and 2) there exist disjoint (up to common vertices) unions
$\gamma=\tilde{\gamma}\cup\gamma_1,\;\gamma'=\tilde{\gamma}\cup\gamma'_1$
where $\tilde{\gamma}\in\Gamma^\omega_\sigma$ and 
$\gamma_1,\gamma_1'\in\Gamma^\omega_0$. 
\end{Definition}
It is important to notice that condition and 2) is not equivalent 
with 2') that $\gamma'-\gamma\in\Gamma^\omega_0$. This is because 
$\gamma\subset\gamma'$ only means that every edge $e\in E(\gamma)$ is a 
countable composition of edges $e'\in E(\gamma)$ and 
2') then does not exclude the existence 
of either a) at least one edge of $\gamma$ which is a countably infinite 
composition of edges of $\gamma'$ or b) an infinite number of edges
which are composed of at least two edges of $\gamma'$. Both possiblities
a) and b) are excluded by condition 2) which, in addition, implies
2'). 
\begin{Lemma} \label{la5.2a}
The relation $\sqsubset$ of definition \ref{def5.3b} is a partial order.
\end{Lemma}
Proof of Lemma \ref{la5.2a} :\\
Only transitivity is nontrivial to prove. If
$\gamma\sqsubset\gamma'\sqsubset\gamma^\dprime$ then first of all
$\gamma\subset\gamma'\subset\gamma^\dprime$ so that 
$\gamma\subset\gamma^\dprime$. Secondly, if 
$\gamma=\tilde{\gamma}\cup\gamma_1,\gamma'=\tilde{\gamma}\cup\gamma'_1
=\tilde{\gamma}'\cup\gamma'_2,
\gamma^\dprime=\tilde{\gamma}'\cup\gamma^\dprime_1$
are the corresponding disjoint unions with $\tilde{\gamma},\tilde{\gamma}'
\in\Gamma^\omega_\sigma$ and $\gamma_1,\gamma'_1,\gamma'_2,\gamma^\dprime_1
\in\Gamma^\omega_0$ then we may define 
$\tilde{\gamma}^\dprime:=
\gamma'-(\gamma_1'\cup\gamma_2')=\tilde{\gamma}-\gamma'_2=
\tilde{\gamma}'-\gamma'_1$ 
which is obviously a subgraph
of both $\tilde{\gamma},\tilde{\gamma}'$ and an element of 
$\Gamma^\omega_\sigma$ since $\gamma_1'\cup\gamma_2'\in\Gamma^\omega_0$
(recall that $\Gamma^\omega_0$ is closed under unions due to piecewise
analyticity and compact support of all its edges). It follows that there
exist $\gamma_2,\gamma_2^\dprime\in\Gamma^\omega_0$ such that 
$\gamma=\tilde{\gamma}^\dprime\cup\gamma_2$ and
$\gamma^\dprime=\tilde{\gamma}^\dprime\cup\gamma_2^\dprime$ are disjoint 
unions.\\
$\Box$\\
It is easy to see that $\Gamma^\omega_\sigma$ equipped with
this partial order is not a directed set. This motivates to construct
directed subsets.
\begin{Definition} \label{def5.3c}
Two graphs $\gamma,\gamma'\in\Gamma^\omega_\sigma$ are said to be finitely 
related, $\gamma\sim\gamma'$, provided that $\gamma,\gamma'\sqsubset
\gamma\cup\gamma'$.
\end{Definition}
\begin{Lemma} \label{la5.2b}
Finite relatedness is an eqivalence relation.
\end{Lemma}
Proof of Lemma \ref{la5.2b} :\\
Reflexivity and symmetry are trivial to check. To see transitivity 
notice that $\gamma,\gamma'\sqsubset\gamma\cup\gamma'$ implies the 
existence of $\tilde{\gamma},\tilde{\gamma}'\in\Gamma^\omega_\sigma$
and of $\gamma_1,\gamma^\dprime_1,\gamma'_1,\gamma^\dprime_2
\in\Gamma^\omega_\sigma$ such that we obtain disjoint unions
$\gamma=\tilde{\gamma}\cup\gamma_1,\gamma'=\tilde{\gamma}'\cup\gamma'_1,
\gamma\cup\gamma'=
\tilde{\gamma}\cup\gamma^\dprime_1=\tilde{\gamma}'\cup\gamma^\dprime_2$.
The last equality demonstrates that we may write the disjoint union
$\gamma\cup\gamma'=
\tilde{\gamma}^\dprime\cup(\gamma^\dprime_1\cup\gamma^\dprime_2)$
with 
$\Gamma^\omega_\sigma\ni \tilde{\gamma}^\dprime
=\gamma\cup\gamma'-(\gamma^\dprime_1\cup\gamma^\dprime_2)
=\tilde{\gamma}-\cup\gamma^\dprime_2=\tilde{\gamma}'-\gamma^\dprime_1$.
This in turn implies that we may actually write also disjoint unions
$\gamma=\tilde{\gamma}^\dprime\cup\gamma_2,\gamma'=\tilde{\gamma}^\dprime
\cup\gamma'_2$ for some $\gamma_2,\gamma'_2$. In other words,
$\gamma\sim\gamma'$ guarantees property 2) of definition \ref{def5.3b}.
Transitivity now follows from the transitivity part of the proof of lemma
\ref{la5.2a}.\\
$\Box$\\
We conclude that $\Gamma^\sigma_\omega$ 
decomposes into equivalence classes $(\gamma_0)$, called {\it clusters} and 
labelled by representants $\gamma_0$, called {\it sources}. Now, by 
construction, each cluster is directed by $\sqsubset$. 
Moreover, since also by construction for any $\gamma,\gamma'\in(\gamma_0)$
the three conditions 1), 2) and 3) 
observed at the beginning of this subsection are not met, we find,
in particular, that 
the operator (\ref{5.25c}) for $\gamma\sqsubset\gamma'$ is now indeed a 
unitary operator which obviously satisfies the 
consistency condition 
$\hat{U}_{\gamma'\gamma^\dprime}\hat{U}_{\gamma\gamma'}=  
\hat{U}_{\gamma\gamma^\dprime}$ for 
$\gamma\sqsubset\gamma'\sqsubset\gamma^\dprime$. The general results
of section \ref{s4.3} now reveal the existence of the inductive
limit Hilbert space ${\cal H}^\otimes_{(\gamma_0)}$ for any cluster
$(\gamma_0)$ and corresponding unitarities 
$\hat{U}_\gamma:\;{\cal H}_\gamma^\otimes\mapsto
{\cal H}^\otimes_{(\gamma_0)}$ for any $\gamma\in (\gamma_0)$ such that
$\hat{U}_\gamma=\hat{U}_{\gamma'}\hat{U}_{\gamma\gamma'}$.

It would be a very pretty result if one could establish that the Hilbert 
spaces ${\cal H}_{(\gamma_0)}$ corresponding to different clusters are 
mutually orthogonal with respect to (\ref{5.25}). But this is certainly not 
the case, just take any $\gamma\in(\gamma_0)\not=(\gamma_0')\ni\gamma'$ and 
consider the $C_0$-vectors $f_\gamma:=\otimes_{e\in E(\gamma)} 1$ and 
$f_{\gamma'}:=\otimes_{e'\in E(\gamma')} 1$. Then trivially
$<\otimes_{f_\gamma},\otimes_{f_{\gamma'}}>=1$.

Denote by ${\cal C}^\omega_\sigma$ the set of clusters in 
$\Gamma^\omega_\sigma$. Then we have the following equivalent definition
of the full Hilbert space 
\be \label{5.25d}
{\cal H}^\otimes=\overline{\cup_{(\gamma_0)\in{\cal C}^\omega_\sigma}
{\cal H}^\otimes_{(\gamma_0)}}
\ee
which displays it as a kind of cluster decomposition. The decomposition is,
however, not a direct sum decomposition. Obviously, the cluster Hilbert
spaces ${\cal H}^\otimes_{(\gamma_0)}$ are mutually isomorphic and therefore 
in particular isomorphic with the original Ashtekar Lewandowski 
Hilbert space ${\cal H}_{AL}\equiv {\cal H}^\otimes_{(\emptyset)}$ 
based on $\Gamma^\omega_0$ obtained by choosing as the source the empty 
graph. The fact that the cluster Hilbert spaces can be written as inductive 
limits is then not any more surprising because they are isomorphic with 
${\cal H}_{AL}$ of which we knew already that it is an inductive limit.

\subsubsection{Rigging Triple Structures}
\label{s5.1.4}

Finally we can equip the space Cyl with a topology in analogy with
the one defined in \cite{23}, the difference coming from the fact that
we do not know an explicit orthonormal basis for ${\cal H}$.
\begin{Definition} \label{def5.3a}
Choose for any $\gamma\in \Gamma^\omega_\sigma$ once and for 
all a von Neumann basis $T_{\gamma s\beta}$ over $\gamma$ where 
$s\in{\cal S}_\gamma$ 
runs through the strong equivalence classes
in ${\cal H}_\gamma$ and $\beta$ through the set of functions 
${\cal F}_\gamma$ defined in (\ref{4.9}). 
Let $f\in${\rm Cyl} be a cylindrical function.  \\
i)\\ 
The family of Fourier semi-norms of $f$ is defined by 
\be \label{5.25a}
|||f|||_\gamma:=\sum_{s\beta} |<T_{s\beta},f>|
\ee
where the inner product in (\ref{5.25a}) is defined by (\ref{5.25}). 
Notice that 
indeed $|||f+g|||_\gamma\le |||f|||_\gamma+|||g|||_\gamma,\;|||z f|||_\gamma
=|z|\;|||f|||_\gamma$ for all $\gamma\in\Gamma^\omega_\sigma,\;z\in\Co,\;
f,g\in${\rm Cyl}. Obviously the family separates the points of {\rm Cyl} 
since
$|||f|||_\gamma<\infty$ for all $\gamma$ implies $f\in{\cal H}$ and 
$|||f|||_\gamma=0$ for all $\gamma$ implies $||f||=0$, that is,
$f$ is the zero $C_0$ sequence and so $f=0$, the zero $C$ function
in {\rm Cyl}.\\
ii)\\
Consider the subspace $\Phi$ of {\rm Cyl} consisting of elements 
which are finite linear combinations of $C$ functions $f$ with the 
property that 
\be \label{5.25b}
|||f|||:=\sup_{\gamma}|||f|||_\gamma<\infty
\ee
iii)\\ 
Item i) displays $\Phi$ as a locally convex vector space. Upon equipping
it with the natural topology (the weakest topology such that all the 
$|||.|||_\gamma$ and addition are continuous) it becomes a topological 
vector space $\Phi$.\\
\end{Definition}
An alternative choice for a topology for $\Phi$, upon which it would become
a normed (but not necessarily complete, Banach) topological vector space, 
is given by the norm (\ref{5.25b}).
The natural topology defined in iii) is not generated by a countable
set of seminorms, therefore it is not metrizable \cite{51} and 
(upon completion) cannot be a Fr\'echet space. On the other hand, the norm
$|||.|||$ certainly defines a metric. The two topologies are therefore 
not equivalent, clearly the natural topology is weaker than the norm 
topology. 
\begin{Lemma} \label{la5.3}
We have $\Phi\subset{\cal H}$ and $\Phi$ is dense in ${\cal H}$. 
\end{Lemma}
Proof of Lemma \ref{la5.3} :\\
To see this, consider an arbitrary element $f=\sum_{n=1}^N f_{\gamma_n}$
where $f_{\gamma_n}$ is a $C$ function over $\gamma_n$in $\Phi$ and 
$N<\infty$. Therefore, 
$|||f_{\gamma_n}|||_{\gamma}<\infty$ for any $\gamma,\;n=1,..,N$. 
In particular,
$|||f_{\gamma_n}|||_{\gamma_n}<\infty$ for any $\gamma$. Thus,
$$
|||f_{\gamma_n}|||_{\gamma_n}=
\sum_{s,\beta}|<T_{\gamma s\beta},f_{\gamma_n}>|<\infty
$$ 
and therefore
$$
||f_{\gamma_n}||^2=
\sum_{s,\beta} |<T_{\gamma s\beta},f_{\gamma_n}>|^2<
(|||f_{\gamma_n}|||_{\gamma_n})^2<\infty
$$ 
from which we see that $f_{\gamma_n}$ is a $C_0$ vector. It follows that
$f$ is a finite linear combinations of $C_0$ vectors and thus an element 
of ${\cal H}$. As the finite linear combinations of $C_0$ vectors form
a dense subset of $\cal H$ which we just showed to be contained in
$\Phi$, we conclude that $\Phi$ is dense in $\cal H$.\\
$\Box$\\
With the natural topology on $\Phi$ we are equipped with the rigging triple 
$\Phi\subset{\cal H}\subset\Phi'$ 
and can take over the framework of \cite{7} to solve 
constraints also in the context of the ITP.

\subsection{Contact with Semiclassical Analysis}
\label{s5.2}

In this subsection we will make contact with the semi-classical states
of section \ref{s3}. \\
\\
Let $\gamma\in\Gamma^\omega_\sigma$ be an infinite graph, filling 
all of $\Sigma$ arbitrarily densely (in the absence of a background metric
by this we mean simply that for an arbitrary choice of neighbourhoods of 
each point of $\Sigma$, $\gamma$ can be chosen to intersect all of them).
Suppose we are given a solution of the classical field equations
(say the Einstein equations in the absence of matter or the 
Einstein-Yang-Mills equations in the presence of matter, in the
latter case $G$ is the direct product of the gravitational $SU(2)$
with the $SU(3)\times SU(2)\times U(1)$ of the standard model),
that is, for each time slice $\Sigma_t,\;t\in \Rl$ we have an initial data 
set $(A^0_t(x),E^0_t(x)),\;x\in\Sigma$ satisfying the field equations 
and in particular the constraint equations. Moreover, we will have to choose
a certain gauge to write down the solution explicitly. Then, with the 
techniques of section \ref{s3}, for each edge $e\in E(\gamma)$ and given
classicality parameter $s$ we obtain a normalized coherent state 
\be \label{5.26}
\xi^s_{g_e^t(A^0,E^0)}:=
\frac{\psi^s_{g_e^t(A^0,E^0)}}{||\psi^s_{g_e^t(A^0,E^0)}||_{{\cal H}_e}}
\ee
where $g_e^t((A^0,E^0))=\exp(-i\tau_j P^e_j(E^0_t,A^0_t)/(2a^2)) h_e(A^0_t)$
for pure general relativity.
Finally, we consider the $C_0$-vector over $\gamma$ given by
\be \label{5.27}
\xi^s_{\gamma,(A^0_t,E^0_t)}:=\otimes_{e\in E(\gamma)} 
\xi^s_{g_e^t(A^0,E^0)}
\ee
These states comprise a preferred set of {\it coherent states over
the infinite graph $\gamma$} and provide the basic tool with which to
address the following list of fascinating physical problems :
\begin{itemize}
\item[i)] Given one and the same graph $\gamma$ and classicality parameter 
$s$, when are the strong and weak 
equivalence classes of the states (\ref{5.27}) equal to each other ?
What is the physical significance of strong and weak equivalence anyway ?
From experience with model systems one expects that two
different weak equivalence classes correspond to drastically different
physical situations such as an infinite difference in ground state energies 
or topologically different situations while the general analysis 
of section \ref{4.1} (lemma \ref{4.14}) states that two incomplete
ITP's corresponding to different strong equivalence classes within the same 
weak one are unitarily equivalent. 

Of course, different topological
situations can be described within the same complete ITP only
if we get rid off the embedding spacetime that one classically started with.
One way to do this would be roughly as follows : \\
Consider $\gamma\in\Gamma^\omega_\sigma$ not as an
embedded graph but merely as a countable, combinatorical one.
A countable combinatorical graph is simply a countable collection of edges 
(which 
are analytic curves when embedded into any given $\Sigma$), and vertices
together with its connectivity relations, that is, information telling us at 
which vertices a given edge ends. 
Now recall that the spectrum 
$\mbox{spec}_\Sigma(\hat{O})$ of important operators $\hat{O}$ of the 
theory such as the area operator (see \cite{15}), as obtained on
the Hilbert spaces corresponding to graphs embedded in a concrete
$\Sigma$, depends on the 
topology of $\Sigma$ and it should be true that a complete set 
of operators encodes full information about the topology of $\Sigma$
via the range of their respective spectra. 
We can now define the {\it universal operator} $\hat{O}$ acting on Hilbert 
spaces 
over combinatorical graphs by a new kind of {\it summming over topologies}, 
namely, one allows the spectrum of $\hat{O}$ to take 
all possible values, that is, $\mbox{spec}(\hat{O})=
\cup_\Sigma \mbox{spec}_\Sigma(\hat{O})$. One would then say that
a given closed subspace of the Hilbert space, carrying a representation of 
the operator algebra describes a concrete topology of
$\Sigma$ provided the spectra of the operators $\hat{O}$ resricted to that
subspace are compatible with th spectra $\mbox{spec}_\Sigma(\hat{O})$.

Now the Infinite Tensor Product Hilbert space as obtained from combinatorical
graphs space comes in as follows. It is 
expected that closed $[f]$-adic subspaces of that ITP corresponding to 
different topologies of $\Sigma$ in the way just described also
correspond to strong equivalence classes within different weak equivalence 
classes. Now while elementary operators of the theory will leave 
these subspaces invariant, there are densely defined operators on 
the complete ITP which mediate between the two. Thus, the ITP might be used 
to describe {\it topology change} in Quantum General Relativity and 
would then wipe away one of the main criticisms directed towards the 
whole programme.

A related interesting question is,
whether classical states ($C_0$-vectors) corresponding to Minkowski
and Kruskal spacetime respectively are orthogonal, more generally, 
whether one can superimpose classical states corresponding to 
globally different spacetimes within the same strong equivalence class. 
Interestingly, all this can be 
analyzed by performing relatively straightforward calculations of the 
type outlined in \cite{31,32}.
\item[ii)] Given one and the same graph $\gamma$ and solution $(A_t^0,E^0_t)$, 
are the $C_0$ vectors (\ref{5.27}) for different values of $s$ in the same 
weak equivalence class for non-compact $\Sigma$ ? Since the parameter 
$s$ plays a role similar to 
a mass parameter in free scalar field theory on Minkowski space 
one might expect this not to be 
the case as the Fock representations over Minkowski space with different 
mass are not unitarily 
equivalent. Indeed, it is easy to see that for Minkowski space in the 
gauge $A_a^j=0,E^a_j=\delta^a_j$ we have 
$<\xi^s_e,\xi^{s'}_e>=|<\xi^s_e,\xi^{s'}_e>|=q<1$
for $s\not=s'$ so that we obtain different weak eqivalence classes.
\item[iii)] Given one and the same $(A_t^0,E_t^0)$ and $s$, what happens
under refinements of the graph $\gamma$ ? Again, since under refinements of
an infinite graph we perform an infinite change on the graph, from expression
(\ref{5.25}) we expect the corresponding $C_0$ vectors to be orthogonal.
This turns out to be correct. 
\item[iv)] Given a $C_0$ vector $\xi$ of the type (\ref{5.27}) we know from 
lemma 
\ref{la4.9} that any other vector in the $[\xi]$-adic incomplete ITP  
can be obtained as a (Cauchy sequence of) linear combinations of
$C_0$ vectors each of which differs from $\xi$ in only a finite number of 
entries $e\in E(\gamma)$. Now consider a deformation of a classical
solution $(A^0_t,E^0_t)$, say the Minkowski metric plus a graviton
or photon.
A plane wave graviton is everywhere excited over $\Sigma$, that is, 
differs everywhere significantly from the Minkowski background, and 
therefore will not lie in the closure of the states just described. 
In the case of the electromagnetic field this is expected because plane
wave solutions have infinite energy.
But
the anyway more physical graviton wave packets, although also
everywhere excited, are Gaussian damped and thus have a chance
to lie in that $[\xi]$-adic strong equivalence class of Minkowski 
space (or any other background). We expect that there is a unitary map
between the usual Fock space description of gravitons and the 
coherent state Gaussian wave packet gravitons of the present 
framework. If that turns out to be correct, we can also describe 
Einstein-Maxwell-Theory
this way and consider photons propagating on quantum spacetimes.
These issues will be examined in \cite{60}. 

The same analysis can, of course, be performed on any background
and this is the way we will try to describe the Hawking effect in this 
approach \cite{65}.
\item[v)] Related to this is the question if we can recover the 
spectacular results of Quantum Field Theory on Curved Backgrounds anyway
\cite{61a}.
The philosophy of that approach is that if the backreaction of matter to 
geometry can 
be safely neglected then treating the metric as a given, classical  
background should be a good approximation to the physics of the system.
Recently, \cite{62} there has been a quantum leap in this field of research
due to a precise formulation of the microlocal spectrum condition on
arbitrary, globally hyperbolic but not necessarily stationary backgrounds.

It is to be expected, and an important consistency check, that to zeroth 
order in 
the Planck length the full quantum gravity calculation should agree
with the predictions of Quantum Field Theory on curved backgrounds, in other
words, QFT on curved backgrounds is a semi-classical limit of quantum 
gravity. The way to check this expectation is of course the following : 
The total Hilbert space of the system matter plus geometry is the tensor
product of the Hilbert spaces for the matter sector and the gravity sector
respectively. Given a classical background metric, we will choose
states $\psi_{total}$ which are tensor products of arbitrary states 
$\psi_{matter}$ from the 
matter Hilbert space with one fixed state $\psi^0_{grav}$ from the gravity 
Hilbert space, namely the coherent $C_0$ vector for the 
metric to be approximated, symbolically $\psi_{total}=
\psi_{matter}\otimes \psi_{grav}^0$. The matter 
Hamiltonian operator of, say, bosonic matter coupled to quantum gravity 
is roughly a linear combination of operators of the form 
$\hat{H}_{total}=\hat{A}_{matter}\hat{A}_{grav}$.
Thus we find for the matrix elements of that operator
$$
<\psi_{total},\hat{H}_{total}\psi'_{total}>_{total}=
<\psi_{matter},\hat{A}_{matter}\psi'_{matter}>_{matter}
<\psi^0_{grav},\hat{A}_{grav}\psi^0_{grav}>_{grav}
$$
which shows that we obtain an {\it effective matter Hamiltonian} given by
$\hat{H}_{total}^{eff}=\hat{A}_{matter}
<\psi^0_{grav},\hat{A}_{grav}\psi^0_{grav}>_{grav}
$
which by the properties of the operator of \cite{20} is {\it finite} !
The corrections to the classical background metric are of course 
contained in the difference
$<\psi^0_{grav},\hat{A}_{grav}\psi^0_{grav}>_{grav}-A_{grav}^{class}$
where $A_{grav}^{class}$ is the classical limit of the gravity operator
evaluated for the given classical background. This quantity is certainly
at least of order $\ell_p$. However, this is not the only correction to 
Quantum Field Theory on curved backgrounds. A second correction comes from
the fact that our theory is non-perturbative which means, in particular,
that the matter Hilbert space is not the usual perturbative Fock space.
Therefore it is not at all obvious that the spectrum of the operator
$\hat{A}_{matter}$ on the non-perturbative Hilbert space coincides with the
spectrum of the usual matter Hamiltonian on the usual perturbative Fock 
space. The spectra better agree, at least modulo corrections of order at 
least $\ell_p$, in order that we can claim to have quantized a theory which
has general relativity plus standard matter as the classical limit.
\item[vi)] A first application of this procedure to discover new physical
effects due to quantum gravity is the exact treatment of the so-called 
{\it $\gamma$-ray-burst effect} \cite{34} which we will do in \cite{35}.

To date the exact astrophysical explanation or source for high energetic
$\gamma$-photons (up to TeV !) is unclear but what is important for us
is that these photons were created billions of years ago, they can come 
from distances comparable to the Hubble radius. The idea is that these
photons on their way to us constantly are influenced by the vacuum 
fluctuations of the gravitational field and although the influence is very, 
very, very small, it can accumulate due to the long travelling time of 
the photons. Now the higher energetic the photon, the more it should 
probe the small scale discreteness of quantum geometry and we thus 
expect an energy-dependent dispersion law. The dispersion law being
energy and therefore (Minkowski) frame dependent, it violates
Poincar\'e invariance. The effect therefore cannot 
come from any perturbative theory (interacting QFT on Minkowski space,
perturbative quantum (super)gravity, perturbative string theory) all of 
whose $S$-matrix elements or $n$-point functions are by definition
(or (Wightman) axiom) Poincar\'e invariant. For observational
purposes it is convenient that the intensity peak of the burst can
have a time width as small as of the order of  1ms. The idea is then
to calibrate the detector to detect events at energies $E_2>E_1$ at times
$t_2>t_1$ due to the energy dependence of the speed of light which 
according to \cite{34} is speculated to be of the form 
$c(E)/c(0)=1-k(E/E_{eff})^\alpha$. Here $k$ is a coefficient 
of the order of $1$, $E_{eff}$ is the effective quantum gravity scale and
$\alpha$ is a power which is hopefully of the order $1$ for the effect
to be detectable. For $\alpha=1$ one finds $t_2-t_1=k[(E_2-E_1)/E_{eff}] 
\;[L/c(0)]$
where $L$ is the distance of the source inside a galaxy and so can be 
determined from its redshift. Inserting the numbers for a burst which is
a billion lightyears away and $E_2-E_1=1$TeV we get $t_2-t_1$ of the 
order of a second (!) if we set $E_{eff}=m_p$ which is large enough 
compared to the width of the signal. Thus, 
for $\alpha=1$ the effect could be indeed observable, say by a 
\v{C}erenkov observatory \cite{63} (but not for $\alpha=2$), 
at least in principle, however, experimentally it is a highly non-trivial 
task to take into 
account all possible errors (dark matter, gravitational lensing, dust, 
atmosphere, ...) and to 
make sure that the measured intensities really came from the same burst.

Our aim in \cite{35} will be to compute $k$ and $\alpha$, or more generally,
the precise dispersion law, {\it exactly} along the lines outlined in 
item v). It is important to realize, that 
the effect is an inevitable theoretical prediction of quantum general 
relativity in 
the present formulation due to the {\it Heisenberg Uncertainty Obstruction}.
Namely, the quantum metric depends on magnetic (connection $A$) and 
electric (conjugate momentum $E$) degrees of freedom which upon quantization
become {\it noncommutative} operators as we have seen in section \ref{s2.2}
and therefore cannot be simultaneously diagonalized. Thus, the best we 
can do is to write down a best approximation eigenstate of the metric
operator, that is, a 
{\it coherent state} which saturates the Heisenberg uncertainty bound.
As we showed in \cite{31,32}, our states of section \ref{s3} have 
precisely these semiclassical properties. However, while a best approximation
state, it is not an exact eigenstate and thus cannot be Poincar\'e
invariant.

It is also important to see that our analysis is more ambitious than
the pioneering work \cite{36}
for three reasons : First of all, instead of coherent states only weaves 
\cite{64} were used, however,
these approximate only half of the degrees of freedom and are more similar
to momentum eigenstates than semiclassical states. Secondly, the matter
field was treated classically and one was computing only the 
dispersion law coming from the changed d'Alembert operator, an option 
which we also have. Thirdly, in contrast to our coherent states,
the weave with the assumed semi-classical behaviour was not proved to 
exist as a normalizable state of the Hilbert space.
\item[vii)] The results of \cite{17,18,19,20,21,22,23} are a small 
indication that quantum gravity plus quantum matter combine to a 
{\it finite} quantum field theory. An elementary particle physicist 
who computes Feynman diagrammes and has to renormalize divergent quantities
all the time will rightfully ask what happened to the ultraviolet 
divergencies
of his everyday life. A short answer seems to be, that in a diffeomorphism 
invariant, background independent theory there is no room for UV 
divergencies since there is no difference between ``large'' and ``small''
distances, the renormalization group gets ``absorbed'' into the 
diffeomorphism group. While plausible, 
to the best of our knowledge nobody has so far investigated these 
speculations in detail. This will be the topic of \cite{61}.
\item[viii)] The result of \cite{17,20} shows that the geometry and matter
Hamiltonian (constraints) are densely defined operators on the 
{\it unextended} Ashtekar-Lewandowski Hilbert space. How does the 
situation change with the huge extension of the Hilbert space performed
in section \ref{s5.1} ? The answer, investigated in detail in \cite{66}
is that, not unexpectedly, these operators are not densely defined 
on all of $\cal H$, however, they are on {\it physical} $[f]$-adic 
Hilbert spaces for $f$ a coherent $C_0$-vector. Here we call a
coherent $C_0$-vector physical if it is labelled by classical field 
configuration that obeys the fall-off conditions at spatial infinity.
To see how this roughly comes about suppose we have a coherent $C_0$
vector $\otimes_f=\otimes_e f_e$ over some infinite graph $\gamma$. The 
Hamiltonian constraint operators applied to $\otimes_f$ are of the form
$\hat{H}\otimes_f=[\sum_v \hat{H}_v] \otimes_f$ where the sum is 
over all vertices of $\gamma$ and $\hat{H}_v$ influences only those 
$f_e$ for which its edge $e$ is incident at $v$, that is, it is a 
local operator. Our first observation is that therefore 
$\hat{H}_v\otimes_f\in {\cal H}_{[f]}$ as guaranteed again by lemma
\ref{la4.9}. So $\hat{H}\otimes_f$ is a countably infinite sum of 
vectors in ${\cal H}_{[f]}$ and the question is whether it is convergent,
that is, whether $||\hat{H}\otimes_f||^2<\infty$. Suppose now that 
$f_e=\xi^s_{g_e^t(A^0,E^0)}$ and $\otimes_f$ is a coherent $C_0$ vector
peaked at $A_t^0,E_t^0$. Then we have by the Ehrenfest theorem of \cite{32} 
$$
||\hat{H}\otimes_f||^2=<\otimes_f,\hat{H}^\dagger\hat{H}\otimes_f>
=|\sum_v H_v(A_t^0,E_t^0)|^2[1+O(s)]
$$
where $H_v(A^0_t,E^0_t)$ is by construction a dicretization of an integral 
over a small region in $\Sigma_t$ of the classical Hamiltonian density
$H(A_t^0,E_t^0)$ and the sum over vertices is a Riemann sum for the 
integral $\int_{\Sigma_t} d^3x H(A_t^0,E_t^0)$. It follows that the 
norm exists if and only if the field configuration $A_t^0,E_t^0$ 
satisfies the fall-off conditions, that is, {\it if it is a point in the 
classical phase space} (no constraints or field equations yet being imposed).
Next, again from lemma \ref{la4.9} we see that for such coherent $C_0$
sequences $f$ the Hamiltonian constraint is densely defined on the 
$[f]$-adic Hilbert space because a dense domain is given by the vector
space of finite linear combinations of $C_0$-vectors differing from 
$\otimes_f$ in at most a finite number of entries $f_e$ and the 
convergence proof just outlined certainly goes through since the finite
number of changes made affect a finite number of vertices only and these 
are of $d^3x$ measure zero in the limit of infinitely fine graphs.

This result is rather pretty because it tells us that the classical 
theory still has some effect on the quantum theory, not every $[.]$-adic
incomplete ITP carries a representation of the operator algebra. The 
set of $C_0$ vectors whose $[.]$-adic ITP's do carry a representation 
includes the physical coherent $C_0$ vectors but excludes the 
non-physical ones. Since coherent states form an overcomplete basis on 
the complete ITP, this statement seems to give a rather complete 
classification of $[.]$-adic ITP's carrying a representation. 

However, there are also other $[.]$-adic ITP's which are not of that form :
An example is provided by the Ashtekar-Lewandowski state $\omega_{AL}$ 
which on the complete ITP $\cal H^\otimes$ is given by the GNS cyclic vector 
$\Omega_{AL}=\otimes_{e\in C_\infty} 1$
where as before $C_\infty$ denotes a supergraph. Now it is easy to check,
using the overcompleteness of Hall's coherent states,
that $1=1_e=\int_{G^\Co} d\nu_s(g_e) \psi^s_{g_e}$ where $\nu_t$ is 
Hall's measure \cite{44}, that is, the Ashtekar-Lewandowski $C_0$-vector
is an infinite-fold superposition of {\it all} coherent states formally 
given by the (kind of functional integral)
$$
\Omega_{AL}=\int [\otimes_e d\nu_s(g_e)] \otimes_e \psi^s_{g_e}
$$
and so includes non-physical coherent $C_0$-vectors which, however,
come with the appropriate weight enabling it to carry a representation of 
the observable algebra. This way to write $\Omega_{AL}$ makes it
obvious that it is not peaked on a particular metic at all although it
is annihilated by all momentum operators which might lead one to assume that
$\Omega_{AL}$ approximates the zero metric ! This is clearly not the case.

Another way to write $\Omega_{AL}$ is 
$\Omega_{AL}=\lim_{s\to\infty} \otimes_{e\in C_\infty} \psi^s_{g_e}$ 
which displays it as a
$C_0$-vector, but only in the {\it anticlassical limit} $s\to\infty$ !
Both ways to write $\Omega_{AL}$ 
reaffirm one more time the impression that in the spatially non-compact case 
the $[1]$-adic incomplete ITP, or, in other 
words, the original, unextended Ashtekar-Lewandowski Hilbert space is a 
{\it pure quantum representation} of the observable algebra which seems  
to have no obvious
semi-classical correspondence. All the solutions found in \cite{18}
are (diffeomorphism invariant versions of) states in that Hilbert space
and thus have presumably no (semi)classical relevance, as speculated by many, 
it is indeed correct that not only infinite linear combinations
but indeed {\it infinite products} are necessary to catch the physically 
relevant sector of the complete ITP. This could also explain the discrepancy
with respect to the number of degrees of freedom in 2+1 gravity pointed out
in \cite{19} : There the $[1]$-adic incomplete ITP was used and gave 
rise to an infinite number of physical states each of which describes  
zero volume almost everywhere. As in that case $\Sigma$ was assumed to
be compact, the $[1]$-adic incomplete ITP coincides in fact with 
${\cal H}^\otimes$, however, the true physical states are only obtained 
as infinite linear combinations of zero volume states which presumably
builds up an $[f]$-adic incomplete ITP describing an everywhere 
non-degenerate metric. We will come back to this in a future publication
\cite{63a}. 
\item[ix)] A heuristic method to use the coherent states 
in order to derive 
a Hamiltonian constraint operator with the correct classical limit
is as follows :
We choose a point $(A^0_{t=0}(x),E^0_{t=0}(x)),\;x\in\Sigma$ on the 
constraint surface of the phase space of,
say, general relativity in some gauge 
(that is, a field configuration on $\Sigma_0$) and  
obtain its trajectory under the Einstein evolution to first order in
the time parameter $t$ as $A^0_t(x)=A^0_0(x)+t\{H_0(N),A^0_0(x)\}$ 
(that is, a field configuration on $\Sigma_t$) and likewise 
for $E_t(x)$. Here, $H_0(N)$ is the Hamiltonian constraint on $\Sigma_0$.
We thus obtain coherent states as in (\ref{5.27}) over any 
$\gamma\in\Gamma^\omega_\sigma$ and now define a family of 
operators $\hat{H}_\gamma(N)$, each of which is densely defined on the 
$[.]$-adic incomplete ITP corresponding to the strong equivalence class of
the $C_0$-sequence associated with the coherent $C_0$-vector (\ref{5.27})
by the definition 
\be \label{5.28} 
\hat{H}_\gamma(N)\xi^s_{\gamma,(A^0_0,E^0_0)}
:=\frac{1}{i} (\frac{d}{dt})_{t=0} \xi^s_{\gamma,(A^0_t,E^0_t)}
\ee
Notice that time evolution preserves the kinematical constraints of the 
phase space, in particular, the fall-off conditions and so the strong
equivalence classes of the $C_0$ sequences defined by the $C_0$-vectors
$\xi^s_{\gamma,(A^0_0,E^0_0)},\xi^s_{\gamma,(A^0_t,E^0_t)}$ are equal
to each other. Since the map $t\mapsto (A^0_t,E^0_t)$ is smooth
on the continuum phase space $M$, it induces a smooth map
on the subset $\bar{M}_{\gamma|M}$ of the graph phase space of section 
\ref{s2.1}.
Therefore, $t\mapsto\xi^s_{\gamma,(A^0_t,E^0_t)}$ is strongly
continuous since 
$||\xi^s_{\gamma,(A^0_t,E^0_t)} -\xi^s_{\gamma,(A^0_0,E^0_0)}||$
depends smoothly on $\bar{M}_{\gamma|M}$. If we could verify also that 
$||\xi^s_{\gamma,(A^0_t,E^0_t)}||=||\xi^s_{\gamma,(A^0_0,E^0_0)}||$
up to terms of order $t^2$ then time evolution would be given 
to first order in $t$ by a one-parameter group of unitary operators
and the existence of $\hat{H}_\gamma(N)$ would follow from 
Stone's theorem \cite{51}. However, due to the complicated classical
constraint algebra $\{H(N),H(M)\}=\int d^3x (M N_{,a}-M_{,a} N)
q^{ab} V_b\not=0$, where $V_b$ is the infinitesimal generator of the 
vector constraint, the quantum evolution is better not unitary in order
that it has the correct classical limit. Therefore, Stone's theorem
will not apply and if we interprete (\ref{5.28}) as a strong limit then 
it might be ill-defined.
\item[x)] A great surprise of quantum theory 
was the resolution of a classical paradoxon, the explanation for the 
stability of atoms. According to classical electrodynamics the 
electrons orbiting the nucleus should emit radiation and fall into
it after a finite time. The discreteness of the bound state energy 
spectrum bounded from below
prevents this from happening and displays a mechanism for
the avoidance of a classical {\it singularity}. It is an 
interesting speculation that something like this could also
happen in quantum gravity, that the classical singularities predicted
by the singularity theorems of Hawking and Penrose \cite{67}
are actually absent in quantum gravity, providing a resolution of the
information paradox. This question can also be naturally adressed within
the framework of coherent states : given a classical black hole spacetime 
with its singularity, say the Kruskal spacetime, we could compute 
the expectation value, with respect to a coherent state for that 
black hole spacetime, of an operator whose classical counterpart becomes
singular there. If the singularity is quantum mechanically absent, then 
the operator should be bounded from above and the expectation value
should be finite. From the point of view of the Bargmann-Segal Hilbert 
space, the coherent state is peaked at the singular spacetime but 
there is a non-zero probability to be away from it just in the right way
to be square integrable. This is in analogy with the eigenstates
of the electron energy operator of the hydrogenium atom whose probability
density at the origin is finite and which are also square integrable
(notice that coherent states are approximate eigenstates of any operator).
More generally, one would like to treat quantum black holes with the 
new semi-classical input provided by coherent states which {\it come out
of the quantum theory} and are not a purely classical input 
such as classically encoding the 
presence of an isolated horizon into the topology of $\Sigma$, thus
inducing corresponding boundary conditions, in quantum general relativity
\cite{25,26,27,28,29,30} or the identification of classical supergravity
black hole solutions with D-brane configurations protected against quantum 
corrections due to the BPS nature of the corresponding states   
in string theory (see, e.g. \cite{68} and references therein). These and 
related questions will be examined in \cite{65}.
\item[xi)] The coherent state framework of \cite{31,32} so far is 
worked out in full detail only for the compact groups of rank one 
and direct products of those. As argued there, the extension to groups of 
higher rank should be straightforward given the strategy for $U(1),SU(2)$
but it is yet a lot of work. It would be important to give the full details
at least for the physically important case of $SU(3)$. The analysis 
will be given in \cite{69}.
\item[xii)] The exposition in section \ref{s4.2} underlines the relevance 
of von Neumann algebras and operator theory for the Infinite Tensor Product.
This provides a pretty interface with the methods of Algebraic Quantum
Field Theory. In particular, it would be interesting to work out the 
Tomita Takesaki Theory for the appearing operator algebras as it was done
for scalar field theory on Minkowski space for diamond regions 
(the Bisognano Wichmann theorem) where the challenge in our context is that 
we have only a spacetime background topology and differentiable 
structure but not a spacetime background metric. Modular theory
is the basic tool to determine the precise type of the type III factors
which from experience with scalar field theory seem to be the most
relevant types of factors in quantum field theory and will be studied
in \cite{70}.

At this point the careful reader will wonder how we can apply the 
theory outlined in section (\ref{4.2}) which was geared only
at bounded operators. However, our basic operators are $\hat{h}_e,
\hat{p}^e_i,\;e\in E(\gamma)$ and the latter is unbounded although
essentially self-adjoint on ${\cal H}_e$, a property which trivially
extends to the ITP. This mismatch can be cured by considering instead
of $\hat{p}^e_i$ the Weyl kind of operator 
\be \label{5.34}
\hat{H}_e:=e^{\tau_j \hat{p}^e_j/2}
\ee
which takes values in the set of group valued bounded operators
and transforms as $\hat{H}_e\mapsto\mbox{Ad}_{g_e(0)}\hat{H}_e$ under 
gauge transformations.
Its boundedness, for instance for $G=SU(2)$, is evident from the formula
\be \label{5.35}
\hat{H}_e=
\frac{e^{is\Delta_e/2}\hat{h}_e e^{-is\Delta_e/2}
\widehat{(h_e)^{-1}}}{e^{is}}
\ee
which can be proved from a similar formula in the first reference 
of \cite{32} by analytical continuation of the classicality parameter $s$.
Here, $\Delta_e$ is the Laplacian on the copy of $G$ corresponding to 
$h_e$. Formula (\ref{5.35}) displays $\hat{H}_e$ as a product of four 
bounded operators. Taking the operator adjoint of (\ref{5.35}) one 
finds that the operators $\hat{H}_e,\hat{h}_e$ obey the kind of Weyl algebra
\be \label{5.36}
\hat{H}_e\hat{h}_e=e^{-2is} \epsilon (\hat{h}_{e^{-1}}\hat{H}_{e^{-1}})^T
\ee
where $(.)^T$ denotes transposition, $\epsilon$ is the skew symmetric spinor
of second rank of unit determinant and $\hat{H}_{e^{-1}}=(\hat{H}_e)^{-1}
=((\hat{H}_e)^\dagger)^T$ (matrix and operator inverse but only operator 
adjoint) clarifies the adjointness relations. 
Notice that one could also consider the objects 
$\hat{H}_e^j:=e^{i\hat{p}^e_j}$ which satisfy the simpler Weyl algebra
$$
\hat{H}_e^j\hat{h}_e=e^{-s\tau_j/2}\hat{h}_e\hat{H}_e^j \epsilon^{-1}
$$
but the $\hat{H}_e^j$ do not transform covariantly under gauge 
transformations.

Although it is slightly inconvenient to work with $\hat{H}_e$ in place of 
$\hat{p}^j_e$ since physical operators are more easily expressed in terms 
of the latter, it can be done and will be useful to prove abstract theorems.
In that respect it is worthwhile mentioning that one could also try to 
work directly with the unbounded operators but the general theory for 
this does not yet exist due to complications associated with the fact
that domains do not interact with the linear structure, efforts have, to the 
best of our knowledge, so far been restricted to the 
discussion of essential self-adjointness of (infinite) sums of operators 
restricted to one $[f]$-adic Hilbert space, where $[f]$ is a strong 
eqivalence class \cite{70a}.
\item[xiii)] The present framework could also be employed to
make contact with the so-called spin foam models \cite{23a,23b,23c,23d,23e}. 
These are a class of state sum models including the one that one 
obtains by studying the transition amplitudes associated with the 
Hamiltonian constraint constructed in \cite{17} and which since then have 
attracted a large amount of researchers. The procedure will be to
exploit the fact that coherent states are the most convenient
(over)complete bases to construct a formal Feynman path integral
(see e.g. \cite{72} and references therein). Certainly, a lot of work 
will be necessary to make that formal path integral rigorous but
coherent states provide a definite starting point. The coherent state
path integral should then be equivalent to a spin foam model which can
be considered as a path integral using ``momentum eigenstate bases''.
These issues will be worked out in \cite{73}.
\item[xiv)] Finally, the coherent states that we have constructed are 
coherent over a fixed graph only, they are {\it pure states}. 
However, our techniques readily combine 
with the random lattice approach developed in \cite{74} to produce 
{\it mixed coherent states}, that is, trace class operators on
$\cal H$ (in physical terms : density matrices). We can outline
some of the ideas already here : Given a density parameter
$\lambda$, an infinite volume cut-off parameter $r$ and a spatial metric 
$q_{ab}$, to be approximated by a mixed coherent state 
$\hat{\rho}^{s\lambda r}_{A^0,E^0}$,
we choose a number of $1\ll N^r<\infty$ points at random in $\Sigma^r$ 
where $\Sigma^r$ is a compact subset of $\Sigma$ which tends to $\Sigma$
as $r\to\infty$. This is done in such a way that the density of points as 
measured
by $q_{ab}$ is roughly constant and equal to $\lambda$. More precisely, 
a region $R\subset\Sigma^r$ is macroscopic if its volume satisfies 
$V_R(q)=\int_R d^3x \sqrt{\det(q(x))}\gg\ell_p^3$. Then we find 
approximately $N_R(q)=\lambda V_R(q)$ points inside this region
(it will be convenient to choose $\lambda=1/a^3$ where $a$ is the 
length parameter of equation (\ref{3.14})). We will also set 
$V^r:=V_{\Sigma^r}(q)$. It follows that 
\be \label{5.37}
d\mu_q^r(x):=\frac{\sqrt{\det(q)(x)}}{V^r}d^3x
\ee
is a probability measure on $\Sigma_r$ (the necessity for the cut-off $r$
is evident). The probability to find the $N^r=\lambda V^r$ points $p_k$ in the 
infinitesimal volumes $\sqrt{\det(q)(p_k)}d^3p_k$ is given by
$\prod_{k=1}^{N^r} d\mu_q^r(p_k)$. For each such random distribution of
points we can construct a four-valent lattice $\gamma^q_{p_1,..,p_{N^r}}$
by the generalized Dirichlet-Voronoi construction \cite{75} which
depends on $q$. Automatically, also a dual lattice is generated
which we can use for the polyhedronal decomposition of $\Sigma^r$ dual
to $\gamma^q_{p_1,..,p_{N^r}}$ and which goes into the definition 
of the momenta $p^e_j,\;e\in E(\gamma^q_{p_1,..,p_{N^r}})$. 
For this lattice, let 
$\hat{P}^{s\lambda r}_{\gamma^q_{p_1,..,p_{N^r}}}(A_0,E_0)$ be the 
one-dimensional projector on the coherent $C_0$-vector
$\xi^s_{\gamma^q_{p_1,..,p_{N^r}},(A_0,E_0)}$ as in  
(\ref{5.27}). Then, the task is to show that for the following operator,
(which is trace class at $r<\infty$),
\be \label{5.38}
\hat{\rho}^{s\lambda r}_{A^0,E^0}:=\int_{(\Sigma^r)^{N^r}}
\prod_{k=1}^{N^r} d\mu^q_r(x_k)
\hat{P}^{s\lambda r}_{\gamma^q_{p_1,..,p_{N^r}}}(A_0,E_0)
\ee
the limit $r\to\infty$ exists as a trace class operator 
on $\cal H$ which can presumably be proved by invoking inductive limit
methods. That (\ref{5.38}) is trace class at finite $r$ follows from
$$
\mbox{tr}[\hat{P}^{s\lambda r}_{\gamma^q_{p_1,..,p_{N^r}}}(A_0,E_0)]
=||\xi^s_{\gamma^q_{p_1,..,p_{N^r}},(A_0,E_0)}||^2=1
$$
by construction
so that actually $\mbox{tr}(\hat{\rho}^{s\lambda r}_{A^0,E^0})=1$
for all $r$ as it should be for a mixed state. This is a strong indication 
that the limit within the trace class ideal of the set of bounded operators 
exists. Practically, it might even be unnecessary to actually perform 
the limit as long as one measures only local operators : if the surfaces 
and paths with respect to which the operator is smeared lie within
$\Sigma_{r_0}$ then the measurement should be the same for all $r>r_0$.

This state is an average over a huge class of graphs and
should have an improved semi-classical behaviour as compared to the pure 
ones. Notice that it is here that the possibility to compute inner products 
between $C_0$ vectors over different graphs becomes important.
The details of this construction will appear in \cite{76}.
\end{itemize}
As the above list reveals there exists a plethora 
of fascinating and challenging open question and a huge 
programme is to be performed. In particular, the formalism is expectedly 
rather complicated as far as computations are concerned. The development
of approximation techniques and error estimates as outlined in \cite{32}
will become important. The coherent states together with the Infinite
Tensor Product beautifully combine three main research streams in 
general relativity :\\
A) Quantum Gravity, since these are states of a quantum theory of general
relativity,\\
B) Classical Mathematical General Relativity, since the states are labelled
by classical solutions of Einstein's equations and\\
C) Numerical Relativity, since the computations will need the help of 
{\it supercomputers}, the 
stage is prepared for {\it numerical canonical quantum general relativity}.
In fact, since the graphs that we are using are not too different from the
grids employed in numerical general relativity and lattice gauge theory,
some codes in classical numerical relativity or lattice gauge theory
might be easily adaptable 
to our purposes, although many new codes have to be written as well,
for instance a fast diagonalization code for the volume operator.\\
\\
Remark :\\
In \cite{77} the authors observe that the quantum fluctuations for the 
holonomy operator of a macroscopic loop, being the product of a large 
number of holonomies along ``plaquettes'' or elementary loops, are
always large and it seems that there is no state that can approximate 
such holonomy operators. First of all, this ``problem'' is not tied to,
say, lattice gauge theories but applies to any theory in which operators
that are  
products of a large (or infinite) number of elementary operators play a role.
Next, while the observation is certainly correct, given a large loop $\alpha$
on a graph $\gamma$ we can trade it for a single plaquette loop $\beta$
while keeping the number of holonomically independent loops constant.
With this relabelling of our degrees of freedom over $\gamma$ the loop
$\alpha$ is now elementary and we can write down a coherent state which
approximates it arbitrarily well. From the point of view of the 
ITP, while the coherent states with either $\alpha$ or $\beta$ considered as
elementary are defined over the same $\gamma$, they correspond to 
{\it different} regroupings of Hilbert spaces labelled by edges in the 
infinite tensor product. Thus, we see once more that the observation of 
\cite{77} is directly related to the fact that the associative law
is generally wrong for the infinite tensor product of Hilbert spaces.

\subsection{Dynamical Framework}
\label{s5.3}

So far our discussion has not touched the question whether it is possible 
to construct
coherent states which are at the same time physical, that is, annihilated by
the constraint operators in an appropriate (generalized) sense. At least 
with respect to the gauge -- and diffeomorphism constraint one might 
think that the answer should be given by the group averaging proposal
applied in \cite{7} to finite graphs. This section is intended to point 
out in which sense this can be carried over to infinite graphs. We first 
consider the averaging of general functions and after that averaging
of coherent states.

\subsubsection{Gauge Group Averaging}
\label{s5.3.1}

The following trivial example demonstrates that the group averaging proposal 
requires due modification in the ITP context already at the level of the 
Gauss constraint :\\
\\ 
Recall that 
the group $\cal G$ of local (generalized, i.e. without continuity 
requirements) gauge transformations 
$g:\;\Sigma\mapsto G;x\mapsto g(x)$ is unitarily represented on 
${\cal H}$ by extending its action on $C_0$-vectors over $\gamma$
\be \label{5.38a}
\hat{U}(g)\otimes_f=\otimes_{e\in E(\gamma)} f_e(g(e(0))h_e g(e(1))^{-1})
\ee
to a dense subset of ${\cal H}$ by linearity and to all of ${\cal H}$
by continuity for any $g$. 

Consider once more for $\gamma$ the $x$-axis in $\Rl^3$ subdivided into unit 
intervals $e_n=[n-1,n],\;n\in\Zl$. On this graph we can introduce the 
non-gauge invariant $C_0$-vector
\be \label{5.39}
\chi_\pi:=\prod_n \chi_\pi(h_{e_n})
\ee
where each edge carries the same irreducible representation $\pi$.
Group averaging this vector with respect to the Gauss constraint means to
compute the infinite dimensional integral
\be \label{5.40}
\eta_G\cdot \chi_j:=\prod_n \int_G d\mu_H(g(n)) 
\delta(g(-\infty),1)\delta(g(\infty),1)
\prod_m \chi_\pi(g(m-1)h_{e_m}g(m)^{-1})
\ee
where the $\delta$ distributions are due to the boundary condition 
that $g(\pm\infty)=1$. We consider (\ref{5.40}) as the limit as $N\to\infty$
of 
\be \label{5.41}
\eta^N_G\cdot \chi_j:=\prod_{n=-N}^N \int_G d\mu_H(g(n)) 
\delta(g(-N),1)\delta(g(N),1)
\prod_{m=-N+1}^N \chi_\pi(g(m-1)h_{e_m}g(m)^{-1})
\ee
which can be readily computed and gives 
\be \label{5.42}
\eta^N_G\cdot \chi_j=
\frac{\chi_j(h_{e_{-N+1}\circ..\circ e_N})}{d_\pi^{2N-1}}
\ee
Thus the norm of this vector is 
$||\eta^N_G\cdot \chi_j||=1/d_\pi^{2N-1}$ and so (\ref{5.40}) vanishes unless
$d_\pi=1$. 

In order to cure this we must obviously factor out the power of $d_\pi$.
This can be achieved by requiring that group averaging should produce 
a norm one vector from a norm one vector, that is, we propose
(see \cite{32})
\be \label{5.43}
\chi_G\cdot f=
\frac{\prod_{v\in V(\gamma)} \int_G d\mu_H(g_v)\prod_{e\in E(\gamma)} 
f_e(g(e(0)) h_e g(e(1))^{-1})}{||\prod_{v\in V(\gamma)} \int_G d\mu_H(g_v)
\prod_{e\in E(\gamma)} f_e(g(e(0)) h_e g(e(1))^{-1})||}
\ee
where one makes sense of the formally zero numerator and denominator 
through a limiting procedure as outlined above. (One does not need to check 
that the result is independent of the way one performs the limit, if one 
gets different gauge invariant answers one simply gets different gauge 
invariant vectors
which is all that we want from the group averaging machinery anyway for 
the case of the gauge group since, due to its finiteness, we can still
use the extended Ashtekar-Lewandowski measure on group averaged states). 
This makes group averaging
a non-(anti)linear procedure. It means, in particular, that we produce 
completely new Infinite Tensor Product Hilbert Spaces. Namely, in the 
case of the example the procedure (\ref{5.43}) gives us the gauge invariant
vector $\Xi_j=\chi_j(h_e)$ where $e$ is the $x$-axis, the prototype
of a tangle \cite{22}. So in this case the original graph $\gamma$ with 
its countable number of edges has collapsed to a graph with a single edge,
a finite tensor product Hilbet space. Following definition \ref{def5.2}
to compute ITP inner products for $C_0$ vectors over different graphs
we see that the scalar product between $\chi_j$ and $\Xi_j$ vanishes,
the averaged and unaveraged vectors are orthogonal to each other.

That this happens is not an accident but generic : Consider a graph
$\gamma$ which is the union of an infinite number of mutually 
disjoint, finite graphs
$\gamma_n,\;n=1,2,..$. Then a $C_0$ vector over $\gamma$ is of the form
$$
f=\otimes_n [\otimes_{e\in E(\gamma_n)} f_e],\;||f_e||=1
$$
and defines an element of ${\cal H}_\gamma=\otimes_{e\in E(\gamma)}
{\cal H}_e$.
Group averaging evidently turns this $C_0$-vector into a new $C_0$-vector
of the form
$$
\eta_G\cdot f=\otimes_n f_{\gamma_n},\;||f_{\gamma_n}||=1
$$
which now is an element of ${\cal H}'_\gamma=\otimes_n {\cal H}_{\gamma_n}$.
This once more demonstrates the source of the trouble : the 
associative law does not hold on the ITP and the latter vector simply
cannot be written, in general, as a linear combinations of vectors of the 
former Hilbert space.

\subsubsection{Diffeomorphism Group Averaging}
\label{s5.3.2}

Next we turn to group averaging with respect to the diffeomorphism
constraint. Recall \cite{7,23} that this is done by relying explicitly on 
the spin-network basis. This is necessary because only if a function 
cylindrical over a graph $\gamma$ depends on
each of its edges through {\it non-trivial} irreducible representations
does group averaging over the diffeomorphism group produce a well-defined 
distribution, the complication being due to the infinite volume of the 
diffeomorphism group with respect to the ``counting measure'' which
produces a singularity each time we sum over diffeomorphisms which 
do not modify the graph on which a function depends. Another complication
associated with so-called graph symmetries can be satisfactorily dealt with,
see \cite{23} for details.

However, as we have seen in section \ref{s5.1.3} spin-network
functions do not provide a basis in the ITP context. It follows that 
not all functions of the ITP Hilbert space can be group averaged
with respect to the diffeomorphism group.

More precisely, recall that  
the group Diff$(\Sigma)$ of analyticity preserving diffeomorphisms 
of $\Sigma$ is unitarily represented on 
${\cal H}$ by extending its action on $C_0$-vectors over $\gamma$
\be \label{5.44}
\hat{U}(\varphi)\otimes_f=\otimes_{e\in E(\gamma)} f_e(h_{\varphi(e)})
\ee
to a dense subset of ${\cal H}$ by linearity and to all of ${\cal H}$
by continuity for any $\varphi\in\mbox{Diff}(\Sigma)$. Given a $C_0$ 
sequence $f$ we 
define its orbit $\{f\}$ to be the set of $C_0$ sequences given by 
\be \label{5.45}
\{f\}=\{f';\;\exists \;\varphi\in\mbox{Diff}(\Sigma)\;\ni\;
\otimes_{f'}=\hat{U}(\varphi)\otimes_f \}
\ee
The following $C_0$ sequences lie in the range of the group average map.
\begin{Definition} \label{def5.4}
A $C_0$ sequence $f$ is called a {\it spin-network $C_0$ sequence} over 
$\gamma$ if and only if $<1,f_e>=0$ for all $e\in E(\gamma)$. A 
spin-network $C_0$ sequence is called finite if its graph symmetry group
is finite. For finite spin-network $C_0$ sequences we define 
the group average of its associated $C_0$-vector with respect to the 
diffeomorphism group by 
\be \label{5.46}
\eta_{Diff}\cdot\otimes_f:=[\otimes_f]:=\sum_{f'\in\{f\}}\otimes_{f'}
\ee
where we have assumed that the graph symmetry group of $\gamma$ is trivial,
otherwise we modify the procedure as in \cite{23} or \cite{39}. 
The object (\ref{5.46}) lies in $\Phi'$, the topological dual of
$\Phi$ as follows from results of \cite{39}.
\end{Definition}
Graphs with infinite graph symmetry group are 
excluded from the domain of the average map, similar as in \cite{39}. Notice 
that for the typical graphs that we have in mind (e.g. cubic lattices) the 
graph symmetry group is in fact infinite due to the infinite number of 
translations which leave the graph invariant, but in order to cure this
it is enough to replace a single edge by a kink.
With this problem out of the way, this defines $\eta_{Diff}$ on finite
spin-network $C_0$ vectors over typical lattices and can be extended
by linearity to finite linear combinations of those. 
That this indeed defines a linear operation is granted due to our 
treatment of graph symmetries.

\subsubsection{Averaging of Coherent States}
\label{s5.3.3}

The interesting question is, of course, whether the coherent states 
that we defined are in the domain of the average map. \\
\\
A) Gauge group averaging.\\
Returning to the example graph already 
discussed in equation (\ref{5.39})
above, consider the (non-normalized) coherent state over the graph with $2N$ 
adjacent unit intervals as edges symmetrically around the origin along the 
$x$-axis, that is,
\be \label{5.47}
\psi^s_{g_N}(A):=\otimes_{n=-N+1}^N \psi^s_{g_{e_n}}(h_{e_n}(A))
\ee
where $\psi^s_g$ was defined in (\ref{3.1}). Under a gauge transformation,
represented by the unitary operator $\hat{U}(g)$,
the tensor product factor with label $n$ is transformed into
$$
\psi^s_{g_{e_n}}(g(e_n(0))h_{e_n}g(e_n(1))^{-1})
=\psi^s_{g(e(0))^{-1}g_{e_n}g(e_n(1))}(h_{e_n})
$$
and integrating over $g(e_n(1)),n=-N+1,..,N-1$ with the Haar measure
produces the state
\be \label{5.48}
\Psi^s_{g_N}(A):=\psi^{2Ns}_{g_{e_N}}(h_{e_N}(A))
\ee
where $e_N=e_{-N+1}\circ..\circ e_N$ and 
$g_{e_N}=g_{e_{-N+1}}..g_{e_N},\;h_{e_N}=h_{e_{-N+1}}..h_{e_N}$.
In other words, the finite number of integrations produce a coherent state
with the correct dependence on $h_e, g_e$, however, the classicality
parameter gets augmented from $s$ to $2Ns$ which in the limit $N\to\infty$,
of course, does not show any classical behaviour any longer. Thus,
in order to produce a gauge invariant coherent state form a non-gauge
invariant one on the ITP by group averaging not only do we have to go through 
a limiting procedure as $N\to\infty$ as already discussed above with an 
associated ``renormalization'' of the norms of the vectors before and after 
averaging, but also one has to rescale the classicality parameter $s$ 
appropriately.

Thus, gauge group averaging becomes very difficult to perform if the graph 
$\gamma$ has at least one infinite connected component. At the opposite
extreme are the infinite cluster graphs which 
are infinite graphs obtained by the infinite disjoint union of finite 
graphs, called clusters. Obviously, each of these finite graphs can be gauge 
group 
averaged (and renormalized) individually. In particular, if all clusters 
are diffeomorphic then group averaging reduces to the infinite repetition
of one averaging for functions over one finite graph. An example to keep
in mind is a cubic lattice in which we remove some edges to obtain
disjoint finite cubic sub-lattices. \\
\\
B) Diffeomorphism group averaging.\\
Recall (see the first reference in \cite{32})
$$
\sqrt{\frac{\sinh(p^e)}{2\sqrt{\pi}p^e}s^{3/2}(1-K_s)}
\le <1,\xi^s_{g_e}>=1/||\psi^s_{g_e}||
\le \sqrt{\frac{\sinh(p^e)}{2\sqrt{\pi}p^e}s^{3/2}(1+K'_s)}
$$
where the constants $K_s,K'_s$ tend to zero exponentially fast with $s\to 0$.
Since
$p^e$ is bounded, tending to zero for ever and ever finer lattice
at least for classical configurations we see that for sufficiently fine 
lattices at given (small) $s$ we have not only\\ 
$|<1,\xi^s_{g_e}>|<1$
as granted by the Schwarz inequality but moreover that there exist
numbers $0<q,q'<1$ with $q<c_e:=|<1,\xi^s_{g_e}>|\le q'$ for all $e$ for 
sufficiently 
fine lattices which is precisely the application that we are aiming at.

Splitting $\xi^s_{g_e}=\delta\xi^s_{g_e}+c_e\cdot 1$ we may want to write 
for given $\gamma\in\Gamma^\omega_\sigma$ the state 
\be \label{5.49}
\xi^s_{\gamma g_\gamma}:=\otimes_{e\in E(\gamma)} \xi^s_{g_e}
\ee
as
\be \label{5.50}
\xi^s_{\gamma g_\gamma}=
\sum_{N=0}^\infty \; \sum_{\{e_1,..,e_N\}\subset E(\gamma)} \;
[\prod_{k=1}^N c_{e_k} ]
[\otimes_{e\in E(\gamma)-\{e_1,..,e_N\}} \delta\xi^s_{g_e}]
\ee
or as 
\be \label{5.51}
\xi^s_{\gamma g_\gamma}=
\sum_{N=0}^\infty \; \sum_{\{e_1,..,e_N\}\subset E(\gamma)} \;
[\prod_{e\in E(\gamma)-\{e_1,..,e_N\}} c_e]
[\otimes_{k=1}^N \delta\xi^s_{g_{e_k}}]
\ee
However, both attempts fail since in (\ref{5.50}) all appearing vectors
have zero norm (in fact $||\delta\xi^s_{g_e}||^2=1-c_e^2<1-q^2<1$)
and in (\ref{5.51}) all coefficients vanish identically.
Thus, the vector (\ref{5.49}) does not lie in the domain of the average 
map.

A substitute for averaging and to deal with the diffeomorphism group is to 
work with 
representatives, i.e. from each diffeomorphism class $\{f\}$ we choose an 
element $f^0_{\{f\}}$. Thus $f^0:\;\{f\}\mapsto f^0_{\{f\}}$ is a choice 
function, 
its existence being granted by the lemma of choice. We specify this choice 
function further by choosing from each graph diffeomorphism 
class $\{\gamma\}$ a representative $\gamma^0_{\{\gamma\}}$. Given a function
$f$, let $\gamma_f$ be the minimal graph on which it depends non-trivially.
Then $f^0_{\{f\}}$ can be chosen to depend on $\gamma^0_{\{\gamma_f\}}$.
If $\gamma_f$ has graph symmetries then this prescription does not yet fix
$f^0_{\{f\}}$ uniquely and we must further choose from one of the 
$\hat{U}(\varphi_n)f^0_{\{f\}}$ where $\varphi_n$ is a symmetry of
$\gamma^0_{\{\gamma_f\}}$. A kind of group averaging map is now defined by
$\eta_{Diff}\circ f:=f^0_{\{f\}}$ which obviously satisfies 
$\eta_{Diff}\circ \hat{U}(\varphi)=\eta_{Diff}$ and the inner 
product on these ``solutions to the diffeomorphism constraint'' is just
the usual inner product between representatives. This makes the whole 
proposal unfortunately very choice dependent and thus less attractive. 
Notice, however, that 
diffeomorphism invariant operators which are defined on the kinematical
Hilbert space obviously keep their adjointness properties.  

A different way to deal with
diffeomorphism invariance is by gauge fixing 
(alternatively, one has to construct gauge and diffeomorphism invariant
coherent states from scratch) : \\
Given a collection $g_\gamma=\{g_e\}_{e\in E(\gamma)}$, a local gauge 
transformation $g\in {\cal G}$ and a diffeomorphism 
$\varphi\in \mbox{Diff}(\Sigma)$ we define 
$g^g_\gamma:=\{g^g_e\}_{e\in E(\gamma)}$ with 
$g^g_e:=g(e(0))^{-1} g_e g(e(1))$ and 
$g^\varphi_\gamma:=\{g^\varphi_e\}_{e\in E(\gamma)}$ with 
$g^\varphi_e:=g_{\varphi^{-1}(e)}$. It is then easy to see, using unitarity
(invariance of norms) that  
\be \label{5.52}
\hat{U}(g)\psi^s_{\gamma g_\gamma}=\psi^s_{\gamma g^g_\gamma}
\mbox{ and }
\hat{U}(\varphi)\psi^s_{\gamma g_\gamma}=
\psi^s_{\varphi(\gamma) g^\varphi_{\varphi(\gamma)}}
\ee
Given classical initial value data $(A^0,E^0)$ in a certain gauge the 
$g_\gamma=g_\gamma((A^0,E^0))$ are fixed and we require that
$g_\gamma^g=g_\gamma=g^\varphi_\gamma$ for all $\gamma$ which (generically) 
trivializes the residual gauge freedom to $g=1,\varphi=$id. 

Thus, as far as the gauge and diffeomorphism constraints are concerned, we
can fix a gauge to take care of gauge and diffeomorphism invariance. 
The issue lies much harder with respect to the Hamiltonian constraint 
because its action \cite{17} is much more complicated than the action
of the kinematical constraints, and almost no Hamiltonian invariant 
observables are known with respect to which one would need to construct
the invariant coherent states. Fortunately, there are certain ``simple''
solutions to the Hamiltonian constraint \cite{17} corresponding to states 
whose underlying graph is out of the range of graphs that the Hamiltonian 
constraint produces. If we build (non-distributional) coherent states
on such graphs, then they lie in the kernel of the Hamiltonian constraint
in the sense of generalized eigenvectors with eigenvalue zero.
Thus, at least for these simple solutions, together with fixing of
gauge and diffeomorphism freedom, we can incorporate the quantum dynamics
of general relativity.\\
\\
\\
All these observations reveal that group averaging non-gauge and/or
non-diffeomorphism invariant coherent states over the gauge or
diffeomorphism group is a non-trivial task, at least not if $\Sigma$ is 
non-compact and applied naively without some sort of renormalization 
leads to meaningless results. More work is needed in order to construct 
rigorous solutions to all constraints which at the same time behave 
semi-classically. However, at the moment we are not so much interested
in obtaining semi-classical solutions to all constraints. Rather,
besides the applications already mentioned in section \ref{s5.2},
it is of paramount importance to test the consistency of a quantum 
representation of the classical constraint algebra and the verification
of its correct classical limit \cite{66}. 
In order to do this one obviously 
{\it must not} have semi-classical states which solve the constraints.\\
\\
\\
\\
{\large Acknowledgements}\\
\\
O. W. thanks the ``Studienstiftung des Deutschen Volkes'' for financial
support.

\end{document}